\documentclass[a4paper,11pt]{article}
\pdfoutput=1
\usepackage{jheppub} 
\usepackage{amsmath}
 \usepackage{amssymb}
 \usepackage{slashed}
\usepackage{dcolumn}
\usepackage{bm}
\usepackage{color}
\usepackage{multirow}

\allowdisplaybreaks

 \newcommand{\lsim}{{\;\raise0.3ex\hbox{$<$\kern-0.75em\raise-1.1ex\hbox{$\sim$}}\;}}
\newcommand{\gsim}{{\;\raise0.3ex\hbox{$>$\kern-0.75em\raise-1.1ex\hbox{$\sim$}}\;}}
\def\bea{\begin{eqnarray}}
\def\eea{\end{eqnarray}}
\def\bec{\begin{center}}
\def\ec{\end{center}}

\def\beq{\begin{equation}}
\def\eeq{\end{equation}}

\def\bea{\begin{eqnarray}}
\def\eea{\end{eqnarray}}
\def\beq#1\eeq{\begin{align}#1\end{align}}
\def\beqnn#1\eeq{\begin{align*}#1\end{align*}}
\def\ba{\begin{array}}
\def\ea{\end{array}}
\def\bc{\begin{center}}
\def\ec{\end{center}}

\title{Axion Quality in Warped Extra-Dimension}
\author[a]{Kiwoon Choi}
\author[b]{Chang Hyeon Lee}
\author[c,a,d]{Chang Sub Shin}
\affiliation[a]{Particle Theory and Cosmology Group, Center for Theoretical Physics of the Universe,\\
  Institute for Basic Science (IBS), Daejeon 34126, South Korea}
\affiliation[b]{Department of Physics and Institute of Quantum Systems,
Chungnam National University, Daejeon 34134, Korea}
\affiliation[c]{Department of Physics and Institute for Sciences of the Universe, Chungnam National University, Daejeon 34134, Korea}
\affiliation[d]{School of Physics, Korea Institute for Advanced Study, Seoul 02455, Korea}

\emailAdd{kchoi@ibs.re.kr}
\emailAdd{changhyeonlee@cnu.ac.kr}
\emailAdd{csshin@cnu.ac.kr}

 \preprint{CTPU-PTC-26-10}

\abstract{We investigate the axion quality problem in warped extra-dimensional models in which the QCD axion arises as the Wilson-line mode of a five-dimensional $U(1)$ gauge field compactified on an $S^1/\mathbb{Z}_2$ orbifold. 
Higher-dimensional gauge invariance severely constrains possible sources of Peccei--Quinn symmetry breaking, implying that non-QCD contributions to the axion potential are predominantly generated by nonlocal effects mediated by $U(1)$-charged fields propagating along the compact dimension. 
We systematically compute these contributions and examine how both the warped geometry and the orbifold fixed points (branes) affect the resulting axion quality. 
Finally, we classify the parametric suppression of the induced axion potential, thereby identifying the conditions under which warped extra-dimensional axions can achieve sufficiently high quality.
}

\vspace{4cm}

\usepackage{placeins}
\begin{document} 
\maketitle
\tableofcontents
\flushbottom
\renewcommand\thesection{\arabic{section}}

\section{Introduction}

The strong CP problem arises because the Standard Model allows a P- and CP-violating topological term in QCD \cite{Kim:2008hd,DiLuzio:2020wdo,Choi:2020rgn},
\begin{equation}
\Delta{\cal L}_{\rm QCD}\;=\;\frac{1}{32\pi^2}\,\bar\theta\, G^a_{\mu\nu}\tilde G^{a\,\mu\nu},
\label{eq:theta_term}
\end{equation}
where $G^a_{\mu\nu}$ and $\tilde G^a_{\mu\nu}$ are the gluon field strength and its dual, respectively.
The non-observation of the neutron electric dipole moment, however, implies the stringent bound $|\bar\theta|\lesssim 10^{-10}$.
 A compelling resolution of this puzzle is provided by the Peccei--Quinn (PQ) mechanism, which introduces a nonlinearly realized global $U(1)$ symmetry that is broken predominantly by the QCD anomaly~\cite{Peccei:1977hh,Peccei:1977ur}  . The associated pseudo-Goldstone boson field, the axion field $a(x)$, dynamically relaxes $\bar\theta$ to a value consistent with experimental bounds~\cite{Weinberg:1977ma,Wilczek:1977pj,Kim:1979if,Shifman:1979if,Dine:1981rt,Zhitnitsky:1980tq}.

At scales just above the QCD scale, the PQ symmetry can be realized 
as a shift symmetry of the axion field,
 \begin{equation}
 U(1)_{\rm PQ}:\,\,\, a(x) \,\rightarrow \, a(x) +{\rm constant}.\end{equation}
 The corresponding low energy effective lagrangian takes the form  
\begin{equation}
{\cal L}_{\rm axion}
=\frac12 \partial^\mu a\partial_\mu a +\partial_\mu a J^\mu+ \frac{N}{32\pi^2}\frac{a}{f_a}G^{a\mu\nu}\tilde G^a_{\mu\nu}
\;+\;V_{\rm HE}(a)\,,
\end{equation}
where $f_a$ is the axion decay constant defining the compact axion field range $a \equiv a+2\pi f_a$. Here $J^\mu$ is a model-dependent current built from light quarks and leptons, $N$ is a
nonzero integer characterizing  the breaking of the PQ symmetry by the QCD anomaly (often called the domain-wall number), and $V_{\rm HE}(a)$ represents the axion potential arising from PQ-breaking effects \emph{other than} the QCD anomaly. 
By a constant shift of the axion field, one may choose a field basis in which  the effective parameter $\bar\theta$ in Eq.~(\ref{eq:theta_term}) is identified as $\bar\theta =N\langle a\rangle/f_a$. In this basis,
nonperturbative QCD effects generate an axion potential through the
axion-gluon coupling $aG\tilde G$, 
whose global minimum occurs at $N\langle a\rangle/f_a=0$  \cite{Kim:2008hd,DiLuzio:2020wdo,Choi:2020rgn}:
\bea
V_{\rm QCD}(a) \simeq
-\frac{m_\pi^2 f_\pi^2}{m_u+m_d}\sqrt{m_u^2+ m_d^2+2m_um_d \cos (Na/f_a)},
\label{eq:qcd_axion_potential}\eea
where $m_\pi$ and $f_\pi$ denote the pion mass and decay constant, respectively, and $m_{u}$ and $m_d$ are the up and down quark masses.
On the other hand, the non-QCD axion potential $V_{\rm HE}$ is generically expected to  have a minimum at $a/f_a={\cal O}(1)$.
 Therefore,
the axion solution to the strong CP problem works naturally \emph{only if} $V_{\rm HE}$ is sufficiently suppressed compared to the QCD-induced potential $V_{\rm QCD}$, 
more specifically, \bea
V_{\rm HE}\lesssim 10^{-10}m_\pi^2 f_\pi^2.\eea 
This requirement gives rise to the so-called axion quality problem, which is sharpened by the expectation that quantum gravity generically violates global symmetries and can therefore generate an excessively large $V_{\rm HE}$~\cite{Kim:1988ix,Rey:1989mg,Barr:1992qq,Kamionkowski:1992mf,Holman:1992us,Kallosh:1995hi,Hebecker:2018ofv}.

One approach to addressing the axion quality problem is provided by composite axion models, which were originally proposed to generate the axion scale $f_a$ dynamically~\cite{Kim:1984pt,Choi:1985cb}.  Composite axion models can be endowed with
additional structure~\cite{Randall:1992ut,Redi:2016esr,Lillard:2017cwx,Gavela:2018paw,Bigazzi:2019hav,Ardu:2020fck,Contino:2021udf,Cox:2023lcv,Gherghetta:2025uds,Gherghetta:2025kff,Agrawal:2025mke,Azatov:2025mep} such that the PQ symmetry emerges as an accidental symmetry of sufficiently high quality,  ensuring $V_{\rm HE}\lesssim 10^{-10}m_\pi^2 f_\pi^2$.\footnote{See \cite{Babu:2002ic,Lee:2011dya,Harigaya:2013vwa,Bhattiprolu:2021vdu,Choi:2022jqy,Burgess:2023dow,Babu:2024qzb,Babu:2026yqp,Csaki:2026qjl} for other approaches to realizing a high-quality QCD axion.}  However, the required additional structure is typically quite involved, which  significantly undermines the simplicity and minimality of the model.

Extra-dimensional axions (EDA), originally studied in the context of string theory~\cite{Witten:1984dg,Choi:1985je,Barr:1985hk}, provide a particularly robust realization of a high-quality axion~\cite{Choi:2003wr,Reece:2024qv,Craig:2024cqs}.
In these constructions, the axion arises as the zero mode of a higher-dimensional 
$p$-form gauge field (in the minimal 5D realization, an Abelian 1-form gauge field)~\cite{Svrcek:2006yi,Benabou:2023npn,Reece:2024qv}.
The associated Peccei--Quinn (PQ) symmetry can be understood as a 
$p$-form symmetry generated by a harmonic 
$p$-form on the compact internal space.
Because this PQ symmetry is closely tied to higher-dimensional gauge invariance, possible sources of PQ breaking are highly constrained. 
Once certain discrete parameters of the model, such as the coefficients of potentially dangerous Chern--Simons terms, monodromy terms, or St\"uckelberg mixing terms, are set to vanish, the non-QCD axion potential $V_{\rm HE}$ arises predominantly from nonlocal effects mediated by $(p-1)$-dimensional objects (i.e., $(p-1)$-branes) that couple to the associated $p$-form gauge field~\cite{Choi:2003wr,Reece:2024qv,Craig:2024cqs}.
This provides a physical mechanism for an \emph{exponential} suppression of $V_{\rm HE}$, controlled primarily by the tension of the charged $(p-1)$-branes and the size of the extra dimension, rather than by generic local UV operators\footnote{This mechanism can be generalized to more complicated scenarios in which the physical axion emerges as a linear combination of the extra-dimensional component of a higher-dimensional gauge field and the field-theoretic axion originating from the phase of a charged scalar field~\cite{Cheng:2001ys,Honecker:2013mya,Choi:2014uaa,Buchbinder:2014qca,Petrossian-Byrne:2025mto,Loladze:2025uvf}.}.

The minimal setup realizing an EDA is a 5D model with a $U(1)$ gauge field $C_M$, whose Wilson-line phase along the compact fifth dimension corresponds to a 4D axion~\cite{ArkaniHamed:2003wu,Choi:2003wr},
\begin{equation}
\frac{a}{f_a} \equiv \oint dy\, C_5\,,
\label{eq:Wilson_line}
\end{equation}
where $y$ denotes the coordinate of the fifth dimension.
A realistic 5D model accommodating chiral fermions typically requires orbifold compactification on $S^1/\mathbb{Z}_2$ (or a related generalization), rather than a simple circle~\cite{Cheng:2010pt}. This step introduces additional model-building
ingredients that are directly relevant for the axion quality:
(i) charged bulk matter fields are subject to appropriate $\mathbb{Z}_2$
boundary conditions, and
(ii) localized interactions can appear at the orbifold fixed points (or
branes) and may participate in generating PQ-breaking effects.
In other words, although  it is well known that EDA provides an attractive route to a high-quality axion,
the detailed structure of the non-QCD axion potential   $V_{\rm HE}$ and its parametric suppression in orbifold compactifications  are  more nontrivial and deserve a systematic study.

In this work we study the axion quality of the Wilson-line axion 
Eq.~(\ref{eq:Wilson_line}) 
in a \emph{warped}  background geometry on $S^1/\mathbb{Z}_2$~\cite{Randall:1999ee}.
The warped geometry is motivated by two considerations.
First, warping naturally 
yields an exponentially suppressed $f_a$ relative to the 4D Planck scale,
realizing an intermediate axion scale without contrived parameter choices~\cite{Choi:2003wr,Flacke:2006ad,Choi:2025wog}.
Second, and central to this work, warping can enhance the exponential suppression of nonlocal PQ-violating effects,
thereby improving the axion quality~\cite{Cox:2019rro}.

A key feature of the Wilson-line axion in Eq.~(\ref{eq:Wilson_line}) is that
the non-QCD axion potential $V_{\rm HE}$ arises predominantly from
$U(1)_C$-charged particles propagating along the extra dimension while coupled
to the axion field encoded in the background gauge field $C_5$.
In this setup, $V_{\rm HE}$ receives two leading classes of contributions:
(i) loop-induced contributions, which can be interpreted as arising from closed
 worldlines of charged particles winding around the covering circle
$S^1$; and
(ii) tree-level contributions associated with open worldlines whose
endpoints lie at distinguished spacetime points, with the worldlines either
stretching across the orbifold interval $S^1/\mathbb{Z}_2$ or winding around
the covering circle $S^1$.

Closed worldline configurations winding around $S^1$ are allowed for generic 5D
fields. For the associated loop-induced potential, a central ingredient is the
$\mathbb{Z}_2$ boundary conditions (BCs) imposed on the $U(1)_C$-charged matter
fields, which admit a variety of distinct possibilities. In this paper, we
focus on three relatively simple classes of BCs yielding different
$\mathbb{Z}_2$-parity structures of the $U(1)_C$ gauge couplings and bulk
masses on the covering circle $S^1$: the ordinary parity-type (P-type) BC and
the charge-conjugation-twisted (C-twisted) BC, each imposed on a single 5D
scalar or fermion field, and a C-twisted hypermultiplet BC, which involves
both charge conjugation and an exchange between two scalar fields
$H_\alpha$ $(\alpha=1,2)$ corresponding to the scalar sector of a
C-twisted 5D supersymmetric hypermultiplet.

Since the gauge group $U(1)_C$ is generated by $\mathbb{Z}_2$-odd gauge
parameters on $S^1$, the $U(1)_C$ gauge couplings of ordinary P-type matter
fields are $\mathbb{Z}_2$-odd on $S^1$, whereas those of C-twisted matter
fields are $\mathbb{Z}_2$-even and therefore constant over $S^1$. The bulk
masses of P-type and C-twisted scalars, as well as those of C-twisted
fermions, are $\mathbb{Z}_2$-even and constant over $S^1$, whereas the bulk
masses of P-type fermions are $\mathbb{Z}_2$-odd. The bulk masses of
C-twisted hypermultiplet scalars generically contain both $\mathbb{Z}_2$-even
and $\mathbb{Z}_2$-odd components, and are therefore piecewise constant over
$S^1$.

As we will see, in the absence of $U(1)_C$-violating fixed-point operators,
only winding configurations of $U(1)_C$-charged bulk scalar and fermion fields
with C-twisted BCs generate an axion potential. This follows from the
orbifold parity properties of the $U(1)_C$ gauge interactions. In the
large-mass and strong-warping regime, the resulting non-QCD axion potential
$V_{\rm HE}$ is exponentially suppressed as
\bea
V_{\rm HE}\, \propto\,
e^{-4k\pi R}\,e^{-2M_{\rm eff}\pi R}
\cos\left(q\theta\right),
\eea
up to a systematically calculable prefactor, where $R$ denotes the radius
of $S^1$, $k$ is the AdS curvature scale, $q$ is the $U(1)_C$ charge of the
bulk field involved.
The effective mass $M_{\rm eff}$ appearing in the exponent depends on the
bulk mass, the BCs of the corresponding 5D field, and the AdS curvature scale:
\[
M_{\rm eff}(\phi)=\sqrt{M^2+4k^2}
\]
for a C-twisted scalar field $\phi$ with constant bulk mass $M$,
\[
M_{\rm eff}(\psi)=M
\]
for a C-twisted fermion $\psi$ with constant bulk mass $M$, and
\[
M_{\rm eff}(H_\alpha)=\frac{1}{2}\left(m_++m_-\right),
\qquad
m_\pm=\sqrt{M_0^2\pm\mu^2+4k^2},
\]
for C-twisted hypermultiplet scalar fields $H_\alpha$ $(\alpha=1,2)$ with
piecewise constant bulk masses
\[
M_1^2=M_0^2+\mu^2\epsilon(y),
\qquad
M_2^2=M_0^2-\mu^2\epsilon(y),
\]
where $\epsilon(y)$ is the $\mathbb{Z}_2$-odd sign function on $S^1$, defined
as $\epsilon(y)=1$ for $0<y<\pi R$ and $\epsilon(y)=-1$ for
$\pi R<y<2\pi R$.

\begingroup
By contrast, the P-type scalar $\tilde\phi$ acquires a loop-induced axion potential only in the presence of $U(1)_C$-violating fixed-point mass terms $\tilde b_i\tilde\phi^2$ $(i=0,\pi)$ at $y=0$ and $y=\pi R$.  The resulting contribution is exponentially suppressed as
\bea
V_{\rm HE}
\;\propto\;
|\tilde b_0\tilde b_\pi|\,
e^{-4k\pi R}
e^{-2M_{\rm eff}(\tilde\phi)\pi R},
\qquad
M_{\rm eff}(\tilde\phi)=\sqrt{M^2+4k^2},
\eea
up to the calculable prefactor discussed in Sec.~\ref{subsec:quadratic_loop}.  Analogous fixed-point operators can affect P-type fermions, but their explicit contributions are not analyzed in this work.
\endgroup

\begingroup
Tree-level contributions arise from charged scalar profiles sourced by linear
operators localized at the orbifold fixed points,
$J_i\phi+\tilde J_i\tilde\phi$ $(i=0,\pi)$.
For even-integer $U(1)_C$ charge $q$, such linear operators are allowed at
both fixed points, generating an axion potential associated with open
worldlines of charged scalar fields stretched across the orbifold interval
$S^1/\mathbb{Z}_2$. In the large-mass and strong-warping regime, the leading
contribution is
\bea
V_{\rm HE}
&\propto&
|J_0J_\pi|
e^{-2k\pi R}e^{-M_{\rm eff}\pi R}
\cos \left(\frac{q\theta}{2}\right)
\nonumber \\
&{\rm or}&\,
|\tilde J_0\tilde J_\pi|
e^{-2k\pi R}e^{-M_{\rm eff}\pi R}
\cos\left(\frac{q\theta}{2}-\delta_J\right),
\eea
where $M_{\rm eff}=\sqrt{M^2+4k^2}$ and
$\delta_J={\rm arg}(\tilde J_0\tilde J_\pi)$.
For a C-twisted scalar, the same classical solution also contains
contributions associated with open worldlines whose endpoints lie on the
same fixed point, at either $y=0$ or $y=\pi R$, while winding around the
covering circle $S^1$. These contributions are proportional to
$J_0^2 e^{-2M_{\rm eff}\pi R}\cos(q\theta)$ or
$J_\pi^2 e^{-4k\pi R}e^{-2M_{\rm eff}\pi R}\cos(q\theta)$,
respectively. They are subleading when localized sources are present at both
fixed points with generic coefficients, but become the leading tree-level
contribution when the source at the opposite fixed point is absent. For a
P-type scalar, by contrast, a tree-level $\theta$ dependence requires
sources at both fixed points.
\endgroup

\begingroup
Beyond the leading exponential suppressions summarized above, we determine the
corresponding prefactors and examine the approximations involved. We compute
the loop-induced contributions from the C-twisted scalar $\phi$ and fermion
$\psi$ using three different methods. In the large-mass limit, we obtain
analytic expressions for the leading terms with a Gaussian approximation to the
four-dimensional momentum integral, and these expressions are identical in all
three methods. We test this approximation by evaluating the exact integral numerically in the
Kaluza--Klein (KK) spectral-function approach, and find that the approximation
captures the exponential suppression and the scaling of the prefactor.
\endgroup

The organization of this paper is as follows.
In Section~2, we review the warped extra-dimensional axion in a 5D model compactified on
$S^1/\mathbb{Z}_2$, deriving the axion zero-mode profile and its decay constant.
We also introduce charged matter fields and classify the admissible orbifold boundary
conditions.
In Section~3, we compute the loop-induced axion potential in the absence of fixed-point
localized operators using both the worldline and KK spectral-function
approaches, with a monodromy-matrix-based treatment for fermions included for completeness.
In Section~4, we incorporate the relevant fixed-point localized operators and analyze the
additional contributions to the axion potential.
Section~5 contains the conclusions, together with a brief summary of the parametric scaling
of the various contributions to the axion potential discussed in the previous sections.
Some of the computational details are provided in the Appendices.

\section{Warped Extra-Dimensional Axion Model}
\label{sec:eda_model}

In this section, we present a 5D warped extra-dimensional axion model on $S^1/\mathbb{Z}_2$  within the standard 
Randall--Sundrum (RS) construction~\cite{Randall:1999ee}.
We keep the discussion compact, focusing on the ingredients needed for the subsequent analysis of the induced axion potential.

\subsection{Background geometry}

\begin{figure}[h]
	\centering
	{\includegraphics[width=0.48\textwidth]{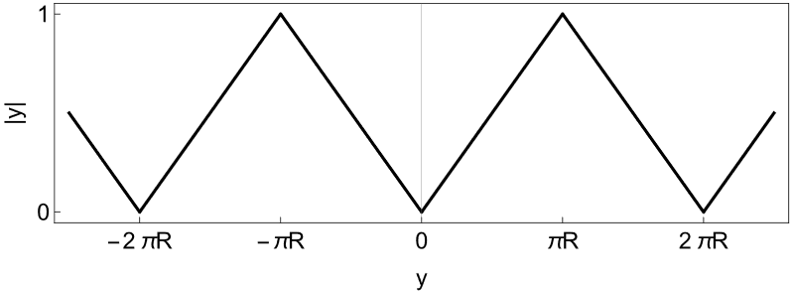}} \\ 
	\vskip 0.5cm 
	{\includegraphics[width=0.5\textwidth]{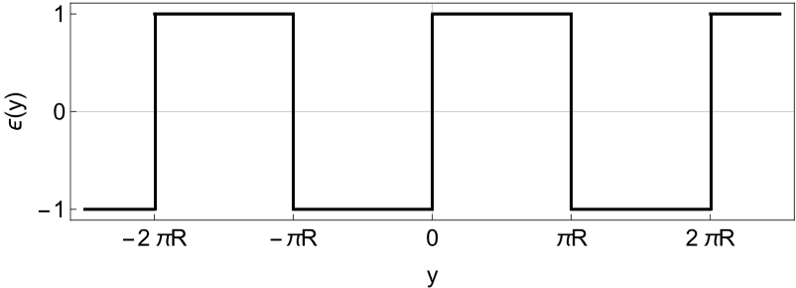}}
	\caption{Periodic $\mathbb{Z}_2$-even linear function $|y|$ and  $\mathbb{Z}_2$-odd sign function $\epsilon(y) = d|y|/dy$.}
	\label{absy_epsilony}
\end{figure}

As the background geometry, we consider a slice of ${\rm AdS}_5$ defined  on  an $S^1/\mathbb{Z}_2$ orbifold.
The covering coordinate $y$ obeys the identifications
\begin{equation}
y \;\equiv\; y+2\pi R, 
\qquad 
y \;\equiv\; -y,
\end{equation}
so that the physical space is the interval $y\in[0,\pi R]$ with boundaries (or orbifold fixed-points) at
$y=0$ (UV) and $y=\pi R$ (IR).
The 5D metric  is
\begin{equation}
ds^2 \;=\;g_{MN}dx^M dx^N
\;=\; e^{-2k|y|}\,\eta_{\mu\nu}dx^\mu dx^\nu + dy^2\,,
\label{eq:RSmetric_y}
\end{equation}
where $k$ is the AdS curvature scale and $|y|$ 
is a  parity-even and $2\pi R$-periodic  function defined by $|y|=y$  for $0\le y \le \pi R$, as illustrated in Fig.~\ref{absy_epsilony}
together with the parity-odd  sign function $\epsilon(y)\equiv d|y|/dy$.
For the background geometry Eq.~(\ref{eq:RSmetric_y}), the 4D reduced Planck mass
$M_{\rm P}\simeq 2.4\times 10^{18}$ GeV is related to the 5D Planck scale $M_5$ as
\begin{equation}
M_{\rm P}^2 \;=\; 2M_5^3 \int_0^{\pi R} dy\, e^{-2k y}
\;=\; \frac{M_5^3}{k}\left(1-e^{-2k\pi R}\right)\,.
\label{eq:MPM5relation}
\end{equation}
In the strong-warping limit $k\pi R\gg 1$, this relation reduces to
$M_{\rm P}^2\simeq M_5^3/k$.
In this limit, the KK spectrum of the 5D graviton is approximately given by
\bea
m_n\,\simeq \,\frac{n\pi k}{e^{k\pi R}-1}\,\simeq n\pi k e^{-k\pi R}
\eea
for large integer-valued KK level $n$.

\subsection{4D axion from 5D $U(1)$ gauge field}

To obtain an extra-dimensional axion (EDA) in the 4D effective theory, 
we introduce a 5D Abelian gauge field $C_M=(C_\mu,C_5)$
obeying the orbifold boundary conditions (BC)\footnote{More generally, there can be multiple 5D $U(1)$ gauge fields
with such BCs, yielding multiple EDAs in the 4D effective theory. Here we restrict the discussion to the simplest case involving a single EDA, since the generalization to multiple EDAs is straightforward.}~\cite{Choi:2003wr}:
\begin{equation}
C_\mu(x,-y) = - C_\mu(x,y)\,,
\qquad
C_5(x,-y) = C_5(x,y)\,.
\label{eq:orbifold_parity_gauge}
\end{equation}
The associated gauge symmetry is 
\begin{equation}
U(1)_C:\,\,\, C_M \;\to\; C_M + \partial_M \Lambda,
\label{eq:gauge_transf}
\end{equation}
where the gauge transformation function $\Lambda$ obeys the BC
\bea
\Lambda(x,y+2\pi R)=\Lambda(x, y) \,\,\, {\rm mod}\,\, 2\pi, \quad \Lambda(x,-y)=-\Lambda(x,y)\,\,\, {\rm mod} \,\, 2\pi
\eea
in the normalization where all $U(1)_C$ charges are integer-valued.
\begingroup
At either orbifold fixed point these conditions allow $\Lambda=0$ or $\pi$ modulo $2\pi$.  The nontrivial residual transformation is therefore generated by
\bea
\Lambda(x,y= 0 \,\,{\rm or}\,\, \pi R) =\pi \quad {\rm mod}\ 2\pi,
\eea
so the bulk $U(1)_C$ gauge symmetry is reduced to its $\mathbb{Z}_{2C}$ subgroup at the fixed points.
\endgroup

The axion is identified with the gauge-invariant Wilson-line phase
\begin{equation}
e^{i\theta(x)} =
\exp\left(i \oint_{S^1} dy \, C_5(x,y)\right)
\quad
\left(\theta \equiv \frac{a}{f_a}\right),
\label{eq:theta_def}
\end{equation}
whose periodicity $\theta\equiv \theta+2\pi$ is guaranteed by the large gauge transformation $\Lambda=y/R$.
The PQ symmetry corresponds to a global 1-form symmetry generated by the harmonic 1-form $dy$ on $S^1$:
\begin{equation}
U(1)_{\rm PQ}:\quad
C=C_M dx^M \,\,\, \rightarrow\,\,\, 
C+\frac{\alpha}{2\pi R}dy
\quad (\alpha=\text{constant}),
\label{eq:1-form_symmetry}
\end{equation}
under which $\theta(x)\rightarrow\theta(x)+\alpha$. 
This PQ transformation is equivalent to a $U(1)_C$ gauge transformation of $C_M$ with parameter $\Lambda=\alpha y/2\pi R$, which is locally well defined but not globally single-valued on $S^1$ unless $\alpha=2\pi n$ with $n\in\mathbb{Z}$.
As emphasized in \cite{Choi:2003wr,Reece:2024qv,Craig:2024cqs}, the close connection between $U(1)_{\rm PQ}$ and the 5D gauge symmetry $U(1)_C$ severely constrains the possible sources of PQ breaking.
With certain discrete parameters set to zero—such as the coefficients of Chern--Simons terms involving hidden Yang--Mills gauge fields, monodromy terms, or St\"uckelberg mixing terms—the potentially dangerous non-QCD contributions to the axion potential arise dominantly from nonlocal effects mediated by \(U(1)_C\)-charged matter fields. Consequently, they are exponentially suppressed when the charged fields are sufficiently heavy.

To compute the decay constant of the Wilson-line axion Eq.~(\ref{eq:theta_def}), one can start with the 5D action 
\begin{equation}
S_C \;=\; - \int d^5x\,\sqrt{|g|}\, \frac{1}{4g_C^2}  C_{MN}C^{MN}\,,
\label{eq:gauge_action}
\end{equation}
where 
$C_{MN}=\partial_M C_N-\partial_N C_M$ is the gauge field strength and
$g_C$ is the 5D $U(1)_C$ gauge coupling.
\label{subsec:zero_mode_fa}
For the axion mode, we can first set $C_{\mu\nu}=0$.  Then the only relevant component of the field strength is
$C_{\mu5}=\partial_\mu C_5-\partial_y C_\mu$ obeying 
the following  equation of motion  derived from Eq.~\eqref{eq:gauge_action}:
\begin{equation}
\partial_y\!\left(\sqrt{|g|}\,C^{\mu 5}\right)=0.
\label{eq:eom_Cmu5}
\end{equation}
For the background metric Eq.~(\ref{eq:RSmetric_y}),
this leads 
to
\begin{equation}
\partial_y\!\left(e^{-2k|y|} C_{\mu 5} \right)=0.
\label{eq:eom_simplified}
\end{equation}
By imposing the matching condition~\eqref{eq:theta_def} on the above equation of motion, 
the extra-dimensional profile of the PQ current is obtained as~\cite{Choi:2003wr}:
\begin{equation}
C_{\mu5}(x,y)
=
\left(
\frac{k\,e^{-2k|\pi R-y|}}{1-e^{-2k\pi R}}
\right)\partial_\mu \theta(x).
\label{eq:Cmu5_profile}
\end{equation}
This shows that the axion kinetic term receives its dominant support toward the IR end of the interval $S^1/\mathbb{Z}_2$.
Substituting Eq.~\eqref{eq:Cmu5_profile} into Eq.~\eqref{eq:gauge_action}, the 4D effective Lagrangian contains the axion kinetic term
\begin{equation}
-2\int_{0}^{\pi R}dy\,\sqrt{|g|}\;\frac{1}{2g_C^2}\,C_{\mu5}C^{\mu5}
=
-\frac12 f_a^2\partial_\mu \theta\partial^\mu \theta,
\label{eq:axion_kin}
\end{equation}
with the axion decay constant
\begin{equation}
f_a^2
=
\frac{1}{g_C^2}\left(\frac{k}{e^{2k\pi R}-1}\right).
\label{eq:fa}
\end{equation}
In the strong-warping limit $k\pi R\gg 1$, this simplifies to
\begin{equation}
f_a \simeq   \left(\frac{k}{g_CM_5^{3/2}}\right) e^{- k \pi R}M_P,
\end{equation}
where $M_{P}\simeq 2.4\times 10^{18}$ GeV  is the 4D reduced Planck mass. 
Because the axion kinetic term is localized near the IR fixed-point $y=\pi R$, the axion decay constant  $f_a$ is red-shifted by the warp factor $e^{- k\pi R}$, allowing it to be exponentially smaller than  $M_P$.

To solve the strong CP problem, one may introduce a Chern–Simons term that
generates the required PQ breaking through the QCD anomaly,
\bea
S_{CS}=\frac{N}{16\pi^2}\int  C\wedge G^a\wedge G^a,
\eea
where $N$ is an integer-valued coefficient, $C=C_M dx^M$, and
$G^a=G_{MN}^a dx^M\wedge dx^N$ is the 2-form 5D gluon field strength.
Such a term can arise, for instance, from integrating out heavy colored
5D fermions charged under $U(1)_C$~\cite{Adachi:2021rjw}, or from more ultraviolet origins such as string constructions~\cite{Witten:1996qb}.
After dimensional reduction, the 4D effective theory contains the
axion–gluon coupling
\begin{equation}
\mathcal{L}_{\rm eff}\;\ni\;\frac{N}{32\pi^2}\frac{a}{f_a}\,
G^{a\mu\nu}\tilde G^a_{\mu\nu}.
\label{eq:axion_QCD_coupling}
\end{equation}
This coupling represents the breaking of the PQ symmetry by the QCD
anomaly, which results in the QCD-induced axion potential
$V_{\rm QCD}(a)$ given in Eq.~(\ref{eq:qcd_axion_potential}).
In the following, we focus on additional PQ-violating contributions to
the axion potential induced by the $U(1)_C$ gauge interactions in the
bulk, as well as possible $U(1)_C$-breaking interactions localized at
the orbifold fixed points.

\subsection{Charged matter fields}
\label{subsec:matteractions}

\begingroup
The electric weak gravity conjecture suggests that the model contain a $U(1)_C$-charged state whose mass $M$ obeys parametrically \cite{Arkani-Hamed:2006emk}
\bea
M \lesssim {\cal O}(q g_C M_5^{3/2}),
\eea
where $q$ is its integer $U(1)_C$ charge, $g_C$ is the 5D gauge coupling, and $M_5$ is the 5D Planck mass.
\endgroup
On $S^1/\mathbb{Z}_2$, matter fields charged under $U(1)_C$ can obey a variety of
distinct $\mathbb{Z}_2$ boundary conditions (BCs). Here we present relatively
simple examples of BCs that are particularly relevant for our later discussion
of the axion potential.

In the case where the orbifold $\mathbb{Z}_2$ action acts separately on a
single $U(1)_C$-charged 5D scalar or Dirac fermion, there are two distinct
types of matter fields. The first type, denoted by
$\tilde\Phi=(\tilde\phi,\tilde\psi)$, consists of 5D scalars $\tilde\phi$ or
Dirac fermions $\tilde\psi$ obeying ordinary parity-type (P-type) BCs
\cite{Bergshoeff:2000zn,Fujita:2001bd}:
\begin{align}
\mbox{* P-type matter } \tilde\Phi=(\tilde\phi,\tilde\psi):\quad
\tilde\phi(x,-y) = \eta_{\tilde\phi}\tilde\phi(x,y),
\quad
\tilde\psi(x,-y) = \eta_{\tilde\psi}\gamma^5\tilde\psi(x,y),
\label{eq:parity_P}
\end{align}
where $\eta_{\tilde\phi,\tilde\psi}=\pm1$.

The second type, denoted by $\Phi=(\phi,\psi)$, obeys
charge-conjugation-twisted (C-twisted) BCs \cite{Abe:2016tfq}:
\begin{align}
\mbox{* C-twisted matter } \Phi=(\phi,\psi):\quad
\phi(x,-y) = \eta_\phi\phi^\ast(x,y),\quad
\psi(x,-y) = \eta_\psi\psi^c(x,y),
\label{eq:parity_CP}
\end{align}
with $\eta_{\phi,\psi}=\pm1$ and
$\psi^c \equiv {\cal C}\,\bar\psi^{\,T}$, where ${\cal C}$ denotes the
4D charge-conjugation matrix. These boundary conditions determine the
Kaluza--Klein mode projections and also whether the corresponding fields can
acquire nonvanishing boundary values at the orbifold fixed points
$y=0$ and $y=\pi R$.

  Under the 5D gauge symmetry $U(1)_C$, these matter fields transform
as
\bea
\tilde\Phi \,\,\rightarrow \,\, e^{i q \epsilon(y)\Lambda}\tilde\Phi,  \quad
\Phi\,\, \rightarrow \,\,  e^{iq\,\Lambda}\Phi,
\eea 
where $q\in \mathbb{Z}$ denotes the $U(1)_C$ gauge charge and
  $\epsilon(y)\equiv d|y|/dy$ is the $\mathbb{Z}_2$-odd periodic sign function  shown in Fig.~\ref{absy_epsilony}. The corresponding covariant derivatives are 
\bea
D_M \tilde \Phi  = (\nabla_M  - i\,q\epsilon(y)\, C_M)\tilde \Phi, \quad
D_M \Phi  = (\nabla_M  - i\,q\, C_M)\Phi,
\label{eq:cov_der}
\eea
where
$\nabla_M$ acting on fermions includes the spin connection.
Note that the P-type matter fields $\tilde\Phi$ 
possess $\mathbb{Z}_2$-odd gauge couplings $q\epsilon(y)$
on the covering circle $S^1$~\cite{Bergshoeff:2000zn,Fujita:2001bd}, whereas the C-twisted matter fields $\Phi$ have $\mathbb{Z}_2$-even (constant)
gauge couplings~\cite{Abe:2016tfq}.
As will be discussed in the next section, these distinct $\mathbb{Z}_2$-parities of the gauge interactions lead to qualitatively different contributions to
the axion potential from loops of charged particles winding around $S^1$.

The 5D action of the above charged matter fields, relevant for our analysis,  is given by
\bea
S_{\rm bulk}
&=&\int d^5x\,\sqrt{|g|}\left[-\,D^M\phi^\ast {D_M\phi}-M_\phi^2\,|\phi|^2 + (\phi\rightarrow \tilde \phi)\right.
\nonumber \\
&&\left.\qquad +\,
\frac{i}{2}\Big(\bar\psi \Gamma^M D_M\psi-(D_M\bar\psi)\Gamma^M\psi\Big)-M_{\psi}\,\bar\psi\psi+(\psi\rightarrow \tilde\psi)
\right],
\label{eq:S_bulk}
\eea
where  $\Gamma^M$ are the curved-space gamma matrices.
The 5D masses of the scalar fields $\phi,\tilde\phi$ and the C-twisted fermion $\psi$ 
can be taken to be constant along the covering circle $S^1$. In contrast, consistency with the $\mathbb{Z}_2$ orbifold symmetry requires that the P-type fermion
$\tilde\psi$ possess  a kink-type $\mathbb{Z}_2$-odd mass  over $S^1$ \cite{Grossman:1999ra,Gherghetta:2000qt},
\bea
M_{\phi,\tilde\phi,\psi}=M, \quad 
M_{\tilde \psi}= M\epsilon(y),
\label{eq:mass_parity}
\eea
where, for simplicity, we use the same mass parameter $M$  for all charged matter fields.

In addition to the bulk action Eq.~(\ref{eq:S_bulk}), the model can contain fixed-point-localized (boundary) interactions consistent with
the symmetries of the theory.  The 5D gauge symmetry $U(1)_C$ is broken by the orbifold BC at the fixed points 
$y_0=0$ and $y_\pi=\pi R$ down to its
$\mathbb{Z}_2$-subgroup generated by the large gauge transformation $\Lambda=\pi$,
\bea
\mathbb{Z}_{2C}:\,\,\,
\Phi \,\,\rightarrow\,\,  (-1)^q\Phi, \quad \tilde\Phi\,\,\rightarrow\,\, (-1)^q \tilde\Phi,\eea
where $q\in\mathbb{Z}$ denotes the $U(1)_C$ gauge charge of $\Phi$ or $\tilde\Phi$. 
As a consequence, fixed-point interactions can have richer structures, in particular allowing operators that preserve  $\mathbb{Z}_{2C}$ but violate  $U(1)_C$.  As we will see, together with the bulk $U(1)_C$ gauge interactions encoded in the covariant derivatives Eq.~(\ref{eq:cov_der}), such $U(1)_C$-violating fixed-point interactions provide the primary sources
of the non-QCD contribution to the Wilson-line axion potential. 
Among the $U(1)_C$-violating fixed-point interactions, the most relevant ones for the axion potential 
are
\begin{equation}
S_{\rm boundary}=-\sum_{i=0,\pi}\int d^4x\,\sqrt{|g(y=y_i)|}\;\Delta {\cal L}_i(y=y_i)\quad (y_0=0,\, y_\pi=\pi R),
\label{eq:S_brane}
\end{equation}
where
\begin{equation}
\Delta {\cal L}_i = J_i \phi  +\tilde J_i\tilde \phi+ \frac{1}{2}b_i \phi^2+\frac{1}{2}\tilde b_i\tilde\phi^2   +\mu_i\bar\psi^c\psi +\tilde\mu_i \bar{\tilde{\psi^c}}\tilde\psi +{\rm h.c}.
\label{eq:appC_brane}
\end{equation}
Additional fixed-point mass-mixing operators such as $\phi\tilde \phi, \phi\tilde\phi^*,\bar\psi\tilde \psi$ and 
$\bar\psi^c\tilde\psi$ can also be present and affect the axion potential. These terms can be incorporated straightforwardly  in the computation of the axion potential, and we therefore omit them for simplicity. 
Which operators in $\Delta {\cal L}_i$ are allowed depends on the orbifold BCs and on the $U(1)_C$ gauge charges of the fields involved. In particular, the linear scalar terms are permitted \emph{only when the scalar fields carry even charge}, i.e.
\bea
J_i, \tilde J_i\neq 0   \,\,\,  \mbox{only for $\phi,\tilde\phi$ with  $q_{\phi,\tilde \phi} \in 2\mathbb{Z}$}.\label{eq:linear_charge} \eea
  Similarly, a bilinear term such as $\phi\tilde\phi$ is allowed 
only when $q_\phi+q_{\tilde\phi}\in 2\mathbb{Z}$.

The above P-type and C-twisted boundary conditions provide the minimal
realizations of the $\mathbb{Z}_2$ orbifold action acting on a single
$U(1)_C$-charged matter field. More generally, when multiple charged fields
are present, the $\mathbb{Z}_2$ orbifold action may act nontrivially in the
field space spanned by those fields, and may also mix fields with their charge
conjugates. This possibility is not merely a formal generalization; it is
realized, for instance, in the scalar sector of a 5D supersymmetric
hypermultiplet with $\mathbb{Z}_2$-even (constant) $U(1)_C$ gauge couplings,
which will be discussed in the next subsection.

To incorporate hypermultiplet matter fields with $\mathbb{Z}_2$-even
(constant) $U(1)_C$ gauge couplings in supersymmetric 5D models, we consider a
pair of complex scalar fields $H_\alpha$ $(\alpha=1,2)$, each carrying
$U(1)_C$ charge $q$, obeying a BC involving both charge conjugation and the
exchange between $H_1$ and $H_2$:
\begin{equation}
\mbox{* C-twisted hypermultiplet scalar $H_{\alpha}$}:\quad
H_1(x,-y)=H_2^*(x,y).
\label{eq:C-twisted-hyper}
\end{equation}
The $U(1)_C$- and $\mathbb{Z}_2$-invariant bulk action of $H_\alpha$ is given
by
\bea
S_{\rm bulk}
&=&-\int d^5x\,\sqrt{|g|}
\left[
D^M H^*_1 D_M H_1
+
D^M H^*_2 D_M H_2
\right.
\nonumber\\
&&\qquad\left.
+\,M_0^2\left(|H_1|^2+|H_2|^2\right)
+\mu^2 \epsilon(y)\left(|H_1|^2-|H_2|^2\right)
\right],
\label{eq:C-twisted-hyper-scalar}
\eea
where the covariant derivatives involve constant $U(1)_C$ gauge couplings:
\begin{equation}
D_M H_\alpha=(\nabla_M-iq C_M)H_\alpha.
\end{equation}
A key difference between a simple C-twisted scalar $\phi$ and the
C-twisted hypermultiplet scalars $H_{\alpha}$ is that the latter generically
have piecewise constant masses over the covering circle $S^1$, containing both
$\mathbb{Z}_2$-even and $\mathbb{Z}_2$-odd components:
\begin{equation}
M_1^2(y)=M_0^2+\mu^2\epsilon(y),\qquad
M_2^2(y)=M_0^2-\mu^2\epsilon(y),
\label{eq:piecewise_scalar}
\end{equation}
whereas the bulk mass of $\phi$ is simply constant over $S^1$.

As in the case of the P-type and C-twisted matter fields
$\tilde\Phi$ and $\Phi$, one may introduce appropriate fixed-point
interactions for the C-twisted hypermultiplet scalar $H_\alpha$, including
$U(1)_C$-violating (but $\mathbb{Z}_2$-invariant) operators such as
$H_1^2+H_2^{*2}$ and $H_1+H_2^*$ (for $q\in 2\mathbb{Z}$). The effects of
such terms on the axion potential are essentially the same as those of the
fixed-point interactions for $\Phi$ and $\tilde\Phi$ given in Eq.~(\ref{eq:S_brane}), and therefore will not
be discussed separately.

\subsection{Matter fields in supersymmetric models}
\label{subsec:susy_matter}

{We now make the supersymmetric embedding of the gauge and matter fields introduced above explicit.  Supersymmetric five-dimensional theories provide a natural extension of the extra-dimensional setup and are also relevant to ultraviolet constructions in which supersymmetry is present at high scales.}

In supersymmetric 5D models, the gauge field
$C_M=(C_\mu, C_5)$ obeying the BCs in
Eq.~(\ref{eq:orbifold_parity_gauge})
may arise from the 5D supergravity multiplet
involving the graviphoton and radion \cite{Bergshoeff:2000zn,Fujita:2001bd,Linch:2002wg},
or more generally from an ordinary non-gravitational vector multiplet.
To avoid restrictions on charged matter fields,
here we consider the case in which $C_M$ corresponds to the vector field
component of an ordinary 5D vector multiplet,
which consists, in 4D ${\cal N}=1$ superspace language
\cite{ArkaniHamed:2001tb,Marti:2001iw},
of a vector superfield ${\cal V}$ and a chiral superfield $\chi$
satisfying the orbifold BCs
\begin{equation}
{\cal V}(x,-y,\theta)=-{\cal V}(x,y,\theta),\qquad
\chi(x,-y,\theta)=\chi(x,y,\theta),
\end{equation}
where $\theta$ denotes the Grassmann coordinate of the 4D ${\cal N}=1$
superspace. The $\mathbb{Z}_2$-odd gauge field component $C_\mu$
resides in ${\cal V}$, while the $\mathbb{Z}_2$-even component $C_5$ is
contained in $\chi$ as
\bea
C_5(x,y) ={\rm Im}\big[\chi(x,y,\theta =0)\big].
\eea

On the other hand, charged matter fields appear in the form of
hypermultiplets, each consisting of two ${\cal N}=1$ chiral multiplets.
In the minimal setup, there are two distinct types of BCs for
hypermultiplet matter fields: a P-type BC, which results in a
$\mathbb{Z}_2$-odd $U(1)_C$ gauge coupling $q\epsilon(y)$ on the
covering circle, and a C-twisted BC, which yields a
$\mathbb{Z}_2$-even (constant) gauge coupling on the covering circle.
As already mentioned in the previous subsection, this difference in the
$\mathbb{Z}_2$ parity of the gauge coupling leads to a crucial difference
in the axion potential induced by the corresponding matter fields.
{In the unbroken supersymmetric limit, the total axion potential must vanish: the bosonic and fermionic contributions either vanish separately or cancel in the sum.}

A P-type hypermultiplet consists of the two chiral multiplets
\bea
\tilde {\cal H}=(\tilde \phi, \tilde \psi_L), \qquad
\tilde {\cal H}^c=(\tilde \phi^c,\tilde \psi_R^c)
\eea
with opposite $U(1)_C$ charges, obeying the BCs imposed on the corresponding
chiral superfields:
\bea
\tilde {\cal H}(x,-y,\theta)=\tilde {\cal H}(x,y,\theta),\qquad
\tilde {\cal H}^c(x,-y,\theta)=-\tilde {\cal H}^c(x,y,\theta).
\eea
This implies that a P-type hypermultiplet consists of two P-type scalar
fields, $\tilde \phi$ and $\tilde \phi^c$, with opposite
$\mathbb{Z}_2$ parities, together with a P-type Dirac fermion
$\tilde\psi$, as defined in the previous subsection.

In a warped background geometry with AdS curvature scale $k$, the bulk action
of a P-type hypermultiplet can be written as
\cite{Gherghetta:2000qt,Marti:2001iw}
\begin{align}
S_{\rm bulk}
&=
\int d^5x
\left[
\int d^4\theta\,e^{-2k|y|}
\Big(
\tilde {\cal H}^\dagger e^{-q\epsilon(y){\cal V}}\tilde{\cal H}
+
\tilde{\cal H}^c e^{q\epsilon(y){\cal V}}\tilde {\cal H}^{c\dagger}
\Big)
\right.
\nonumber\\
&\left.
\qquad\qquad\quad +
\int d^2\theta\,e^{-3k|y|}
\,
\tilde{\cal H}^c
\left(
\partial_y-\frac{q\epsilon(y)}{\sqrt2}\chi
+\Big(M-\frac{3k}{2}\Big)\epsilon(y)
\right)\tilde{\cal H}
+\text{h.c.}
\right],
\end{align}
where $M$ is a bulk mass parameter.
{After integrating over the Grassmann coordinates and performing the appropriate
integration by parts, the bulk mass parameters appearing in
Eq.~(\ref{eq:S_bulk}) are} \cite{Gherghetta:2000qt,Marti:2001iw}
\bea
M_{\tilde \phi}^2
=
M^2+Mk-\frac{15}{4}k^2,
\qquad
M_{\tilde \phi^c}^2
=
M^2-Mk-\frac{15}{4}k^2,
\eea
and
\bea
M_{\tilde \psi}
=
M\epsilon(y),
\eea
{where the scalar bulk masses are constant on the covering circle, whereas the P-type fermion has the kink mass required by the orbifold parity.}

We now turn to a C-twisted hypermultiplet involving the chiral
multiplets\footnote{For convenience, we use here a somewhat unconventional
notation for the chiral multiplet ${\cal H}^c$, in which its scalar component
is written as a complex-conjugated scalar field.}
\bea
{\cal H}=(H_1, \psi_L),\qquad
{\cal H}^c=(H_2^*, \psi_R^c),
\eea
where $H_1$ and $H_2$ carry a common
$U(1)_C$ charge $q$.
The associated chiral superfields obey the BC
\begin{equation}
{\cal H}(x,-y,\theta)={\cal H}^c(x,y,\theta).
\label{eq:C-twisted-hyper-SUSY}
\end{equation}
This BC involves both charge conjugation and the exchange between $H_1$ and
$H_2$.
{Thus, this C-twisted hypermultiplet consists of the C-twisted
scalar pair $H_{\alpha}$ $(\alpha=1,2)$ defined in
Eq.~(\ref{eq:C-twisted-hyper}), together with a C-twisted Dirac fermion
$\psi$.} As noted in the previous subsection, their $U(1)_C$ gauge couplings
and the bulk mass of $\psi$ are constant along the covering circle $S^1$,
while the scalar masses of $H_\alpha$ contain both $\mathbb{Z}_2$-even and
$\mathbb{Z}_2$-odd components, and therefore are piecewise constant over
$S^1$.

In a warped background geometry, the bulk action of a C-twisted
hypermultiplet can be written as
\begin{align}
S_{\rm bulk}& =
\int d^5x
\left[
\int d^4\theta\,e^{-2k|y|}
\Big(
{\cal H}^\dagger e^{-q{\cal V}}{\cal H}
+
{\cal H}^c e^{q{\cal V}}{\cal H}^{c\dagger}
\Big)
\right.
\nonumber\\
&\left.
\qquad\qquad\quad +
\int d^2\theta\,e^{-3k|y|}
\,
{\cal H}^c
\left(
\partial_y-\frac{q}{\sqrt2}\chi
+M-\frac{3k}{2}\epsilon(y)
\right){\cal H}
+\text{h.c.}
\right].
\end{align}
This action is invariant under the $\mathbb{Z}_2$ transformation involving
$y\rightarrow -y$ together with the exchange between ${\cal H}$ and
${\cal H}^c$, once total derivative terms are dropped.
A crucial point is that the C-twisted realization allows a constant
$U(1)_C$ gauge coupling $q$ and a constant fermion bulk mass $M$, in contrast
to the P-type case, where $\mathbb{Z}_2$ invariance requires kink-type
profiles.

{For the C-twisted hypermultiplet, integrating over the Grassmann coordinates and performing
appropriate integration by parts gives the bulk fermion mass appearing in
Eq.~(\ref{eq:S_bulk}) as}
\bea
M_\psi = M.
\eea
{The scalar bulk mass terms take the form of
Eq.~(\ref{eq:C-twisted-hyper-scalar}) with}
\begin{equation}
M_1^2(y)=M^2-\frac{15}{4}k^2+Mk\,\epsilon(y), \qquad
M_2^2(y)=M^2-\frac{15}{4}k^2-Mk\,\epsilon(y),
\label{eq:C-twisted-piecewisemass}
\end{equation}
{The scalar masses therefore realize the piecewise form in Eq.~\eqref{eq:piecewise_scalar}, with $M_0^2=M^2-15k^2/4$ and $\mu^2=Mk$.  Unlike the simple C-twisted scalar $\phi$ with a constant bulk mass, the two hypermultiplet scalars experience different bulk masses on the two halves of the covering circle.  This distinction will be important for the winding action discussed in Sec.~\ref{worldline}.  The full component action also contains scalar terms localized at the orbifold fixed points, with coefficients fixed by the same superfield action.  Only the bulk mass profile above enters the worldline calculation presented below; a direct KK analysis of the complete supersymmetric spectrum, including the localized terms, is left for separate study.}


\section{Axion potential in the absence of fixed-point localized operators }
\label{sec:potential_loop}

In the model of Sec.~\ref{sec:eda_model}, the Wilson-line axion 
$\theta$ defined in Eq.~(\ref{eq:theta_def}) can acquire a potential through bulk $U(1)_C$ gauge interactions as well as through 
$U(1)_C$-violating operators localized at the fixed points.
To evaluate the induced axion potential, one may introduce a constant axion background $\theta$,  and compute the axion-dependent vacuum energy density in the gauge
\begin{equation}
C_5=\frac{\theta}{2\pi R}.
\label{eq:background_axion}
\end{equation}
In this section, we focus on the contributions in the absence of fixed-point interactions, which arise solely from the bulk 
$U(1)_C$ gauge interactions.

It is straightforward to see that P-type matter fields
$\tilde\Phi=(\tilde\phi,\tilde\psi)$ do not generate a nontrivial axion
potential in the absence of $U(1)_C$-violating fixed-point operators.
The bulk gauge interactions are determined by the covariant derivative
\bea
D_5\tilde \Phi
=
\left(\partial_y - \frac{iq\theta}{2\pi R}\epsilon(y)\right)\tilde\Phi,
\eea
which involves a $\mathbb{Z}_2$-odd gauge coupling
$q\epsilon(y)$ over the covering circle $S^1$.
The constant axion background $\theta$ can then be removed by the field
redefinition
\bea
\tilde \Phi \,\, \rightarrow \,\,
e^{iq\theta|y|/2\pi R}\tilde \Phi,
\label{eq:field_red_Phitilde}
\eea
which is single-valued on the covering circle $S^1$ for generic values of
$\theta$ (see the plot of $|y|$ depicted in
Fig.~\ref{absy_epsilony}).
Therefore, physical quantities such as the mass spectrum and the vacuum energy
density are independent of $\theta$, implying that no axion potential is
generated.

For C-twisted matter fields $\Phi=(\phi,\psi, H_\alpha)$, whose covariant derivative is
\bea
D_5\Phi
=
\left(\partial_y - \frac{iq\theta}{2\pi R}\right)\Phi
\eea
with a $\mathbb{Z}_2$-even (constant) gauge coupling $q$,
the analogous field redefinition
\bea
\Phi \,\, \rightarrow \,\,
e^{iq\theta y/2\pi R}\Phi
\label{eq:field_red_Phi}
\eea
is not single-valued on $S^1$.
Consequently, the background $\theta$ cannot be gauged away, leading to a
$\theta$-dependent mass spectrum and vacuum energy, and thus to a nontrivial
axion potential.

The above  conclusion can be understood by tracking the phase acquired by a charged particle propagating along the fifth dimension in the background gauge field $C_5=\theta/2\pi R$.
For a particle of charge $q$, propagation from $y=0$ to $y=\pi R$ yields
\begin{equation}
\exp\left(i q\int_0^{\pi R}dy \,  C_5\right)
=\exp\left(i\frac{q\theta}{2}\right).
\label{eq:appC_half_phase}
\end{equation}
A winding configuration around $S^1$ consists of two such traversals, $y=0\rightarrow y=\pi R$ and $y=\pi R\rightarrow y=2\pi R$.
Because the $U(1)_C$ gauge coupling carries a definite $\mathbb{Z}_2$ parity, the second traversal may contribute either the same phase or the opposite phase. 
For P-type matter fields, the coupling $q\epsilon(y)$ is $\mathbb{Z}_2$-odd. The
second traversal therefore produces the opposite phase, $e^{i q\theta/2}\cdot e^{-i q\theta/2} =1$,
 so the total winding phase is trivial and the vacuum energy is independent of $\theta$.
For C-twisted matter fields,
the coupling is $\mathbb{Z}_2$-even. The
phases then add, $e^{i q\theta/2}\cdot e^{i q\theta/2} = e^{i q\theta}$, yielding a nontrivial total winding phase.
Consequently, the vacuum energy depends on $\theta$,
generating an axion potential whose leading harmonic is proportional to 
 $\cos(q\theta)$.
 
 In the remainder of this section, we compute the axion potential induced by
C-twisted matter fields $\Phi=(\phi,\psi,H_\alpha)$ winding around $S^1$ in
the warped background geometry. We present three complementary approaches.
We first employ a Euclidean worldline (proper-time)
representation~\cite{Strassler:1992zr,Schubert:2001he,Bastianelli:2002fv},
which makes the winding interpretation manifest and directly exhibits the
exponential suppression arising from the worldline-instanton action in warped
space.
We then perform an independent Casimir-energy computation based on the
$\theta$-dependent Kaluza--Klein (KK)
spectrum~\cite{Hosotani:1983xw,ArkaniHamed:2003wu}, providing both a
cross-check and a practical method for evaluating the potential.
Third, for fermions we discuss a formulation based on the monodromy matrix
associated with a single winding around
$S^1$~\cite{Forman:1987xk,Kirsten:2003py}, which captures the full
$\theta$ dependence of the fermion determinant.
Some technical details of these computations are presented in the Appendices.

\subsection{Axion potential in the worldline approach}
\label{worldline}
\begingroup

Consider a $C$-twisted complex scalar $\phi$ with $U(1)_C$ charge $q$ and bulk mass $M$.
Integrating out $\phi$ generates the one-loop effective potential
\bea
\int d^4x\,V_\phi(\theta)={\rm Tr}\ln {\cal O}_\phi(\theta),
\label{eq:potential_from_phi}
\eea
where
\bea
{\cal O}_\phi(\theta)
\equiv
-\frac{1}{\sqrt{|g|}}D_M\!\left(\sqrt{|g|}\,g^{MN}D_N\right)+M^2.
\label{eq:Gamma1_scalar_worldline}
\eea
The axion dependence enters through the constant background $C_5=\theta/(2\pi R)$.
It is convenient to perform the warp-dependent field redefinition
\begin{equation}
\phi\rightarrow |g|^{-1/4}\phi .
\label{eq:field_redefinition}
\end{equation}
For the RS metric, the quadratic operator, including the fixed-point term generated by this field redefinition, is
\begin{equation}
{\cal O}_\phi(\theta)
=-e^{2k|y|}\eta^{\mu\nu}\partial_\mu\partial_\nu
-\left(\partial_y-iqC_5\right)^2
+M_{\rm eff}^2
-2k\,\partial_y\epsilon(y),
\qquad
M_{\rm eff}^2=M^2+4k^2.
\label{eq:Ophi_warped_worldline}
\end{equation}
The last term is localized at the orbifold fixed points and changes the matching of the bulk solutions there.  Its effect can be separated from the bulk propagation: Appendix~\ref{app:common_determinant} shows that it leaves the one-winding action unchanged and modifies the prefactor.  For $M\gg k$ the scalar prefactor receives its first correction at order $(k/M)^2$, while for $k$ comparable to or larger than $M$ the modification can be more important.  We therefore begin with the bulk worldline contribution, which makes the origin of the winding suppression transparent.  After Fourier transformation along the four-dimensional directions, the bulk operator is
\begin{equation}
{\cal O}_\phi(\theta;p)
=-\left(\partial_y-iqC_5\right)^2+M_{\rm eff}^2+e^{2k|y|}p^2 .
\label{eq:Ophi_1d_p}
\end{equation}
Using the Schwinger proper-time representation and evaluating the orbifold trace on the covering circle, its effective potential can be written as
\begin{equation}
V_{\phi}(\theta)
=
-\pi R\int\frac{d^4p}{(2\pi)^4}
\int_0^\infty\frac{dT}{T}
\sum_{n\in\mathbb Z}e^{inq\theta}
\int_{y(1)=y(0)+2n\pi R}{\cal D}y(\tau)\,
\exp\!\left[-S_E[y(\tau);T]\right],
\label{eq:V_worldline_master}
\end{equation}
where
\begin{equation}
S_E[y(\tau);T]
=
\int_0^1d\tau\left[
\frac{\dot y^2}{4T}
+T\left(M_{\rm eff}^2+e^{2k|y|}p^2\right)
\right].
\label{eq:SE_warped}
\end{equation}
The phase $e^{inq\theta}$ is the Aharonov--Bohm phase of a trajectory winding $n$ times around $S^1$, so only $n\neq0$ contributes to the axion potential.
The path integral in each winding sector includes the integral over the starting point $y(0)$ on the covering circle.

In the flat limit, the path integral can be evaluated exactly.  Subtracting the $\theta$-independent part gives
\begin{align}
V^{\rm flat}_\phi(\theta)
&=-\frac{1}{64\pi^6R^4}
\sum_{n=1}^\infty
\left(\frac{4\pi^2n^2M^2R^2+6\pi nMR+3}{n^5}\right)
 e^{-2nM\pi R}\cos(nq\theta).
\label{eq:scalar_flat_worldline}
\end{align}
This result will also provide a useful normalization check for the KK spectral-function calculation.

We now turn to the warped case.  For $M\pi R\gg1$, the leading harmonic is dominated by the $n=\pm1$ winding sectors.  After the rescaling $p\to e^{-k\pi R}p$, the worldline action takes the form
\begin{align}
S_E
=\int_0^1d\tau\left[
\frac{\dot y^2}{4T}+TM_{\rm eff}^2
+Tp^2e^{-2k(\pi R-|y|)}
\right].
\label{eq:SE_warped_rescaled}
\end{align}
Performing the four-dimensional momentum integral first gives
\begin{align}
V^{\rm warped}_\phi(\theta)
&=-2\pi R\,e^{-4k\pi R}\cos(q\theta)
\int_0^\infty\frac{dT}{T}\frac{1}{(4\pi T)^2}
\left\langle\frac{1}{I[y]^2}\right\rangle
\int_{n=1}{\cal D}y\,e^{-S_0},
\label{eq:potential_expectation_value}
\end{align}
with
\begin{equation}
S_0=\int_0^1d\tau\left(\frac{\dot y^2}{4T}+TM_{\rm eff}^2\right),
\qquad
I[y]=\int_0^1d\tau\,e^{-2k(\pi R-|y(\tau)|)}.
\end{equation}
The expectation value is the corresponding normalized $n=1$ winding average.
We use the classical approximation
\begin{equation}
\left\langle\frac{1}{I[y]^2}\right\rangle\simeq\frac{1}{I_0^2},
\qquad
I_0=\frac{1-e^{-2k\pi R}}{2k\pi R},
\label{eq:classical_approx_I}
\end{equation}
Since $I_0$ is independent of the starting point for a complete winding, it may be evaluated on the representative classical path $y(\tau)=2\pi R\tau$.
The approximation becomes exact in the limits $M_{\rm eff}\to\infty$ or $k\to0$.  Appendix~\ref{app:path_integral_expansion} evaluates the fluctuations of $I[y]$ with the starting point averaged over the winding circle.  The expansion is controlled by $1/(M_{\rm eff}\pi R)$; in the strongly warped limit its leading term is proportional to $k/M_{\rm eff}$, while near the flat limit it is further suppressed by $k\pi R$.  This correction belongs to the bulk worldline calculation itself and is distinct from the fixed-point matching discussed in Appendix~\ref{app:common_determinant}.  In the analytic worldline result below we use the classical value $I_0$; the KK calculation in Sec.~\ref{subsec:KKmethod} provides the corresponding prefactor with the fixed-point matching included.

The remaining Gaussian path integral and the proper-time integral then give
\begin{align}
V^{\rm warped}_\phi(\theta)
\simeq
-\frac{(M^2+4k^2)k^2}{4\pi^2}
\frac{e^{-4k\pi R}}{(1-e^{-2k\pi R})^2}
 e^{-2\pi R\sqrt{M^2+4k^2}}\cos(q\theta),
\label{eq:pot_warped_worldline_scalar}
\end{align}
for $M\pi R\gg1$, within the bulk worldline and classical-$I[y]$ approximations described above.  We keep the full $M_{\rm eff}=\sqrt{M^2+4k^2}$ in the exponential because it is the mass entering the winding action.  In the flat limit this reduces to
\begin{equation}
V^{\rm flat}_\phi(\theta)
\simeq-\frac{M^2}{16\pi^4R^2}e^{-2 M \pi R}\cos(q\theta),
\end{equation}
which agrees with the leading harmonic of Eq.~\eqref{eq:scalar_flat_worldline}.
The exponential $e^{-2M_{\rm eff}\pi R}$ is the one-winding worldline action, while the factor $e^{-4k\pi R}$ reflects the IR-redshifted four-dimensional measure in the warped geometry.

For a $C$-twisted Dirac fermion $\psi$ with charge $q$ and bulk mass $M$,
\begin{equation}
\int d^4x\,V_\psi(\theta)=-\frac12{\rm Tr}\ln{\cal O}_\psi(\theta),
\label{eq:one_loop_fermion}
\end{equation}
where one may use a second-order representative of the squared Dirac operator.  The corresponding quadratic operator contains the fixed-point term $-(k/2)\partial_y\epsilon(y)$ generated by the warp-dependent field redefinition.  As for the scalar, this term affects the matching at the fixed points but not the winding exponent; Appendix~\ref{app:common_determinant} treats it explicitly.  To display the bulk winding action, we first use the operator between the fixed points,
\begin{equation}
{\cal O}_\psi(\theta;p)
=\left[-\left(\partial_y-iqC_5\right)^2+M_\psi^2(y)+e^{2k|y|}p^2\right]{\bf I}_4,
\label{eq:Opsi_worldline}
\end{equation}
with
\begin{equation}
M_\psi(y)=M+\frac{k}{2}\epsilon(y)=
\begin{cases}
M+k/2,&0<y<\pi R,\\[2pt]
M-k/2,&\pi R<y<2\pi R.
\end{cases}
\label{eq:Meff_fermion}
\end{equation}
The $p$ integration again produces the same functional $I[y]$.  We use
\begin{equation}
\left\langle\frac{1}{I[y]^2}\right\rangle
\simeq\frac{1}{I[y_{\rm cl}]^2},
\label{eq:approx_I_cl}
\end{equation}
where $y_{\rm cl}$ is the classical winding path for the piecewise constant mass.  We use the same classical approximation for $I[y]$ as in the scalar case.

The classical trajectory is piecewise linear, crossing the IR fixed point at a junction time $\tau_0$.  Its action can be written as
\begin{align}
S_{\rm cl}(T,\tau_0)
&=\frac{\pi^2R^2}{4T\tau_0}+T\tau_0\left(M+\frac{k}{2}\right)^2
+\frac{\pi^2R^2}{4T(1-\tau_0)}+T(1-\tau_0)\left(M-\frac{k}{2}\right)^2.
\end{align}
At the joint saddle point of $T$ and $\tau_0$,
\begin{equation}
T_*=\frac{4\pi MR}{4M^2-k^2},
\qquad
\tau_0(T_*)=\frac{2M-k}{4M},
\qquad
S_{\rm cl}(T_*,\tau_0)=2M\pi R,
\end{equation}
which exists for $M>k/2$.  Moreover $I[y_{\rm cl}]=I_0$, independent of the junction time.  For unequal masses on the two half-circles, the prefactor also receives the matching factor associated with the change of mass at the midpoint.  Including this factor together with the Gaussian fluctuations of $T$ and $\tau_0$ gives
\begin{align}
V^{\rm warped}_\psi(\theta)
&\simeq
\frac{M^2k^2}{2\pi^2}
\left(1-\frac{k^2}{4M^2}\right)^3
\left(\frac{e^{-2k\pi R}}{1-e^{-2k\pi R}}\right)^2
 e^{-2M\pi R}\cos(q\theta).
\label{eq:worldline_fermion_warped_pot}
\end{align}
The displayed prefactor is controlled when both segment masses are large, in particular when $(M-k/2)\pi R\gg1$.  The resulting fermion winding action is independent of the AdS curvature,
\begin{equation}
\int_0^{2\pi R}dy\,M_\psi(y)=2M\pi R,
\end{equation}
as illustrated in Fig.~\ref{worldline4}.  Appendix~\ref{app:common_determinant} shows that the fixed-point matching does not alter this exponent.

\begin{figure}[t]
\centering
\includegraphics[width=0.5\textwidth]{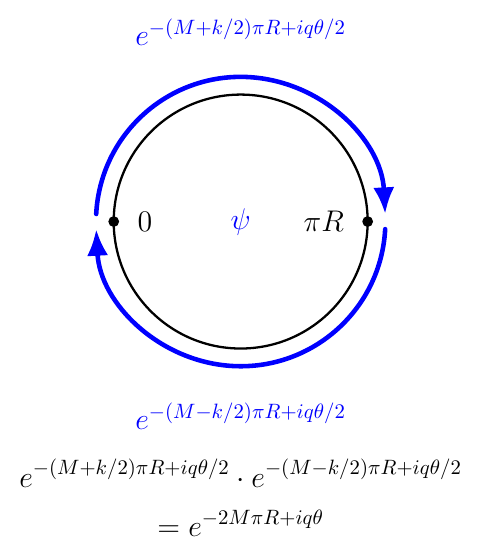}
\caption{Euclidean worldline for a C-twisted fermion $\psi$, whose effective bulk mass differs between the regions $0<y<\pi R$ and $-\pi R<y<0$.  The instanton action is the sum of the contributions from the two path segments with effective masses $M+k/2$ and $M-k/2$.}
\label{worldline4}
\end{figure}

The same saddle-point argument applies to a $C$-twisted scalar pair $H_\alpha$ ($\alpha=1,2$) with piecewise constant effective masses
\begin{equation}
m_\pm=\sqrt{M_0^2\pm\mu^2+4k^2}.
\end{equation}
A one-winding path traverses a half-circle with mass $m_+$ and the other half with mass $m_-$, giving
\begin{equation}
S_{\rm inst}=\pi R(m_++m_-).
\end{equation}
For unequal $m_+$ and $m_-$, the prefactor receives the corresponding matching factor at the point where the mass changes.  Including it in the same saddle-point calculation gives, for each $H_\alpha$,
\begin{align}
V^{\rm warped}_{H_\alpha}(\theta)
&\simeq
-\frac{4k^2}{\pi^2}
\frac{m_+^3m_-^3}{(m_++m_-)^4}
\left(\frac{e^{-2k\pi R}}{1-e^{-2k\pi R}}\right)^2
 e^{-\pi R(m_++m_-)}\cos(q\theta).
\label{eq:piecewise_scalar_warped_pot}
\end{align}
The displayed prefactor assumes $m_+\pi R\gg1$ and $m_-\pi R\gg1$.  For the scalar sector of a supersymmetric hypermultiplet,
\begin{equation}
M_0^2=M^2-\frac{15}{4}k^2,
\qquad
\mu^2=Mk,
\end{equation}
so that, for $M>k/2$, $m_\pm=M\pm k/2$.  The two halves of the winding trajectory then give
\begin{equation}
S_{\rm inst}=\pi R\left(M+\frac{k}{2}\right)+\pi R\left(M-\frac{k}{2}\right)=2M\pi R.
\end{equation}
Thus the $k$-dependent parts cancel directly in the scalar winding action, and the exponential suppression is the same $e^{-2M\pi R}$ as for the Dirac fermion.  This is qualitatively different from the simple constant-mass scalar, whose winding exponent involves $M_{\rm eff}=\sqrt{M^2+4k^2}$.

Using the unequal-mass prefactor in Eq.~\eqref{eq:piecewise_scalar_warped_pot} and the supersymmetric bulk mass relation, each scalar gives one half of Eq.~\eqref{eq:worldline_fermion_warped_pot} with the opposite sign.  The two scalar bulk contributions therefore cancel the fermion contribution.  This relative cancellation and the common exponent are unchanged by the matching factor.  A direct verification using the complete KK spectrum would additionally require the localized scalar terms fixed by the superfield action, and we leave that calculation for separate work.

\endgroup

\subsection{Axion potential in the KK spectral-function approach}
\label{subsec:KKmethod}

The worldline result can be cross-checked, and often reproduced more
efficiently, by an independent computation based on the axion-dependent
Kaluza--Klein (KK) spectrum. In this subsection, we evaluate the axion
potential induced by the C-twisted matter field
$\Phi=(\phi,\psi)$ with constant bulk masses using the KK spectral-function
approach. 

The starting point is the standard relation between the one-loop vacuum
energy and the KK mass eigenvalues.
For a matter field $\Phi=(\phi,\psi)$ with a $\theta$-dependent KK spectrum
$\{m_n^2(\theta)\}$, the one-loop axion potential can be written as
\begin{equation}
V_\Phi(\theta)
=
(-1)^{n_F}\frac{1}{2}
\sum_n
\int\!\frac{d^4p}{(2\pi)^4}\,
\ln\!\left(p^2+m_n^2(\theta)\right)
\;-\;(\theta\to0),
\label{eq:Vloop_KK_general}
\end{equation}
where $n_F$ denotes the fermion number, i.e. $n_F=0$ for $\phi$ and
$n_F=1$ for $\psi$, and $p$ is the 4D Euclidean momentum.
The subtraction $(\theta\to0)$ removes the divergent
$\theta$-independent vacuum energy and isolates the calculable
physical axion potential.

The KK mass eigenvalues are determined by the boundary conditions imposed on bulk fields on $S^1/\mathbb{Z}_2$.
In a warped background, however, the KK spectrum depends nontrivially  on the geometry. 
It is therefore convenient to encode the spectrum in a holomorphic spectral function ${\cal N}_\Phi(z;\theta)$ satisfying ${\cal N}_\Phi(z;\theta)={\cal N}_\Phi(-z;\theta)$~\cite{GrootNibbelink:2001bx,Choi:2002ps,Haba:2008sd,Choi:2010xn} and
\begin{equation}
{\cal N}_\Phi(z;\theta)= 
0 \quad {\rm iff}  \quad z= m_n(\theta).
\label{eq:Nfunction_def}
\end{equation}
The pole part of its logarithmic derivative is given by
\bea
\left.\frac{{\cal N}'_\Phi(z;\theta)}{{\cal N}_\Phi(z;\theta)}\right|_{\rm pole}
=\sum_n\frac{2z}{z^2-m_n^2(\theta)}  \,,
\eea
where ${\cal N}_\Phi' = d{\cal N}_\Phi/dz$, and the sum runs over all KK modes (counted with multiplicity). $(\cdots)|_{\rm pole}$ denotes the subtraction of the entire part of ${\cal N}'_\Phi/{\cal N}_\Phi$, leaving only its pole contribution. 
For any even function $f(z)$ analytic inside a contour ${\cal C}$ enclosing all KK eigenvalues on the real axis,
Cauchy's theorem implies
\bea
\sum_n f\!\left(m_n(\theta)\right)
= \frac{1}{2} \oint_{\cal C}\frac{dz}{2\pi i}\, f(z)\,\frac{{\cal N}'_\Phi(z;\theta)}{{\cal N}_\Phi(z;\theta)}\,.
\eea
Applying this identity to Eq.~\eqref{eq:Vloop_KK_general} yields
\begin{align}
V_{\Phi}(\theta)
&= \frac{1}{2}(-1)^{n_F}\int\!\frac{d^4p}{(2\pi)^4}\,
\oint_{\cal C}\frac{dz}{2\pi i}\,
\ln\!\left(p^2+z^2\right)\,
\frac{{\cal N}'_\Phi(z;\theta)}{2\, {\cal N}_\Phi(z;\theta)}
\;-\;(\theta\to 0) \nonumber
\\
&= \frac{1}{2}(-1)^{n_F+1}\int\!\frac{d^4p}{(2\pi)^4}\,
\oint_{\cal C}\frac{dz}{2\pi i}\,
\frac{z}{p^2+z^2}\,
\ln\frac{{\cal N}_\Phi(z;\theta)}{{\cal N}_\Phi(z;0)}\,,
\label{Effpotential}
\end{align}
where the second line follows from integrating by parts with respect to $z$.

\begin{figure}[t]
	\centering
\includegraphics[width=0.85\textwidth]{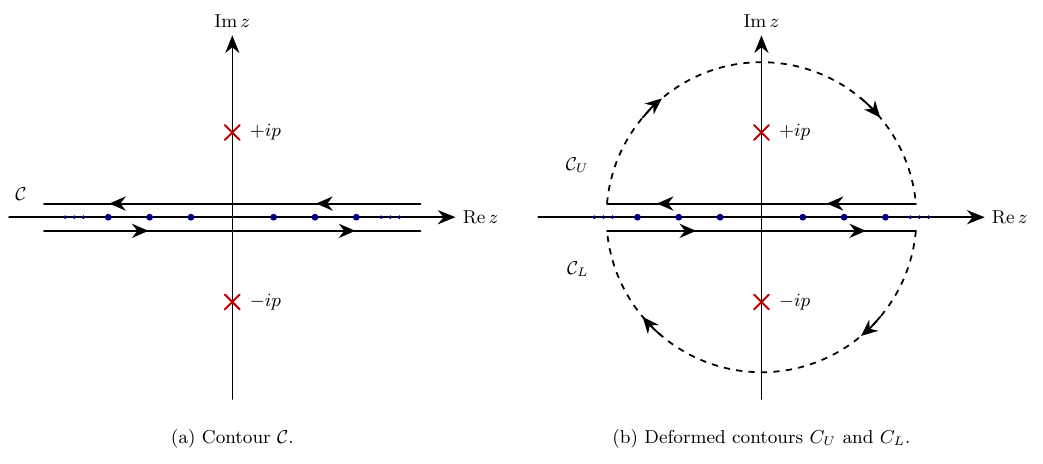}
	\caption{Contours for the integration of Eq.~\eqref{Effpotential}.} 
	\label{fig:contour_deformation}
\end{figure}

In Eq.~\eqref{Effpotential}, the factor
$1/(p^2+z^2)$ has simple poles at $z=\pm i p$.
The contour ${\cal C}$ may therefore be deformed into a pair of lines running just above ($+\infty\to-\infty$) and just below
($-\infty\to+\infty$) the real axis, which are then closed by large semicircles in the upper and lower half-planes (see Fig.~\ref{fig:contour_deformation}).
The resulting contours enclose the poles at $z=+ip$ and $z=-ip$, respectively.
To justify this deformation, we impose the following asymptotic boundary condition on the spectral function as $|z|\to\infty$:
\begin{equation}
\ln\frac{{\cal N}_\Phi(z;\theta)}{{\cal N}_\Phi(z;0)} \;\longrightarrow\; 0
\quad{\rm for} \quad |z|\to\infty .
\label{eq:Nratio_asymptotic}
\end{equation}
For a given KK spectrum $\{m_n(\theta)\}$, one can always choose the
corresponding spectral function so that it satisfies this asymptotic
condition. In particular, since the $\theta$ dependence originates from
nonlocal effects along the extra dimension, the ratio
\[
\frac{{\cal N}_\Phi(z;\theta)-{\cal N}_\Phi(z;0)}{{\cal N}_\Phi(z;0)}
\]
for such a spectral function is typically exponentially suppressed in
the limit $|z|\to\infty$.
Consequently, only the residues at $z=\pm ip$ contribute to the $\theta$-dependent part of the potential, yielding
\begin{equation}
V_{\Phi}(\theta)
= \frac{1}{2}(-1)^{n_F}\!
\int\!\frac{d^4p}{(2\pi)^4}\,
\ln\!\left[\frac{{\cal N}_\Phi(ip;\theta)}{{\cal N}_\Phi(ip;0)}\right],
\label{eq:Vloop_spectral}
\end{equation}
where ${\cal N}_\Phi$ is defined so as to count the real physical degrees of freedom in the spectrum, including the appropriate spin multiplicities.

A generic spectral function ${\cal N}_\Phi(z;\theta)$ ($\Phi=\phi,\psi$) can be written as  
\begin{equation}
{\cal N}_\Phi(z;\theta)
=\left[\hat{\cal N}_\Phi(z;\theta)\right]^{g_\Phi},
\end{equation}
with  
\begin{equation}
\hat{\cal N}_\Phi(z;\theta)
= N_\Phi(z)+{\cal A}_\Phi(z;\theta),
\end{equation}
where $g_\phi=1$ and $g_\psi=2$ denote the spin factor, and
\begin{equation}
N_\Phi(z)\equiv \hat{\cal N}_\Phi(z;0), 
\qquad
{\cal A}_\Phi(z;\theta)\equiv
\hat{\cal N}_\Phi(z;\theta)-\hat{\cal N}_\Phi(z;0).
\end{equation}
As shown in Appendix~\ref{app:Kaluza-Klein_spectral_functions}, for all cases considered in this paper one can choose a spectral function such that \({\cal A}_\Phi(z;\theta)\) is \emph{independent} of \(z\), i.e.,
\begin{equation}
{\cal A}_\Phi(z;\theta) = {\cal A}_\Phi(\theta),
\end{equation}
where \({\cal A}_\Phi(\theta)\) is a periodic function of \(\theta\).
The one-loop axion potential can then be written as
\begin{equation}
V_{\Phi}(\theta)
= (-1)^{n_F}\frac{g_\Phi}{2}  \int\!\frac{d^4p}{(2\pi)^4}\,
\ln\!\left(1+\frac{{\cal A}_\Phi(\theta)}{N_\Phi(i p)}\right),
\label{eq:spectral_master}
\end{equation}
which serves as the master formula for the axion potential in the
spectral-function approach.

\begingroup
The same formula provides an independent check of the absence of an axion potential from P-type matter when the $U(1)_C$-violating fixed-point operators are switched off.  In that case the P-type scalar spectral function is $\theta$ independent,
\begin{equation}
{\cal N}_{\tilde\phi}(z;\theta)\big|_{\tilde b_0=\tilde b_\pi=0}
={\cal N}_{\tilde\phi}(z;0),
\end{equation}
so Eq.~\eqref{eq:spectral_master} gives $V_{\tilde\phi}(\theta)=0$.  The explicit P-type spectral function in the presence of the fixed-point masses is given in Sec.~\ref{subsec:quadratic_loop} and Appendix~\ref{app:Kaluza-Klein_spectral_functions}.
\endgroup

In the flat limit $k=0$, it is straightforward to apply the
spectral-function formula Eq.~\eqref{eq:spectral_master} to compute the
axion potential. For this purpose, we consider a C-twisted charged
scalar field $\phi$ with bulk mass $M$, charge $q$, and the orbifold
boundary conditions
\bea
\phi(x, y+2\pi R)=\phi(x, y),\qquad
\phi(x, -y) = \pm \phi^*(x, y).
\eea
The 5D scalar field can be expanded as
\bea
\phi(x, y)=\sum_{n\in \mathbb{Z}} \chi_n(x) e^{iny/R},
\eea
where $\chi_n=\pm\chi_n^*$. The 5D equation of motion in the presence of
a constant background gauge field $C_5=\theta/2\pi R$ yields
\bea
\left[\left(\partial_y-iq\frac{\theta}{2\pi R}\right)^2 -M^2+m_n^2\right] e^{iny/R}=0,
\eea
where $m_n$ denotes the KK mass of $\chi_n$, which is either a real or
purely imaginary 4D scalar field depending on the $\mathbb{Z}_2$-parity
of $\phi$. One then finds
\bea
m_n^2 = M^2 +\left(\frac{n}{R}-\frac{q\theta}{2\pi R}\right)^2 \quad (n\in \mathbb{Z}) .
\eea
One also obtains the same KK spectrum for a C-twisted charged fermion
satisfying the boundary conditions
\bea
\psi(x, y+2\pi R)=\psi(x, y),\qquad
\psi(x, -y) = \pm \psi^c(x, y),
\eea
where $\psi^c={\cal C}\bar\psi^T$ with ${\cal C}$ denoting the 4D
charge-conjugation matrix.

The above KK spectrum can be encoded in the spectral function
\begin{equation}
{\cal N}_\Phi(z;\theta)=\left[N_\Phi(z) + {\cal A}_\Phi(\theta)\right]^{g_\Phi},
\label{eq:n_phi_z}
\end{equation}
where
\begin{equation}
N_\Phi(z) = -2\sin^2\!\left(\pi R \sqrt{z^2 - M^2}\right),\qquad
{\cal A}_\Phi(\theta)= 2\sin^2\!\left(\frac{q\theta}{2}\right).
\label{eq:flat_spectral_function}
\end{equation}
Plugging this spectral function into Eq.~\eqref{eq:spectral_master}, we obtain
\begin{align}
\hskip 0.2cm V_{\Phi}^{\rm flat}(\theta)
&=(-1)^{n_F}\frac{g_{\Phi}}{2}
\int\!\frac{d^4p}{(2\pi)^4}\,
\ln\!\left[
\frac{\cosh\Bigl(2\pi R\sqrt{p^2+M^2}\Bigr)-\cos(q\theta)}
{\cosh\Bigl(2\pi R\sqrt{p^2+M^2}\Bigr)-1}
\right] \nonumber \\
&\hskip -1.3cm =(-1)^{n_F}g_{\Phi} \sum_{n=1}^\infty \frac{1-\cos(nq\theta)}{n}
\int\!\frac{d^4p}{(2\pi)^4}\,
\exp\!\Bigl[-2n\pi R\sqrt{p^2+M^2}\Bigr] \nonumber \\
&\hskip -1.3cm =(-1)^{n_F}
\frac{g_{\Phi}}{64\pi^6 R^4}
\sum_{n=1}^\infty
\left(\frac{4\pi^2n^2M^2R^2 + 6\pi nMR+3}{n^5}\right)
e^{-2M\pi R n}\big(1-\cos(nq\theta)\big).
\end{align}
Here, $g_\Phi$ denotes the spin factor, with $g_\phi=1$ and $g_\psi=2$.
This result agrees with Eq.~\eqref{eq:scalar_flat_worldline}
obtained in the previous subsection using the worldline approach.

One may also apply the spectral-function formula Eq.~\eqref{eq:spectral_master} to compute the axion potential in the warped case. In this case, the spectral functions encoding the KK spectrum can be constructed from Bessel functions of the first and second kinds, as described in Appendix~\ref{app:Kaluza-Klein_spectral_functions}. However, when the warping is non-negligible, the resulting spectral functions are typically too complicated to yield a useful closed-form expression for the axion potential.

Here we are primarily interested in the behavior of the axion potential
in the large bulk-mass regime, $M\pi R\gg 1$, which is the regime most
relevant to the axion quality problem. In this regime, the first harmonic
of the axion potential can be determined directly from the spectral
function, yielding
\bea V_\Phi(\theta)=C^{\rm exact}_{\Phi,1}\cos(q\theta)+\cdots,\eea
where  
\begin{equation}
C^{\rm exact}_{\Phi,1}
=(-1)^{n_F+1}\frac{g_\Phi}{8\pi^2}
\int_0^\infty dp\,
\frac{p^3}
{N_\Phi(ip)+1+\sqrt{N_\Phi(ip)\big[N_\Phi(ip)+2\big]}} .
\label{eq:exact_first_harmonic_spectral}
\end{equation}
$N_\Phi(z)$ is the function appearing in the spectral function
defined in \eqref{eq:n_phi_z}.

For $M\pi R\gg1$, $N_\Phi(0)$ is exponentially large and the leading
winding contribution may be written in terms of the small-momentum
coefficients
\bea
N_0=N_\Phi(ip=0), \qquad
N_2=\left.\frac{dN_\Phi(ip)}{dp^2}\right|_{p^2=0}.
\label{eq:n0n2}
\eea
Keeping the leading term in the large-$N_0$ expansion and approximating its
momentum dependence by
\begin{equation}
\frac{1}{N_\Phi(ip)}\,
\simeq\,
\frac{1}{N_{0}}
\exp\!\left(-\frac{N_{2}}{N_{0}}\,p^2\right),
\label{eq:spectral_gaussian}
\end{equation}
gives
\begin{equation}
V_{\Phi}(\theta)
\simeq (-1)^{n_F}
\frac{g_\Phi}{32\pi^2}
\left(\frac{N_0}{N_2}\right)^{2}
\frac{{\cal A}_\Phi(\theta)}{N_0}.
\label{eq:spectral_pot}
\end{equation}
Equation~\eqref{eq:spectral_pot} is useful for obtaining analytic large-mass
expressions, while Eq.~\eqref{eq:exact_first_harmonic_spectral} provides a
direct numerical test of the finite-momentum approximation.

The spectral functions in warped geometry for a \(C\)-twisted charged scalar \(\phi\) and Dirac fermion \(\psi\) are derived in Appendix~\ref{app:Kaluza-Klein_spectral_functions}. Here, we present the result for \(\phi\) and defer the corresponding expression for \(\psi\) to Appendix~\ref{app:Kaluza-Klein_spectral_functions}. Following the procedure outlined there, the spectral function for a $C$-twisted scalar field $\phi$ is given by
\begin{equation}
{\cal N}_\phi(z; \theta)= N_\phi (z) +{\cal A}_\phi(\theta),
\end{equation}
where
\begin{equation}
N_\phi (z)
= \frac{\pi^2}{2}N_{++}^{\alpha}(z)N_{--}^{\alpha}(z), \qquad
{\cal A}_\phi(\theta)= 2\sin^2\!\left(\frac{q\theta}{2}\right).
\label{eq:N_phi}
\end{equation}
Here $N^{\alpha}_{++}(z)$ and $N^{\alpha}_{--}(z)$ are given by combinations of the Bessel functions $J_\alpha$ and $Y_\alpha$:
\begin{align}
N^{\alpha}_{++}(z)
&=\left[2 Y_\alpha \!\left(\frac{z}{k}\right)
+ \frac{z}{k} Y'_\alpha \!\left(\frac{z}{k}\right)\right]
\left[2 J_\alpha\!\left(\frac{z}{k e^{-k\pi R}}\right)
+\frac{z}{k e^{-k\pi R}} J'_\alpha\!\left(\frac{z}{k e^{-k\pi R}}\right)\right]
\nonumber\\
&\quad-
\left[2 J_\alpha\!\left(\frac{z}{k}\right)
+\frac{z}{k} J'_\alpha\!\left(\frac{z}{k}\right)\right]
\left[2 Y_\alpha\!\left(\frac{z}{k e^{-k\pi R}}\right)
+\frac{z}{k e^{-k\pi R}} Y'_\alpha\!\left(\frac{z}{k e^{-k\pi R}}\right)\right],
\nonumber\\[4pt]
N^{\alpha}_{--}(z)
&=J_{\alpha}\!\left(\frac{z}{k}\right)
Y_{\alpha}\!\left(\frac{z}{k e^{-k\pi R}}\right)
- J_{\alpha}\!\left(\frac{z}{k e^{-k\pi R}}\right)
Y_{\alpha}\!\left(\frac{z}{k}\right).
\label{eq:N++_N--}
\end{align}
The parameter $\alpha$ is defined as
\begin{equation}
\alpha=\sqrt{\frac{M^2}{k^2}+4},
\end{equation}
where $M$ denotes the bulk mass of $\phi$ and $k$ the AdS curvature scale.
In the flat limit $k\to 0$, the above spectral function reduces to Eq.~(\ref{eq:flat_spectral_function}).

\begingroup
The small-momentum coefficients $N_0$ and $N_2$ can be evaluated without first taking the strong-warping limit.  Keeping the full $k\pi R$ dependence and taking the large-mass winding limit $M\pi R\gg1$, Eq.~\eqref{eq:spectral_pot} gives
\begin{align}
V^{\rm warped}_{\phi}(\theta)
&\simeq
-\frac{k^2 M_{\rm eff}^2(M_{\rm eff}^2-4k^2)(M_{\rm eff}^2-k^2)^2}
{4\pi^2\left(e^{2k\pi R}-1\right)^2
\left[M_{\rm eff}(M_{\rm eff}^2-7k^2)+6k^3\coth(k\pi R)\right]^2}
\,e^{-2M_{\rm eff}\pi R}\cos(q\theta),
\label{eq:pot_general_warping_scalar}\\[2mm]
V^{\rm warped}_{\psi}(\theta)
&\simeq
\frac{k^2(4M^2-k^2)^4}
{32\pi^2\left(e^{2k\pi R}-1\right)^2
\left[M(4M^2-3k^2)+k^3\coth(k\pi R)\right]^2}
\,e^{-2M\pi R}\cos(q\theta),
\label{eq:pot_general_warping_fermion}
\end{align}
where $M_{\rm eff}=\sqrt{M^2+4k^2}$.  The fermion expression in Eq.~\eqref{eq:pot_general_warping_fermion} is the fixed-$k/M$ large-$M\pi R$ limit for $k/M\neq2$.  At the isolated point $k=2M$, the order of limits is nonuniform; evaluating the quadratic coefficient before taking $M\pi R\gg1$ gives
\begin{equation}
V^{\rm warped}_{\psi}(\theta)\big|_{k=2M}
\simeq
\frac{32M^4}{9\pi^2}e^{-10M\pi R}\cos(q\theta).
\label{eq:fermion_k2M}
\end{equation}

Unlike the strong-warping expressions used previously, Eqs.~\eqref{eq:pot_general_warping_scalar} and \eqref{eq:pot_general_warping_fermion} retain the full dependence on $k\pi R$.  In particular, their flat-space limits are
\begin{align}
V^{\rm warped}_{\phi}(\theta)&\xrightarrow{k\to0}
-\frac{M^2}{16\pi^4R^2}e^{-2M\pi R}\cos(q\theta),\nonumber\\
V^{\rm warped}_{\psi}(\theta)&\xrightarrow{k\to0}
+\frac{M^2}{8\pi^4R^2}e^{-2M\pi R}\cos(q\theta),
\label{eq:KK_flat_limits}
\end{align}
in agreement with the leading large-$M\pi R$ harmonic of the exact flat-space result above and with the worldline calculation.

For $k\pi R\gg1$, the same expressions reduce to the strong-warping forms
\begin{align}
V^{\rm warped}_{\phi}(\theta) &\simeq  -
\frac{(M^2+4k^2)k^2}{4\pi^2}
{\cal G}(M,k)\,
e^{-4k\pi R} e^{-2M_{\rm eff}\pi R}\,
\cos(q\theta), \label{eq:pot_strong_warping_scalar}\\
V^{\rm warped}_{\psi}(\theta) &\simeq
 \frac{\left(M+\frac{1}{2}k\right)^2 k^2}{2\pi^2}
\left(\frac{M+\frac{1}{2}k}{M+k}\right)^2
e^{-4k\pi R} e^{-2M\pi R}\,
\cos(q\theta),
\label{eq:pot_strong_warping_fermion}
\end{align}
where
\begin{equation}
{\cal G}(M,k)=
\frac{(M_{\rm eff}+2k)^2(M_{\rm eff}+k)^2}
{M^2(M_{\rm eff}+3k)^2}.
\end{equation}
\endgroup

\begingroup
The Gaussian approximation in Eq.~\eqref{eq:spectral_gaussian} replaces the full momentum dependence relevant for Eq.~\eqref{eq:exact_first_harmonic_spectral} by its quadratic expansion around $p^2=0$.  We test this approximation by evaluating Eq.~\eqref{eq:exact_first_harmonic_spectral} numerically and comparing it with the analytic large-$M\pi R$ coefficients in Eqs.~\eqref{eq:pot_general_warping_scalar} and \eqref{eq:pot_general_warping_fermion}.  In Fig.~\ref{fig:numerical_spectral_coefficients} we fix $M\pi R=6\pi$ and vary $10^{-2}\leq k/M\leq10^2$.

\begin{figure}[t]
\centering
\begin{minipage}{0.48\textwidth}
\centering
\includegraphics[width=\textwidth]{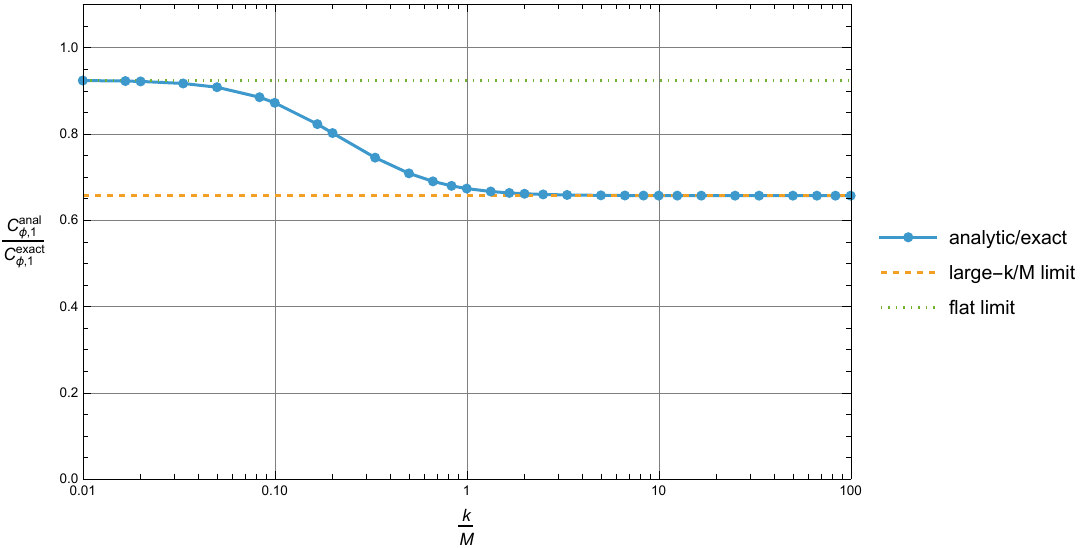}
\\[-2mm]
(a) C-twisted scalar
\end{minipage}
\hfill
\begin{minipage}{0.48\textwidth}
\centering
\includegraphics[width=\textwidth]{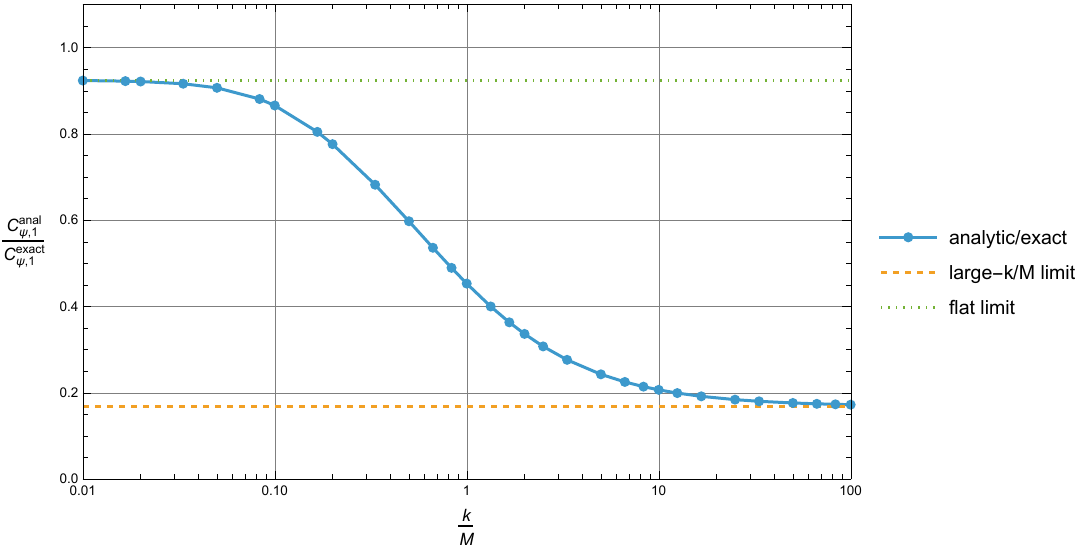}
\\[-2mm]
(b) C-twisted fermion
\end{minipage}
\caption{Ratio $C_{\Phi,1}^{\rm anal}/C_{\Phi,1}^{\rm exact}$ for the C-twisted scalar and fermion, with $M\pi R=6\pi$ and $10^{-2}\leq k/M\leq10^2$.  The coefficient $C_{\Phi,1}^{\rm exact}$ is obtained from the full first-harmonic momentum integral in Eq.~\eqref{eq:exact_first_harmonic_spectral}, while $C_{\Phi,1}^{\rm anal}$ uses the Gaussian large-mass expressions.  The dotted and dashed lines denote the flat-space and large-$k/M$ limiting values, respectively.}
\label{fig:numerical_spectral_coefficients}
\end{figure}

The flat-space reference value is fixed analytically.  Using the exact flat-space result and its leading large-$M\pi R$ limit gives
\begin{equation}
\left.\frac{C_{\Phi,1}^{\rm anal}}{C_{\Phi,1}^{\rm exact}}\right|_{k/M\to0}
=\frac{4(M\pi R)^2}{4(M\pi R)^2+6M\pi R+3}
=0.9244807\ldots
\qquad (M\pi R=6\pi),
\label{eq:flat_ratio_numerical}
\end{equation}
for both the scalar and fermion.  Thus the fact that the curves approach a value below unity at small $k/M$ reflects the subleading finite-$M\pi R$ terms omitted from the analytic large-mass expression.  Away from the flat limit, the ratio remains finite and approaches a constant also for large $k/M$, although the finite coefficient correction is considerably larger for the fermion.  The analytic result therefore captures the winding exponent and the parametric dependence throughout the scan, while the numerical integral determines the remaining finite coefficient.

For the scalar, the opposite hierarchy $k\gg M$ gives a result that is not visible in the bulk worldline expression.  For $M\pi R\gg1$ and $k\pi R\gg1$, Eq.~\eqref{eq:pot_strong_warping_scalar} reduces to
\begin{equation}
V^{\rm warped}_{\phi}(\theta)
\simeq
-\frac{144}{25\pi^2}\frac{k^6}{M^2}
 e^{-4k\pi R}e^{-2M_{\rm eff}\pi R}\cos(q\theta),
\qquad k\gg M.
\label{eq:scalar_large_k_scaling}
\end{equation}
The first corrections to the displayed coefficient are of order $M^2/k^2$.  The additional factor $k^2/M^2$ relative to the $k^4$ behavior suggested by the bulk worldline expression arises from the fixed-point term generated by the scalar field redefinition and is reproduced by the numerical momentum integral.
\endgroup

\begingroup
For $M\gg k$, the KK result including the fixed-point terms reduces to the bulk worldline prefactor at leading order in $k/M$, for arbitrary $k\pi R$ and $M\pi R\gg1$.  When $k$ is comparable to or larger than $M$, the fixed-point matching can instead change the prefactor appreciably, while leaving the winding exponent unchanged.  The general expressions in Eqs.~\eqref{eq:pot_general_warping_scalar} and \eqref{eq:pot_general_warping_fermion} include this effect.  Appendix~\ref{app:common_determinant} shows directly how the same modification appears in the proper-time calculation.  The separate error associated with the Gaussian momentum approximation is measured by the numerical integral in Fig.~\ref{fig:numerical_spectral_coefficients}.

For fermions, the spectral-function method also determines the winding exponent for $M<k/2$, where the bulk semiclassical saddle used in the previous subsection is not applicable.  The suppression remains $\exp(-2M\pi R)$ on both sides of $M=k/2$, in agreement with the monodromy calculation discussed below.
\endgroup

\subsection{Axion potential in the monodromy-matrix approach for fermions}
\label{subsec:fermion_det}

For fermions, it is useful to complement the worldline approach with a determinant-based formulation, in which the dependence on the background axion field $\theta$ becomes more transparent. After reducing the 5D Dirac operator to a first-order differential operator along the fifth dimension, the $\theta$-dependence enters only through a finite-dimensional monodromy matrix~\cite{Forman:1987xk,Kirsten:2003py}. The resulting determinant can be written in terms of the $2\times2$ monodromy matrix and compared directly with the results of the previous two subsections.


Consider a $C$-twisted Dirac fermion $\psi$ with charge $q$ and mass $M$.
In the RS geometry Eq.~\eqref{eq:RSmetric_y} and in the presence of a constant
background axion field $C_5=\theta/2\pi R$, the Euclidean Dirac operator for $\psi$ in the 4D momentum basis is given by
\begin{equation}
{D}_\theta(p)
= - e^{k|y|} \gamma^\mu p_{\mu}
-\gamma^5\left(\partial_y-iq \frac{\theta}{2\pi R}- 2 k\,\epsilon(y) \right)
+ M,
\end{equation}
where $p^\mu$ denotes the Euclidean four-momentum.
In a chiral basis, this operator takes the form
\begin{equation}
{D}_\theta(p)=
\begin{pmatrix}
\partial_y -iq\frac{\theta}{2\pi R}- 2k\,\epsilon(y)+ M  & - e^{k|y|} \sigma\!\cdot p \\
- e^{k|y|} \bar\sigma\!\cdot p & -\partial_y +iq\frac{\theta}{2\pi R} + 2k\,\epsilon(y) + M
\end{pmatrix},
\end{equation}
where
$\sigma\!\cdot p= \sigma^\mu p_\mu$ and
$\bar\sigma\!\cdot p= \bar\sigma^\mu p_\mu$
for the standard Euclidean sigma matrices $\sigma^\mu$ and $\bar\sigma^\mu$.

Upon replacing $\sigma\!\cdot p$ and $\bar\sigma\!\cdot p$ by their eigenvalues $\pm |p|$, the Dirac operator ${D}_\theta(p)$ reduces to two identical $2\times 2$ blocks:
\begin{equation}
{D}_\theta(p)=
\begin{pmatrix}
 \partial_y -iq\frac{\theta}{2\pi R} -2 k\,\epsilon(y)  + M & -e^{k|y|} |p| \\
 e^{k|y|} |p| & -\partial_y +iq\frac{\theta}{2\pi R} +2 k\,\epsilon(y) + M
\end{pmatrix} \otimes \mathbf{1}_2 .
\end{equation}
This shows that the $4\times 4$ matrix-valued 5D Dirac operator $D_\theta(p)$ reduces to a $2\times 2$ matrix-valued 1D first-order operator of the form
\begin{equation} 
{\cal D}_H=\partial_y - H_\theta(y;p),
\label{eq:evol_chi}
\end{equation}
where $H_\theta(y;p)$ can be decomposed as
\begin{equation}
H_\theta(y;p)
=
\Big(2k\,\epsilon(y)+iq\,\frac{\theta}{2\pi R}\Big)\mathbf{1}_2
+ h(y;|p|),
\label{eq:Htheta_decomp}
\end{equation}
with the traceless part
\begin{equation}
h(y;|p|)
=
\begin{pmatrix}
- M & e^{k|y|}|p| \\
 e^{k|y|}|p| & +M
\end{pmatrix}.
\label{eq:h_matrix}
\end{equation}
As a consequence, the determinant ratio for $D_\theta(p)$ reduces to that of ${\cal D}_H$, yielding
\begin{equation}
\left.\ln\frac{\det D_\theta(p)}{\det D_0(p)}\right|_{S^1/\mathbb{Z}_2}
=
\ln
\frac{\det (\partial_y-H_\theta)}{\det (\partial_y-H_0)}.
\end{equation}

The solution of the 1D  equation
\begin{equation}
\left(\partial_y - H_\theta(y;p)\right)\chi(y)=0,
\qquad
\chi(y+2\pi R)=\chi(y),
\label{eq:1d_dirac_eq}
\end{equation}
can be written as
\begin{equation}
\chi(y)=U_\theta(y;p)\,\chi(0),
\end{equation}
{where $U_\theta(y;p)$ is the transfer matrix from $0$ to $y$, satisfying}
\begin{equation}
\Big(\partial_y - H_\theta(y;p)\Big)\, U_\theta(y;p)=0,
\qquad
U_\theta(0;p)=\mathbf{1}_2.
\end{equation}
{The transfer matrix can be written as}
\begin{align}
U_\theta(y;p)
&=
\exp\!\left[\int_0^{y} dy'\,
\Big(2k\,\epsilon(y')+iq\frac{\theta}{2\pi R}\Big)\right]
\nonumber\\
&\qquad\times
\mathcal{P}\exp\!\left[\int_0^{y} dy'\,
h(y';|p|)\right],
\label{eq:Utheta_Uh}
\end{align}
where $\mathcal{P}$ denotes path ordering.
The corresponding one-period monodromy matrix is
\begin{equation}
U_\theta(2\pi R;p)=e^{iq\theta}\,U_h(p),
\end{equation}
where the axion background appears as an overall phase, while the effects of the warped geometry and the bulk mass are fully encoded in
\begin{equation}
U_h(p)
\equiv
U_0(2\pi R;p)
=
\mathcal{P}\exp\!\left[\int_0^{2\pi R} dy\, h(y;|p|)\right].
\label{eq:U_hp}
\end{equation}
A nontrivial solution of Eq.~\eqref{eq:1d_dirac_eq} exists only if the monodromy matrix satisfies
\begin{equation}
\det \bigl(\mathbf{1}_2 -e^{iq\theta} U_h(p)\bigr)=0.
\end{equation}
This implies that the determinant ratio of the 1D first-order operator ${\cal D}_H=\partial_y - H_\theta$
 on the covering circle $S^1$ is determined by the monodromy matrix as~\cite{Forman:1987xk,Kirsten:2003py}
\begin{equation}
\ln
\frac{\det (\partial_y-H_\theta)}{\det (\partial_y-H_0)}
=
\ln\frac{\det \bigl(\mathbf{1}_2 - e^{iq\theta}U_h(p)\bigr)}{\det \bigl(\mathbf{1}_2 - U_h(p)\bigr)}.
\end{equation}

The axion potential induced by a $C$-twisted fermion $\psi$ is then given by
\begin{align}
V_\psi(\theta)
&=
-\frac{1}{2}\int\!\frac{d^4p}{(2\pi)^4}\,
\ln
\frac{\det \bigl(D_\theta(p)D^\dagger_\theta(p)\bigr)}{\det \bigl(D_{0}(p)D_{0}^\dagger(p)\bigr)}
\nonumber \\
&=
-\int\!\frac{d^4p}{(2\pi)^4}\,
\ln\left|\frac{\det\!\left(\mathbf{1}_2 - e^{iq\theta}U_h(p)\right)}
{\det\!\left(\mathbf{1}_2 - U_h(p)\right)}\right|.
\end{align}
Since ${\rm tr}\,h(y;|p|)=0$, it follows that $\det U_h(p)=1$, and the
eigenvalues of $U_h(p)$ can be parameterized as
\begin{equation}
{\rm Diag}\,(U_h) =
\begin{pmatrix}
e^{+\ell(p)} & 0 \\
0 & e^{-\ell(p)}
\end{pmatrix},
\qquad \ell(p)\ge 0.
\label{eq:spec_Uh}
\end{equation}
Using this parameterization, one finds
\begin{equation}
\left|\frac{\det\!\left(\mathbf{1}_2 - e^{iq\theta}U_h(p)\right)}
{\det\!\left(\mathbf{1}_2 - U_h(p)\right)}\right|
=\frac{\cosh \ell(p)-\cos(q\theta)}{\cosh \ell(p)-1},
\label{eq:det_eigen}
\end{equation}
which yields
\begin{equation}
V_\psi(\theta)
=
- \int\!\frac{d^4p}{(2\pi)^4}\,
\ln\!\left[\frac{\cosh \ell(p)-\cos(q\theta)}{\cosh \ell(p)-1}\right].
\end{equation}
Equivalently, after dropping the $\theta$-independent part,
\begin{align}
V_\psi(\theta)
&=
- \int\!\frac{d^4p}{(2\pi)^4}
\ln\!\Big(1-e^{-\ell(p)+i q\theta}\Big)
+\text{h.c.}
\nonumber \\
&=
\sum_{n=1}^{\infty} \frac{2\cos(n q\theta)}{n}
\int\!\frac{d^4p}{(2\pi)^4} e^{-n\ell(p)} .
\label{eq:Vf_log_r}
\end{align}
Thus, in the monodromy-matrix approach each winding sector $n$
is weighted by $e^{-n\ell(p)}$, where $\ell(p)$ represents the
effective Euclidean length extracted from the eigenvalues of the 
matrix $U_h(p)$.

In the flat limit $k=0$, $H_\theta(y;p)$ becomes \emph{$y$-independent},
so that the corresponding monodromy matrix can be obtained straightforwardly as
\bea
U_h(p) =
\exp\!\left[
2\pi R
\begin{pmatrix}
- M & |p| \\
|p| & M
\end{pmatrix}
\right],
\eea
for which
\begin{equation}
\ell(p)=2\pi R\sqrt{M^2+p^2}.
\end{equation}
This leads to
\bea
V_\psi(\theta)
&=&
\sum_{n=1}^{\infty} \frac{2\cos(n q\theta)}{n}
\int\!\frac{d^4p}{(2\pi)^4}\,
e^{-2n\pi R\sqrt{p^2+M^2}},
\eea
which agrees with the results obtained in the previous two subsections using the worldline
and KK spectral-function methods.

 To evaluate $\ell(p)$ in the warped case, we consider a momentum expansion,
which can be obtained from a Dyson expansion of the path-ordered exponential defining $U_h(p)$. 
We begin by decomposing the matrix $h(y;|p|)$ in Eq.~\eqref{eq:h_matrix} as
\begin{equation}
h(y;|p|)=-M\sigma_3 + |p|e^{k|y|}\sigma_1,
\end{equation}
and introducing
\begin{equation}
U_I(y;|p|)\equiv e^{My\sigma_3}\,
{\cal P}\exp\!\left[\int_0^y dy'\, h(y';|p|)\right].
\end{equation}
The matrix $U_I(y;|p|)$ then satisfies
\begin{equation}
\partial_y U_I(y;|p|)=|p|\,W(y)\,U_I(y;|p|),
\end{equation}
with
\begin{equation}
W(y)\equiv
e^{k|y|}
\left(
e^{2My}\sigma_+ + e^{-2My}\sigma_-
\right),
\qquad
\sigma_\pm\equiv \frac{\sigma_1\pm i\sigma_2}{2}.
\end{equation}
The corresponding Dyson expansion of $U_I(2\pi R;|p|)$ is
\begin{equation}
U_I(2\pi R;|p|)=
1+K_1|p|+K_2|p|^2+\mathcal{O}(|p|^3),
\label{eq:dyson_ex_u_I}
\end{equation}
where
\begin{align}
K_1&=\int_0^{2\pi R}dy\, W(y), \\
K_2&=\int_{0<y_2<y_1<2\pi R}dy_1dy_2\, W(y_1)W(y_2).
\end{align}
From this expansion, we obtain
\bea
\frac{1}{2}{\rm Tr} \, U_h(p)
&=&
\cosh \ell(p)\, =\,
\frac{1}{2}{\rm Tr}\!\left(e^{-2M\pi R\sigma_3}U_I(2\pi R;|p|)\right)
\nonumber\\
&&\hskip -2.5cm =
\cosh(2M\pi R)
+
|p|^2
\int_{0<y_2<y_1<2\pi R}dy_1dy_2\,
e^{k|y_1|+k|y_2|}
\cosh\!\left[M\!\left(2\pi R-2(y_1-y_2)\right)\right]\nonumber \\
\hskip -1cm
+\,\mathcal{O}(|p|^4).
\label{eq:TraceUh}
\eea

On the other hand, expanding \(\ell(p)\) as
\bea
\ell(p)=2M\pi R+C_2|p|^2+{\cal O}(|p|^4),
\label{eq:ell_p_expansion}
\eea
and matching to Eq.~\eqref{eq:TraceUh}, we obtain
\bea
C_2
&=&
\int_{0<y_2<y_1<2\pi R}dy_1dy_2\,
e^{k|y_1|+k|y_2|}
\frac{
\cosh\!\left[M\!\left(2\pi R-2(y_1-y_2)\right)\right]
}{
\sinh(2M\pi R)
}
\nonumber\\
&=&
\frac{e^{2k\pi R}}{(1-e^{-4M\pi R})}
\frac{2(k+M)}{k(k+2M)^2}
\Bigg[
\left(1-e^{-(k+2M)\pi R}\right)^2  
\nonumber\\
&& \hskip 2.8cm
+\, e^{-4M\pi R} \frac{(k-M)(k+2M)^2}{(k+M)(k-2M)^2}
\left(1-e^{-(k-2M)\pi R}\right)^2
\Bigg].
\eea
For $M\pi R\gg1$, the coefficient $C_2$ has simple asymptotic forms in the two hierarchies below.  The first retains the full $k\pi R$ dependence for $M\gg k$, while the second also assumes $k\pi R\gg1$ for $k\gg M$:
\begin{equation}
C_2 \simeq
\frac{e^{2k\pi R}-1}{2kM} \quad (M\gg k), 
\qquad
\frac{2  e^{2k \pi R}}{k^2}\left( 1 -\frac{3M}{k}\right) \quad (k\gg M).
\label{eq:C2_limit}
\end{equation}

\begingroup
To evaluate the axion potential in Eq.~\eqref{eq:Vf_log_r}, we use the small-momentum Gaussian approximation
\bea
e^{-\ell(p)}\,\simeq \, e^{-2M\pi R -C_2 |p|^2}.
\label{eq:gaussian_approx_ell}
\eea
For $M\gg k$ and $M\pi R\gg1$, retaining the full $k\pi R$ dependence, the leading harmonic is
\bea
V_\psi(\theta)\;\simeq\; \frac{M^2 k^2}{2\pi^2}\,
\frac{e^{-2M\pi R}}{(e^{2k\pi R}-1)^2}\cos(q\theta)
\qquad (M\gg k).
\label{eq:Vf_warped_asympt_prefactor}
\eea
This agrees with the leading $M\gg k$ limits of the worldline and KK spectral-function results; the corrections to the displayed prefactor are suppressed by powers of $k/M$.

For the opposite hierarchy $k\gg M$, with $M\pi R\gg1$ and $k\pi R\gg1$, Eq.~\eqref{eq:C2_limit} gives instead
\begin{equation}
V_\psi(\theta)\simeq \frac{k^4}{32\pi^2}
 e^{-4\pi kR}e^{-2M\pi R}\cos(q\theta),
\qquad k\gg M.
\label{eq:Vf_monodromy_large_k}
\end{equation}
This is the $k\gg M$ limit of the KK result in Eq.~\eqref{eq:pot_strong_warping_fermion}, which includes the fixed-point term in the second-order equation.  The direct first-harmonic KK momentum integral in Sec.~\ref{subsec:KKmethod} confirms the same winding exponent and $k^4$ scaling.  Higher-order momentum dependence beyond the Gaussian approximation can change the finite numerical coefficient relative to the Gaussian value $1/(32\pi^2)$, as quantified in Fig.~\ref{fig:numerical_spectral_coefficients}.
\endgroup


Compared with the worldline and KK spectral-function approaches presented in
the previous two subsections, the derivation of the fermion-induced axion
potential based on the monodromy-matrix formulation more transparently
isolates the axion dependence of the problem.
It reduces the functional determinant of the Dirac operator to an ordinary
determinant governed by the $2\times2$ monodromy matrix $e^{iq\theta}U_h(p)$.
The $\theta$ dependence of the functional determinant thus enters solely
through the simple phase $e^{iq\theta}$ appearing in the monodromy matrix,
while its coefficient is determined by the eigenvalues of the matrix $U_h(p)$.

\begingroup
The same transfer-matrix method can be applied to the scalar second-order equation and to the second-order fermion equation, with the fixed-point terms imposed through the matching conditions at $y=0$ and $y=\pi R$.  Appendix~\ref{app:completed_monodromy} gives these calculations and shows explicitly how the fermion transfer matrix is related to the first-order Dirac monodromy used above.  The determinant ratios obtained there agree with the KK and proper-time calculations of Appendix~\ref{app:common_determinant}.
\endgroup

\section{Additional potentials in the presence of  fixed-point localized operators}
\label{sec:fixedpoint}

In this section we incorporate the $U(1)_C$-violating fixed-point operators in
Eq.~\eqref{eq:appC_brane} and examine the additional axion potentials that arise from the
interplay between the bulk $U(1)_C$ gauge interactions and the fixed-point localized terms.
As emphasized in Section~\ref{sec:potential_loop}, a $\theta$-dependent potential can arise
only from processes that are nonlocal along the compact direction.
Fixed-point interactions provide new realizations of such nonlocality, since
a localized interaction at one fixed point can communicate with that at the other fixed point
through bulk propagation, generating additional contributions to the axion potential.

As specific examples, we discuss three additional contributions to the axion potential:
(1) a tree-level potential induced by linear scalar terms localized at the fixed points;
(2) a loop-induced potential from a P-type $\mathbb{Z}_2$-even scalar field $\tilde\phi$ in the presence of
$U(1)_C$-violating fixed-point mass terms; and
(3) the modification of the potential induced by loops of the C-twisted scalar $\phi$
in the presence of $U(1)_C$-violating fixed-point mass terms.
Note that P-type matter fields do not generate any axion potential in the absence of
$U(1)_C$-violating fixed-point operators.

\subsection{Tree-level potential from linear scalar terms}
\label{subsec:linear_tree}
\begingroup

For a charged scalar with even-integer charge $q$, the allowed fixed-point localized operators include the linear terms
\begin{equation}
\Delta{\cal L}_i(y=y_i)=J_i\phi+\tilde J_i\tilde\phi+{\rm h.c.},
\qquad (y_0=0,\;y_\pi=\pi R).
\label{eq:Vi_expand_sec5}
\end{equation}
We take the P-type scalar $\tilde\phi$ to be $\mathbb Z_2$ even, and for definiteness consider a $\mathbb Z_2$-even C-twisted scalar satisfying $\phi(x,-y)=\phi^*(x,y)$, for which $J_{0,\pi}$ may be chosen real.

The linear terms induce $\theta$-dependent classical scalar profiles.  We write
\begin{align}
\langle\phi(x,y)\rangle
&=e^{2k|y|}e^{iq\theta y/2\pi R}v(y),
\nonumber\\
\langle\tilde\phi(x,y)\rangle
&=e^{2k|y|}e^{iq\theta|y|/2\pi R}\tilde v(y),
\label{eq:tree_background_ansatz}
\end{align}
for which
\begin{align}
(\partial_y^2-M_{\rm eff}^2)v(y)
&=\delta(y)\big[J_0-4kv(0)\big]
+\delta(y-\pi R)\big[e^{-2k\pi R}e^{-iq\theta/2}J_\pi+4kv(\pi R)\big],
\nonumber\\
(\partial_y^2-M_{\rm eff}^2)\tilde v(y)
&=\delta(y)\big[\tilde J_0^*-4k\tilde v(0)\big]
+\delta(y-\pi R)\big[e^{-2k\pi R}e^{-iq\theta/2}\tilde J_\pi^*+4k\tilde v(\pi R)\big],
\label{eq:tree_eom}
\end{align}
where $M_{\rm eff}^2=M^2+4k^2$.  
Since the original scalar fields are periodic on the covering circle,
the rescaled profiles obey
\begin{eqnarray}
     v(-y)=v^*(y), && \qquad 
 v(y+2\pi R)=e^{-iq\theta}v(y),\\
\tilde v(-y)=\tilde v(y),
&& \qquad \tilde v(y+2\pi R)=\tilde v(y).
\end{eqnarray}
Integrating the equations of motion across the fixed points and
imposing the orbifold and periodicity conditions gives the following
boundary conditions for the C-twisted scalar
\begin{eqnarray}
\operatorname{Re}\!\left[(\partial_y+2k)v\right]_{0^+}
&=& \frac{J_0}{2}, \nonumber \\
\operatorname{Re}\!\left[
e^{iq\theta/2}(\partial_y+2k)v\right]_{\pi R^-}
&=& -\frac{e^{-2k\pi R}J_\pi}{2}.
\end{eqnarray}
For the P-type scalar, the corresponding boundary conditions are
\begin{eqnarray}
\left[(\partial_y+2k)\tilde v\right]_{0^+}
&=& \frac{\tilde J_0^*}{2}, \nonumber \\
\left[(\partial_y+2k)\tilde v\right]_{\pi R^-}
&=& -\frac{e^{-2k\pi R}e^{-iq\theta/2}\tilde J_\pi^*}{2}.
\end{eqnarray}
The bulk solutions satisfying these conditions are
\begin{eqnarray}
v(y)
&=&-\frac{J_0}{2\xi(\theta)}\Bigg\{
\Bigl[
M_{\rm eff}\cosh(M_{\rm eff}\pi R)
+2k\sinh(M_{\rm eff}\pi R)
\Bigr]
\sinh\!\left[M_{\rm eff}(\pi R-y)\right]
\nonumber\\
&&\qquad\qquad
+M_{\rm eff}e^{-iq\theta/2}
\cos\!\left(\frac{q\theta}{2}\right)
\sinh(M_{\rm eff}y)
\Bigg\}
\nonumber\\[1mm]
&&-\frac{e^{-2k\pi R}J_\pi}{2\xi(\theta)}\Bigg\{
M_{\rm eff}\cos\!\left(\frac{q\theta}{2}\right)
\sinh\!\left[M_{\rm eff}(\pi R-y)\right]
\nonumber\\
&&\qquad\qquad
+e^{-iq\theta/2}\Bigl[
M_{\rm eff}\cosh(M_{\rm eff}\pi R)
-2k\sinh(M_{\rm eff}\pi R)
\Bigr]\sinh(M_{\rm eff}y)
\Bigg\},
\\[3mm]
\tilde v(y)
&=&-\frac{M_{\rm eff}\tilde J_0^*}
{2M^2\sinh(M_{\rm eff}\pi R)}
\Bigg[
\cosh\!\left[M_{\rm eff}(\pi R-y)\right]
+\frac{2k}{M_{\rm eff}}
\sinh\!\left[M_{\rm eff}(\pi R-y)\right]
\Bigg]
\nonumber\\[1mm]
&&-\frac{M_{\rm eff}e^{-2k\pi R}e^{-iq\theta/2}\tilde J_\pi^*}
{2M^2\sinh(M_{\rm eff}\pi R)}
\Bigg[
\cosh(M_{\rm eff}y)
-\frac{2k}{M_{\rm eff}}\sinh(M_{\rm eff}y)
\Bigg].
\end{eqnarray}
for $0\leq y\leq\pi R$, where $\xi(\theta)$ is given 
\begin{equation}
\xi(\theta)=\frac12M_{\rm eff}^2
\Big[\cosh(2M_{\rm eff}\pi R)-\cos(q\theta)\Big]
-4k^2\sinh^2(M_{\rm eff}\pi R).
\label{eq:tree_alpha_def}
\end{equation}
Substituting these profiles into the action gives the exact tree-level potentials,  
\begin{align}
V^{\rm tree}_{\phi}(\theta)
&=-\frac{J_0^2}{4\xi(\theta)}
\Big[M_{\rm eff}\sinh(2M_{\rm eff}\pi R)+4k\sinh^2(M_{\rm eff}\pi R)\Big]
\nonumber\\
&\quad
-\frac{J_\pi^2}{4\xi(\theta)}e^{-4k\pi R}
\Big[M_{\rm eff}\sinh(2M_{\rm eff}\pi R)-4k\sinh^2(M_{\rm eff}\pi R)\Big]
\nonumber\\
&\quad
-\frac{J_0J_\pi}{\xi(\theta)}e^{-2k\pi R}
M_{\rm eff}\sinh(M_{\rm eff}\pi R)
\cos\!\left(\frac{q\theta}{2}\right),
\nonumber\\[1mm]
V^{\rm tree}_{\tilde\phi}(\theta)
&=-\frac{M_{\rm eff}e^{-2k\pi R}}{2M^2\sinh(M_{\rm eff}\pi R)}
\left(\tilde J_0\tilde J_\pi^*e^{-iq\theta/2}+{\rm h.c.}\right).
\label{eq:Vtree_full}
\end{align}
The explicit $\theta$ dependence of $\xi(\theta)$ is important for the C-twisted scalar: even when only one of $J_0$ or $J_\pi$ is nonzero, the potential does not vanish.

For $M_{\rm eff}\pi R\gg1$, subtracting the $\theta$-independent part and keeping the leading contribution in each source structure gives
\begin{align}
V^{\rm tree}_{\phi}(\theta)
&\simeq
-\frac{2M_{\rm eff}}{M^2}J_0J_\pi
e^{-2k\pi R}e^{-M_{\rm eff}\pi R}
\cos\!\left(\frac{q\theta}{2}\right)
\nonumber\\
&\quad
-\frac{M_{\rm eff}^2}{M^4}
\Big[J_0^2(M_{\rm eff}+2k)
+J_\pi^2e^{-4k\pi R}(M_{\rm eff}-2k)\Big]
 e^{-2M_{\rm eff}\pi R}\cos(q\theta)
\nonumber\\
&\quad
+{\cal O}\!\left(e^{-2k\pi R}e^{-3M_{\rm eff}\pi R},e^{-4M_{\rm eff}\pi R}\right),
\label{eq:Vtree_phi_large}
\\[1mm]
V^{\rm tree}_{\tilde\phi}(\theta)
&\simeq
-\frac{2M_{\rm eff}}{M^2}|\tilde J_0\tilde J_\pi|
 e^{-2k\pi R}e^{-M_{\rm eff}\pi R}
\cos\!\left(\frac{q\theta}{2}-\delta_J\right),
\end{align}
where $\delta_J={\rm Arg}(\tilde J_0\tilde J_\pi^*)$.

The three source structures in the C-twisted result have a direct propagation interpretation.  The mixed term $J_0J_\pi$ describes a single traversal between the two fixed points and therefore carries $e^{-M_{\rm eff}\pi R}$ and the harmonic $\cos(q\theta/2)$.  The $J_0^2$ and $J_\pi^2$ terms correspond to propagation from one fixed point around the covering circle and back to the same fixed point.  They carry the full-winding factor $e^{-2M_{\rm eff}\pi R}$ and the harmonic $\cos(q\theta)$.  The additional factors $1$, $e^{-2k\pi R}$, and $e^{-4k\pi R}$ distinguish UV--UV, UV--IR, and IR--IR source insertions, respectively.

If both C-twisted sources are present with coefficients of the same natural order, the mixed single-traversal term remains the leading contribution.  Since $M_{\rm eff}>2k$ for $M>0$, its exponential factor $e^{-(M_{\rm eff}+2k)\pi R}$ is less suppressed than $e^{-2M_{\rm eff}\pi R}$.  The full-winding term becomes the leading C-twisted contribution when the source at the opposite fixed point is absent or parametrically suppressed.  By contrast, for the P-type scalar the $\theta$ dependence arises only through the product $\tilde J_0\tilde J_\pi^*$; a single P-type source therefore generates no tree-level axion potential.

This distinction also clarifies the relation to the loop-induced potential.  C-twisted loops involve full winding around the covering circle, whereas the leading tree-level term with sources at both fixed points requires only a single traversal.  For comparable bulk masses and unsuppressed linear sources, the tree-level term is therefore parametrically less suppressed.  This channel is available only for even-charged scalars, and can be absent or subdominant if such states are heavier than the unit-charged states responsible for the loop contribution.

\begin{figure}[t]
\centering
\includegraphics[width=0.45\textwidth]{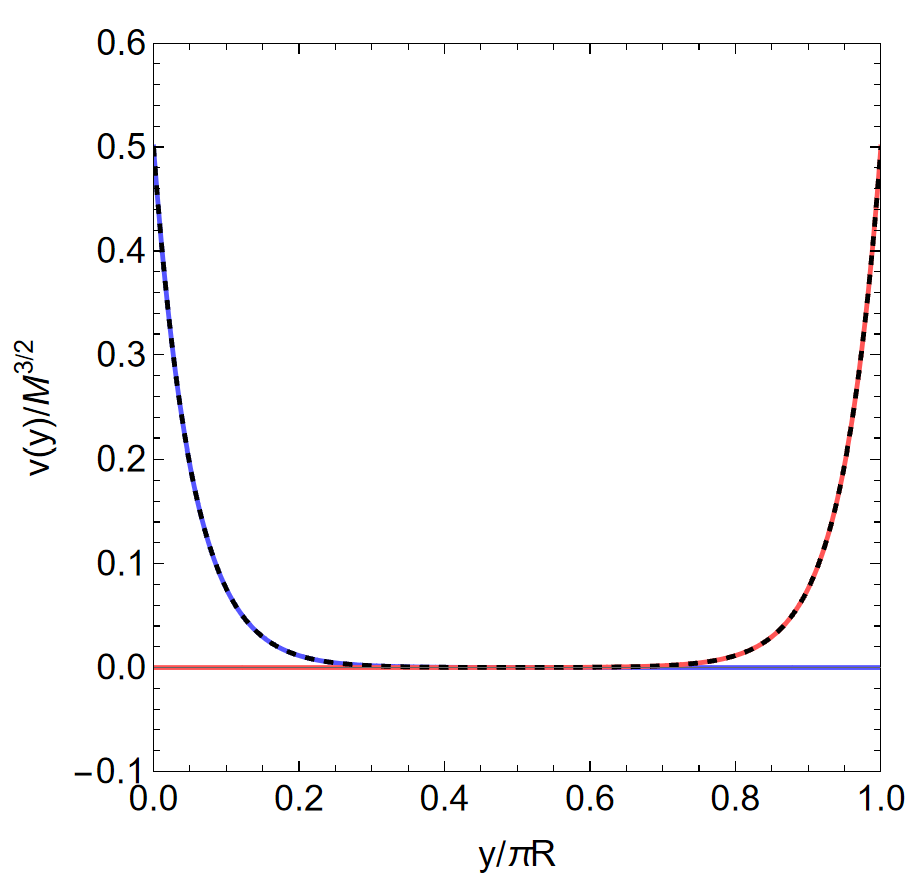}
\caption{An example of the profile for $\tilde v(y)$ with $\theta=0$, induced by the linear scalar terms with coefficients $\tilde J_0$ and $\tilde J_\pi$.  The blue and red solid lines denote the contributions from $\tilde J_0$ and $\tilde J_\pi$, respectively, while the black dashed line shows their sum.  The plot corresponds to the flat case ($k=0$) with $\tilde J_0=\tilde J_\pi=-M^{5/2}$ and $MR=6$.}
\label{fig:v_profile}
\end{figure}

\endgroup

\subsection{Loop-induced potential from P-type scalar fields}
\label{subsec:quadratic_loop}

As noted in the previous section, in the absence of $U(1)_C$-violating
fixed-point operators, loops of P-type matter fields
$\tilde\Phi=(\tilde \phi, \tilde\psi)$ do not generate an axion potential.
However, in the presence of $U(1)_C$-violating mass terms localized at the
fixed points, the boundary conditions for $\tilde\Phi$ are modified,
leading to a $\theta$-dependent KK spectrum and therefore to a nontrivial
axion potential.

As a specific example, in this subsection we consider the axion potential
induced by a P-type charged scalar $\tilde\phi$ in the presence of the
fixed-point mass terms
\begin{equation}
\Delta{\cal L}_i (y=y_i)
=
\frac{1}{2}\tilde b_i\,\tilde\phi^2
+\frac{1}{2}\tilde b_i^\ast\,\tilde\phi^{\ast 2}.
\label{app:fixed_point_tilde_phi}
\end{equation}
Here $\tilde\phi$ is taken to have even $\mathbb{Z}_2$ parity,
$\tilde\phi(x,-y)=\tilde\phi(x,y)$, so that it can have a nonzero field
value at the fixed points.

To examine the spectral function of $\tilde\phi$, we start from the
simplest case: the flat limit without fixed-point mass terms,
$k=\tilde b_i=0$.
In this limit, the KK spectrum is $\theta$-independent and is given by
\bea
m_n^2 = M^2 + \left(\frac{n}{R}\right)^2
\qquad (n=0,1,2,\ldots),
\eea
which can be encoded in the spectral function
\bea
{\cal N}_{\tilde\phi}(z;\theta)
= 4(z^2- M^2)\sin^2\!\left( \pi R \sqrt{z^2- M^2} \right).
\eea

Turning on the fixed-point mass terms Eq.~(\ref{app:fixed_point_tilde_phi})
modifies the boundary conditions, leading to a $\theta$-dependent
shift of the KK spectrum of $\tilde\phi$.
As shown in Appendix~\ref{app:Kaluza-Klein_spectral_functions}, the spectral function in the presence of the
fixed-point mass terms Eq.~(\ref{app:fixed_point_tilde_phi}) in the flat
limit is given by
\bea
{\cal N}_{\tilde\phi}(z; \theta)= N_{\tilde\phi}(z) +{\cal A}_{\tilde\phi}(\theta),
\eea
where
\bea
N_{\tilde\phi}(z)&=&
\frac{\left(4(z^2- M^2) + |\tilde{b}_0|^2 \right)
\left(4(z^2- M^2) + |\tilde{b}_\pi|^2\right)}
{4(z^2- M^2)}
\sin^2\!\left( \pi R \sqrt{z^2- M^2} \right)
\nonumber\\
&&
-\left|\tilde{b}_0 +\tilde{b}_\pi\right|^2,
\nonumber\\
{\cal A}_{\tilde\phi}(\theta)&=&
\left(\tilde{b}_0\tilde{b}_\pi^* (1-e^{-i q\theta})+\mathrm{h.c.}\right).
\eea
This result explicitly shows that the $\theta$ dependence enters through
the nonlocal product of the fixed-point localized mass terms,
$\tilde b_0\tilde b_\pi^*$, as anticipated.
This structure reflects the fact that a nonzero axion potential arises
only from processes connecting the two orbifold fixed points.

Turning on a nonzero AdS curvature scale $k$ further modifies the
spectral function, yielding
\bea
N_{\tilde\phi}(z)
&=& \pi^2 k^2 \bigl(N_{++}^{\alpha}(z)\bigr)^2
-\frac{\pi^2|\tilde {b}_0|^2}{4}\bigl(N_{-+}^{\alpha}(z)\bigr)^2
-\frac{\pi^2|\tilde{b}_\pi|^2}{4}\bigl(N_{+-}^{\alpha}(z)\bigr)^2
\nonumber\\
&&
+\,\frac{\pi^2|\tilde{b}_0|^2|\tilde{b}_\pi|^2}{16k^2}
\bigl(N_{--}^{\alpha}(z)\bigr)^2
-2\,{\rm Re}(\tilde{b}_0\tilde{b}_\pi^{\ast}),
\nonumber \\[2pt]
{\cal A}_{\tilde\phi}(\theta)
&=&
\left(\tilde{b}_0\tilde{b}_\pi^* (1-e^{-i q\theta})+\mathrm{h.c.}\right).
\label{eq:Ntphi_split}
\eea
Here $N_{++}^{\alpha}(z)$ and $N_{--}^{\alpha}(z)$ are given in
Eq.~(\ref{eq:N++_N--}), while
\begin{align}
N^{\alpha}_{+-}(z)
&=\left[2 Y_\alpha \!\left(\frac{z}{k}\right)
+ \frac{z}{k} Y'_\alpha \!\left(\frac{z}{k}\right)\right]
J_\alpha\!\left(\frac{z}{k e^{-k\pi R}}\right)
\nonumber\\
&\quad-
\left[2 J_\alpha\!\left(\frac{z}{k}\right)
+\frac{z}{k} J'_\alpha \!\left(\frac{z}{k}\right)\right]
Y_\alpha\!\left(\frac{z}{k e^{-k\pi R}}\right),
\nonumber\\[4pt]
N^{\alpha}_{-+}(z)
&=
J_\alpha\!\left(\frac{z}{k} \right)
\left[2 Y_\alpha \!\left(\frac{z}{k e^{-k\pi R}}\right)
+ \frac{z}{k e^{-k\pi R}} Y'_\alpha
\!\left(\frac{z}{k e^{-k\pi R}}\right)\right]
\nonumber\\
&\quad-
Y_\alpha\!\left(\frac{z}{k}\right)
\left[2 J_\alpha\!\left(\frac{z}{k e^{-k\pi R}}\right)
+\frac{z}{k e^{-k\pi R}} J'_\alpha
\!\left(\frac{z}{k e^{-k\pi R}}\right)\right],
\end{align}
with $\alpha=\sqrt{4+{M^2}/{k^2}}$, and
$J_\alpha$ and $Y_\alpha$ denoting the Bessel functions of the first and
second kinds, respectively.

One may use the above results to evaluate
\begin{equation}
N_0=N_{\tilde\phi}(ip=0),
\qquad
N_2=\left.\frac{dN_{\tilde\phi}(ip)}{dp^2}\right|_{p^2=0},
\end{equation}
and then apply the Gaussian approximation Eq.~(\ref{eq:spectral_gaussian}) to the spectral function in
order to derive the axion potential in the large-mass limit $M\pi R\gg 1$.
Here we consider two regimes: the flat regime $k=0$
(or a mildly warped regime with $k\pi R\ll 1$), and the strongly warped
regime $k\pi R\gg 1$, while retaining the large-mass limit.

In the flat (or mildly warped) regime,  assuming
\begin{equation}
|\tilde b_{0,\pi}| < 2M,
\end{equation} 
we find
\bea
N_0 \simeq  M^2\left(1-\frac{|\tilde b_0|^2}{4M^2}\right)
\left(1-\frac{|\tilde b_\pi|^2}{4M^2}\right)\, e^{2\pi RM},
\qquad
\left(\frac{N_0}{N_2}\right)^2 \simeq  \frac{M^2}{(\pi R)^2}.
\eea
Substituting this result into Eq.~\eqref{eq:spectral_pot}, the axion
potential in the flat regime and in the large-mass limit is given by
\begin{eqnarray}
\hskip -0.5cm
V_{\tilde\phi}^{\rm flat}(\theta)
\simeq
-\frac{|\tilde b_0\tilde b_\pi|}{16\pi^4 R^2 \Big(1-\frac{|\tilde b_0|^2}{4M^2}\Big)\Big(1 - \frac{|\tilde b_\pi|^2}{4M^2}\Big)}e^{-2M\pi R}\cos(q\theta- \delta_b)
\quad
(\text{up to a constant}),
\end{eqnarray}
where $\delta_b ={\rm Arg}(\tilde b_0 \tilde b_\pi^*)$. 
For the strongly warped regime with $M_{\rm eff} \pi R \gg 1$, we obtain
\begin{eqnarray}
    {N}_0  & \simeq &  \left[ \left(1-\frac{2k}{M_{\rm eff}}\right)^2-\frac{|\tilde{b}_0|^2}{4 M_{\rm eff}^2} \right]  \left[ \left(1+\frac{2k}{M_{\rm eff}}\right)^2- \frac{|\tilde{b}_\pi|^2}{4 M_{\rm eff}^2} \right]\,M_{\rm eff}^2 \,  e^{2 M_{\rm eff} \pi R} \nonumber \\
   & \simeq &  \left(1-\frac{|\tilde b_0|^2}{4M^2_{\rm eff}}\right)
\left(1-\frac{|\tilde b_\pi|^2}{4M^2_{\rm eff}}\right)\, M_{\rm eff}^2  \, e^{2M_{\rm eff}\pi R} \qquad {\rm for}\quad M_{\rm eff} \gg k,
\label{eq:Ntphi_asymptotics}
\end{eqnarray}
and
\begin{eqnarray}
    \left(\frac{{N}_0}{{N}_2}\right)^2 & \simeq &  \left[  \frac{ \left(1+\frac{k}{M_{\rm eff}} \right) \Big(\Big(1+ \frac{2k}{M_{\rm eff}} \Big)^2- \frac{|\tilde{b}_\pi|^2}{4 M_{\rm eff}^2}\Big)}{\Big(1+\frac{2k}{M_{\rm eff}}\Big)\Big(1+\frac{4k}{M_{\rm eff}} \Big)-\frac{|\tilde{b}_\pi|^2}{4 M_{\rm eff}^2}}  \right]^2\left(4k^2 M^2_{\rm eff} e^{-4k\pi R} \right)\nonumber \\ 
    & \simeq & 4k^2 M^2_{\rm eff}\,e^{-4k\pi R}  \qquad {\rm for} \quad M_{\rm eff} \gg k, \label{eq:Ntphi_ratio_asymptotics}
\end{eqnarray}
where $M_{\rm eff}=\sqrt{M^2+4k^2}$. 
Substituting this result into the Gaussian approximation yields the axion potential for $M_{\rm eff}\gg k$
\begin{align}
V_{\tilde{\phi}}^{\rm warped}(\theta)
&\simeq - \frac{|\tilde b_0 \tilde b_\pi|\, k^2}{4\pi^2 \Big(1-\frac{|\tilde b_0|^2}{4M^2_{\rm eff}}\Big)\Big(1 - \frac{|\tilde b_\pi|^2}{4M^2_{\rm eff}}\Big)}e^{-4k\pi R}e^{-2M_{\rm eff}\pi R}\cos (q\theta - \delta_b).
\label{eq:Vtphi_final}
\end{align}

\subsection{Modification of the loop-induced potential from C-twisted scalar fields}

The axion potential induced by loops of C-twisted matter fields
$\Phi=(\phi,\psi)$ in the absence of fixed-point localized operators was
discussed in the previous section.
When $U(1)_C$-violating fixed-point operators are introduced, the axion
potential is modified accordingly.
In this subsection, we examine the modification arising from the
$U(1)_C$-violating scalar mass terms
\begin{equation}
\Delta{\cal L}_i (y=y_i)
=
\frac{1}{2} b_i\,\left(\phi^2+\phi^{\ast 2}\right)
\label{eq:fixed_point_mass_phi}
\end{equation}
with real mass parameters $b_i=(b_0, b_\pi)$

A simple way to see how these mass terms affect the axion potential is to
examine the change in the KK spectrum of $\phi$ induced by them.
As shown in Appendix~\ref{app:Kaluza-Klein_spectral_functions}, these mass terms modify the boundary conditions
for the KK modes, thereby shifting the KK spectrum.

For the flat case \(k=0\), using the result in Eq.~(\ref{eq:appen_Nphi_result}) from Appendix~\ref{app:Kaluza-Klein_spectral_functions} and assuming
\begin{equation}
|b_{0,\pi}| < 2M,
\end{equation}
 we obtain the spectral function of \(\phi\) in the presence of the fixed-point mass terms in Eq.~(\ref{eq:fixed_point_mass_phi}) as
\bea
{\cal N}_\phi(z;\theta)
&=&
-2\left(1-\frac{b_0 b_\pi}{4 (z^2- M^2)} \right)
\sin^2\!\left(\pi R \sqrt{z^2- M^2}\right)
\nonumber \\[3pt]
&&
+\frac{1}{2}(b_0 + b_\pi)
\frac{\sin\!\left(2\pi R \sqrt{z^2-M^2}\right)}
{\sqrt{z^2- M^2}}
+ 2 \sin^2\!\left(\frac{q\theta}{2}\right).
\eea
This expression reduces to the spectral function Eq.~(\ref{eq:flat_spectral_function}) in the
limit $b_{0,\pi}=0$.
Using this result, one can evaluate
$N_0=N_\phi(ip=0)$ and
$N_2=\left.dN_\phi(ip)/dp^2\right|_{p^2=0}$ and then apply the Gaussian
approximation Eq.~(\ref{eq:spectral_gaussian}) to derive the axion
potential in the large-mass limit $M\pi R\gg1$.
This yields
\bea
N_0
\simeq
\frac{1}{2}
\left(1+\frac{b_0}{2M}\right)
\left(1+\frac{b_\pi}{2M}\right)
e^{2\pi RM},
\qquad
\left(\frac{N_0}{N_2}\right)^2\simeq \frac{M^2}{(\pi R)^2}.
\eea
Substituting these expressions into Eq.~(\ref{eq:spectral_pot}), we obtain
the modified axion potential
\begin{align}
V^{\rm flat}_{\phi}(\theta)
&\simeq
-
\frac{M^4}{4\pi^4R^2\left(2M+b_0\right)\left(2 M+ b_\pi\right)}\,
e^{-2\pi RM}
\cos(q\theta).
\end{align}

Note that the fixed-point mass terms of the C-twisted scalar $\phi$
primarily modify the overall prefactor of the axion potential and
therefore do not alter its essential structure.
This contrasts with the case of the P-type scalar $\tilde\phi$, where
the $\theta$ dependence arises entirely through the nonlocal product of
the fixed-point mass parameters $\tilde b_0\tilde b_\pi^*$, as shown in
the previous subsection.

The modified spectral function of $\phi$ in the presence of the
fixed-point mass terms in Eq.~(\ref{eq:fixed_point_mass_phi})
is rather complicated in the warped case, and we therefore leave its
explicit form to Appendix~\ref{app:Kaluza-Klein_spectral_functions}.
For the spectral function given in Eq.~(\ref{eq:appen_Nphi_result}),
in the large mass and strong warping limit we find
\begin{eqnarray}
    { N}_0 \simeq \frac{1}{2}
\left(1 -\frac{2k}{M_{\rm eff}} + \frac{b_0}{2M_{\rm eff}}\right)
\left(1 +\frac{2k}{M_{\rm eff}} + \frac{b_\pi}{2 M_{\rm eff}} \right)
e^{2M_{\rm eff} \pi R}, \label{eq:spectral_phi}
\end{eqnarray}
and
\begin{eqnarray}
    \left(\frac{{ N}_0}{{ N}_2}\right)^2
& \simeq &  \left[ \frac{\left(1+ \frac{k}{M_{\rm eff}}\right) \left(1+\frac{2k}{M_{\rm eff}} + \frac{b_\pi}{2M_{\rm eff}} \right)}{1+\frac{3k}{M_{\rm eff}} + \frac{b_\pi}{2 M_{\rm eff}}} \right]^2 \left(4k^2 M_{\rm eff}^2 \, e^{-4k\pi R}\right) \nonumber \\
& \simeq & 4 k^2  M_{\rm eff}^2\, e^{- 4 k \pi R} \qquad {\rm for}\quad M_{\rm eff} \gg k , 
\end{eqnarray}
where $M_{\rm eff}=\sqrt{M^2+4k^2}$.
Applying the Gaussian approximation Eq.~(\ref{eq:spectral_gaussian}) to the
spectral function and substituting the result into
Eq.~\eqref{eq:spectral_pot}, we obtain the leading harmonic of the axion
potential
\bea
\hskip -0.5cm
V^{\rm warped}_\phi(\theta)
\simeq
-\frac{M_{\rm eff}^2\,k^2(M_{\rm eff} + 2k +\frac{1}{2}b_\pi)^2 (M_{\rm eff}+k)^2\, e^{-4k\pi R}\,e^{-2M_{\rm eff}\pi R}
\cos(q\theta)}
{4\pi^2 \left(M_{\rm eff} - 2k + \frac{1}{2}b_0\right)
\left(M_{\rm eff} + 2 k + \frac{1}{2} b_\pi\right) (M_{\rm eff}+3k +\frac{1}{2}b_\pi )^2},
\label{eq:pot_strong_warping_with_brane}
\eea
which reduces to Eq.~(\ref{eq:pot_strong_warping_scalar}) in the limit
$b_{0,\pi}=0$.

\section{Conclusions}
\label{sec:conclusion}
\begingroup

In this paper, we have studied the axion quality problem for a Wilson-line axion
$\theta=\oint dy\,C_5$ arising from a five-dimensional $\mathbb{Z}_2$-odd $U(1)_C$ gauge field $C_M=(C_\mu, C_5)$ on an $S^1/\mathbb Z_2$ orbifold with warped geometry.  Once local sources of PQ breaking such as dangerous Chern--Simons, monodromy, or St\"uckelberg terms are absent, the non-QCD axion potential is generated by nonlocal propagation of $U(1)_C$-charged states along the compact direction.  Our analysis determines how the warped geometry, orbifold boundary conditions, and fixed-point interactions control these contributions.

In the absence of $U(1)_C$-violating fixed-point operators, P-type matter
fields do not generate an axion potential, whereas C-twisted matter fields
do. For a C-twisted scalar $\phi$ and Dirac fermion $\psi$ with constant
bulk mass $M$, the leading one-winding contributions in the strong-warping ($k\pi R\gg 1$)
and large-bulk-mass ($M\pi R\gg 1$) regime are suppressed as
\begin{equation}
V^{\rm loop}_{\phi}
\propto
e^{-4k\pi R}e^{-2M_{\rm eff}\pi R}\cos(q\theta),
\qquad
V^{\rm loop}_{\psi}
\propto
e^{-4k\pi R}e^{-2M\pi R}\cos(q\theta),
\end{equation}
where $M_{\rm eff}=\sqrt{M^2+4k^2}$.
For C-twisted hypermultiplet scalars $H_\alpha$ ($\alpha=1,2$) with
piecewise constant masses
$M_1^2=M_0^2+\mu^2\epsilon(y)$ and
$M_2^2=M_0^2-\mu^2\epsilon(y)$,
the corresponding one-winding contribution is suppressed as
\bea
V^{\rm loop}_{H_\alpha}
\propto
e^{-4k\pi R}e^{-(m_++m_-)\pi R}\cos(q\theta),
\eea
where
$m_\pm=\sqrt{M_0^2\pm\mu^2+4k^2}$.
In the supersymmetric example, $m_\pm=M\pm k/2$, so that the
curvature-dependent contributions cancel between the two half-circles,
yielding a scalar winding action of $2M\pi R$, identical to that of the
fermion. The worldline calculation makes these winding actions manifest.
For the constant-mass scalar and fermion, the KK spectral-function and
monodromy calculations provide independent checks of the corresponding
one-loop determinants.

The bulk worldline calculation isolates the winding action and gives a simple analytic expression for the prefactor.  The fixed-point terms generated by the warp-dependent field redefinitions do not change the winding action, but they do modify the prefactor.  For $M\gg k$ this modification begins at higher order in $k/M$, whereas for $k$ comparable to or larger than $M$ it can be substantial; for example, the scalar result scales as $k^6/M^2$ rather than $k^4$ for $k\gg M$.  The analytic KK expressions include these fixed-point effects at arbitrary $k\pi R$ and reduce continuously to the flat-space result as $k\to0$, with the strongly warped formulas following as the $k\pi R\gg1$ limits.  The remaining Gaussian momentum approximation has been tested against the exact first-harmonic KK integral and reproduces the winding exponent and parametric dependence, with a finite correction to the overall coefficient.

Fixed-point-localized operators open additional nonlocal channels.   For a C-twisted scalar with allowed linear sources, the axion potential in the strong-warping and large-bulk-mass regime contains 
\begin{align}
J_0J_\pi:&\quad
e^{-2k\pi R}e^{-M_{\rm eff}\pi R}\cos\!\left(\frac{q\theta}{2}\right),
\nonumber\\
J_0^2:&\quad
e^{-2M_{\rm eff}\pi R}\cos(q\theta),
\nonumber\\
J_\pi^2:&\quad
e^{-4k\pi R}e^{-2M_{\rm eff}\pi R}\cos(q\theta),
\label{eq:tree_channels_summary}
\end{align}
up to the source-dependent polynomial coefficients given in Sec.~\ref{subsec:linear_tree}.  The mixed term corresponds to a single traversal between the two fixed points, whereas the $J_0^2$ and $J_{\pi}^2$ terms correspond to full winding from a fixed point back to itself.  Thus a single C-twisted source generates a full-winding axion potential, although the mixed term remains the leading contribution when both sources are present with generic coefficients.  For the P-type scalar, the tree-level $\theta$ dependence requires sources at both fixed points.

The leading channels considered in this paper are summarized in Table~\ref{tab:summary_channels}.  Only the axion harmonic and the leading exponential suppression are displayed in the last two columns.

\begin{table}[t]
\centering
\scriptsize
\setlength{\tabcolsep}{3pt}
\renewcommand{\arraystretch}{1.22}
\begin{tabular}{|p{0.205\textwidth}|p{0.205\textwidth}|p{0.16\textwidth}|p{0.235\textwidth}|}
\hline
Matter/channel & Required fixed-point operator & Harmonic & Leading suppression \\
\hline
P-type matter loop & none & none & no axion potential \\
\hline
C-twisted $\phi$ loop & none & $\cos(q\theta)$ & $e^{-4k\pi R}e^{-2M_{\rm eff}\pi R}$ \\
\hline
C-twisted $\psi$ loop & none & $\cos(q\theta)$ & $e^{-4k\pi R}e^{-2M\pi R}$ \\
\hline
C-twisted $H_\alpha$ loop & none & $\cos(q\theta)$ & $e^{-4k\pi R}e^{-\pi R(m_++m_-)}$ \\
\hline
P-type $\tilde\phi$ loop & $\tilde b_0\tilde b_\pi$ & $\cos(q\theta-\delta_b)$ & $e^{-4k\pi R}e^{-2M_{\rm eff}\pi R}$ \\
\hline
C-twisted $\phi$: UV--IR tree & $J_0J_\pi$ & $\cos(q\theta/2)$ & $e^{-2k\pi R}e^{-M_{\rm eff}\pi R}$ \\
\hline
C-twisted $\phi$: UV--UV tree & $J_0^2$ & $\cos(q\theta)$ & $e^{-2M_{\rm eff}\pi R}$ \\
\hline
C-twisted $\phi$: IR--IR tree & $J_\pi^2$ & $\cos(q\theta)$ & $e^{-4k\pi R}e^{-2M_{\rm eff}\pi R}$ \\
\hline
P-type $\tilde\phi$ tree & $\tilde J_0\tilde J_\pi$ & $\cos(q\theta/2-\delta_J)$ & $e^{-2k\pi R}e^{-M_{\rm eff}\pi R}$ \\
\hline
\end{tabular}
\caption{Leading non-QCD axion-potential channels discussed in the text.  For the C-twisted scalar, a fixed-point mass term $b_i\phi^2$ changes the loop prefactor without changing the winding exponent.  For the supersymmetric relation $m_\pm=M\pm k/2$, the hypermultiplet-scalar winding suppression in the table becomes $e^{-2M\pi R}$, equal to the fermion suppression.}
\label{tab:summary_channels}
\end{table}
\FloatBarrier

The quality requirement can be satisfied over a finite region of parameter space simultaneously with the weak-gravity and effective-field-theory constraints.  To display this region explicitly, consider first the unit-charged C-twisted fermion as the state controlling $V_{\rm HE}$ and define
\begin{equation}
x\equiv\frac{M}{M_5},\qquad \alpha\equiv\frac{k}{M_5},
\end{equation}
where $M_5$ denotes the five-dimensional Planck mass.
Using the strong-warping expression in Eq.~\eqref{eq:pot_strong_warping_fermion}, one obtains
\begin{equation}
|V_{\psi}^{\rm warped}|\simeq
\frac{\alpha^2(x+\alpha/2)^4}{2\pi^2(x+\alpha)^2}M_5^4
\exp\left[-\pi kR\left(4+\frac{2x}{\alpha}\right)\right].
\label{eq:quality_region_fermion}
\end{equation}
Then the axion-quality condition $|V_{\rm HE}|< 10^{-10}m_\pi^2f_\pi^2$ requires
\begin{equation}
kR>\frac{1}{\pi(4+2x/\alpha)}
\ln\left[
\frac{\alpha^2(x+\alpha/2)^4}{2\pi^2(x+\alpha)^2}
\frac{M_5^4}{V_{\rm max}}
\right],
\label{eq:quality_region_bound}
\end{equation}
while the electric weak-gravity conjecture gives the parametric estimate \bea
x\lesssim {\cal O}(qg_C\sqrt{M_5}),\eea
where $g_C$ is the five-dimensional $U(1)_C$ gauge coupling constant.  For the illustrative slice in Fig.~\ref{fig:quality_region}, we take the undetermined order-one coefficient in this estimate to be unity and set $\alpha=0.05$ and $qg_C\sqrt{M_5}=0.30$.  Using the strong-warping relation $M_P^2\simeq M_5^3/k$, this choice of $\alpha$ fixes $M_5\simeq5.4\times10^{17}\,{\rm GeV}$ for the numerical boundary shown in the figure.  Over the displayed region the large-mass and strong-warping approximations are self-consistent.  The finite ${\cal O}(1)$ correction from the exact spectral integral shifts the boundary in Eq.~\eqref{eq:quality_region_bound} only logarithmically.

\begin{figure}[t]
\centering
\includegraphics[width=0.62\textwidth]{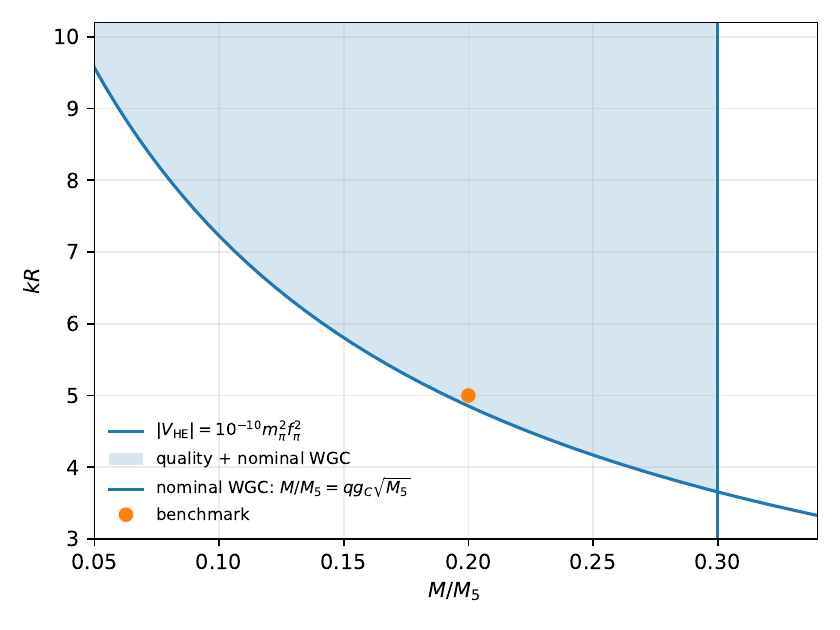}
\caption{Illustrative parameter region for a unit-charged C-twisted fermion with $k/M_5=0.05$ and $g_C\sqrt{M_5}=0.30$.  The solid curve is the analytic quality boundary obtained from Eq.~\eqref{eq:quality_region_bound}.  Taking the order-one coefficient in the electric-WGC estimate to be unity, the vertical line shows the nominal WGC limit; the shaded region satisfies both conditions.  The point $(M/M_5,kR)=(0.20,5.0)$ is the benchmark discussed below.}
\label{fig:quality_region}
\end{figure}
\FloatBarrier

For example, the quality boundary is $kR\simeq7.23$, $4.85$, and $3.66$ for $M/M_5=0.10$, $0.20$, and $0.30$, respectively.  A representative point inside the allowed region is
\begin{equation}
kR=5.0,\qquad
\frac{k}{M_5}=0.05,\qquad
g_C\sqrt{M_5}=0.30,\qquad
\frac{M}{M_5}=0.20.
\label{eq:benchmark_parameters}
\end{equation}
With the nominal order-one WGC coefficient chosen above, the benchmark satisfies the corresponding inequality, $M/M_5=0.20<0.30$.  The ratios $k/M_5=0.05$ and $M/M_5=0.20$ are below the gravitational cutoff scale of order $M_5$, while $g_C^2M_5=0.09$ satisfies the conservative dimensionless perturbativity criterion $g_C^2M_5<1$.  In the strongly warped regime, $M_P^2\simeq M_5^3/k$ gives $M_5\simeq5.4\times10^{17}\,{\rm GeV}$ and Eq.~\eqref{eq:fa} gives $f_a\simeq6.0\times10^{10}\,{\rm GeV}$.  The analytic fermion result gives $|V_{\rm HE}|\simeq6\times10^{-17}\,{\rm GeV}^4$ before the finite ${\cal O}(1)$ correction determined by the numerical spectral integral, 
which is below $10^{-10}m_\pi^2 f_\pi^2\sim10^{-14}\,{\rm GeV}^4$.  
For this illustrative region we take the $q=1$ fermion to be the charged state controlling $V_{\rm HE}$; an even-charged scalar with a linear source may be absent from the light spectrum or taken sufficiently heavier that its tree-level contribution is subleading.

For the scalar channels, the different warp suppressions can also be organized from the AdS/CFT viewpoint.  Using
\begin{equation}
e^{-k\pi R}\simeq\frac{\mu_{\rm IR}}{\mu_{\rm UV}},
\qquad
\Delta=2+\frac{M_{\rm eff}}{k},
\end{equation}
a single traversal and a full winding contribute, respectively,
\begin{equation}
e^{-M_{\rm eff}\pi R}
=\left(\frac{\mu_{\rm IR}}{\mu_{\rm UV}}\right)^{\Delta-2},
\qquad
e^{-2M_{\rm eff}\pi R}
=\left(\frac{\mu_{\rm IR}}{\mu_{\rm UV}}\right)^{2\Delta-4}.
\end{equation}
After normalizing the localized source coefficients to their natural local scales, the three C-twisted tree structures can be written schematically as
\begin{align}
J_0^2:&\quad
\mu_{\rm UV}^4
\left(\frac{\mu_{\rm IR}}{\mu_{\rm UV}}\right)^{2\Delta-4},
\nonumber\\
J_0J_\pi:&\quad
\mu_{\rm UV}^2\mu_{\rm IR}^2
\left(\frac{\mu_{\rm IR}}{\mu_{\rm UV}}\right)^{\Delta-2},
\nonumber\\
J_\pi^2:&\quad
\mu_{\rm IR}^4
\left(\frac{\mu_{\rm IR}}{\mu_{\rm UV}}\right)^{2\Delta-4}.
\end{align}
The scalar loop winding has the same nonlocal full-winding factor and an IR-sized four-dimensional prefactor,
\begin{equation}
V^{\rm loop}_{\phi}\sim
\mu_{\rm IR}^4
\left(\frac{\mu_{\rm IR}}{\mu_{\rm UV}}\right)^{2\Delta-4},
\end{equation}
up to the calculable $k/M$-dependent coefficient.  Thus $\Delta-2=M_{\rm eff}/k$ controls the suppression associated with propagation through the bulk, while the remaining powers of $\mu_{\rm IR}/\mu_{\rm UV}$ reflect the locations of the corresponding source insertions.

\endgroup


\acknowledgments
This work is supported by IBS under the project code IBS-R018-D1. 
CHL and CSS are supported by the Global-Learning \& Academic Research Institution for Master's, PhD Students, and Postdocs (G-LAMP) Program of the NRF grant RS-2025025442707, and by the NRF grant RS-2026-25498521.

\appendix



\begingroup
 \section{Quantum corrections to the worldline functional $\langle I[y(\tau)]^{-2}\rangle$}
\label{app:path_integral_expansion}

In this appendix, we examine the fluctuations of the warp-weight functional $I[y(\tau)]$ that enter the bulk worldline calculation of Sec.~\ref{worldline}.  For the winding contribution, the trace on the covering circle includes an average over the starting point of the path.  It is therefore useful to write
\begin{equation}
y(\tau)=y_0+2\ell\tau+\eta(\tau),
\qquad
\eta(0)=\eta(1)=0,
\qquad
\ell\equiv\pi R,
\label{eq:app_trace_path}
\end{equation}
and to define the normalized trace average
\begin{equation}
\langle F[y]\rangle_{\rm tr}
\equiv
\frac{1}{2\ell\,Z_0(T)}
\int_0^{2\ell}dy_0
\int_{\eta(0)=\eta(1)=0}{\cal D}\eta\,
\exp\!\left[-\frac{1}{4T}\int_0^1d\tau\,\dot\eta^2\right]
F[y_0+2\ell\tau+\eta(\tau)],
\label{eq:app_trace_average}
\end{equation}
where
\begin{equation}
Z_0(T)=
\int_{\eta(0)=\eta(1)=0}{\cal D}\eta\,
\exp\!\left[-\frac{1}{4T}\int_0^1d\tau\,\dot\eta^2\right].
\end{equation}
The integration over $y_0$ is part of the circle heat trace.  Holding $y_0$ fixed instead gives a fixed-starting-point bridge, which is not the trace entering the effective potential.

Using the notation of Sec.~\ref{worldline},
\begin{equation}
I[y]=\int_0^1d\tau\,e^{-2k(\ell-|y(\tau)|)},
\qquad
I_0=\frac{1-e^{-2k\ell}}{2k\ell},
\label{eq:app_trace_I}
\end{equation}
where $|y|$ is understood as the periodic $\mathbb Z_2$-even function on the covering circle.  The scalar one-winding contribution can then be written in the form
\begin{equation}
V_{\phi}^{\rm warped}(\theta)
=-\frac{R}{8\pi}e^{-4k\pi R}\cos(q\theta)
\int_0^\infty\frac{dT}{T^3}\,
\frac{Z_0(T)}{I_0^2}\,
\exp\!\left[-\left(\frac{\ell^2}{T}+M_{\rm eff}^2T\right)\right]
\left\langle\frac{I_0^2}{I[y]^2}\right\rangle_{\rm tr}.
\label{eq:app_prefactor_start}
\end{equation}

To expand the last factor, define the periodic warp weight
\begin{equation}
w(y)\equiv e^{-2k(\ell-|y|)},
\qquad
\delta I_n\equiv
\frac{1}{n!}\int_0^1d\tau\,
w^{(n)}(y_0+2\ell\tau)\,\eta(\tau)^n,
\label{eq:app_deltaI_def}
\end{equation}
so that
\begin{equation}
I[y]=I_0+\delta I_1+\delta I_2+\cdots,
\qquad
\left\langle\frac{I_0^2}{I[y]^2}\right\rangle_{\rm tr}
=1+\Delta_{\rm tr}^{(2)}+\Delta_{\rm tr}^{(4)}+\cdots.
\label{eq:app_trace_expansion}
\end{equation}
The first two even orders are
\begin{align}
\Delta_{\rm tr}^{(2)}
&=-\frac{2\langle\delta I_2\rangle_{\rm tr}}{I_0}
+\frac{3\langle\delta I_1^2\rangle_{\rm tr}}{I_0^2},
\nonumber\\
\Delta_{\rm tr}^{(4)}
&=-\frac{2\langle\delta I_4\rangle_{\rm tr}}{I_0}
+\frac{6\langle\delta I_1\delta I_3\rangle_{\rm tr}}{I_0^2}
+\frac{3\langle\delta I_2^2\rangle_{\rm tr}}{I_0^2}
-\frac{12\langle\delta I_1^2\delta I_2\rangle_{\rm tr}}{I_0^3}
+\frac{5\langle\delta I_1^4\rangle_{\rm tr}}{I_0^4}.
\label{eq:app_trace_Delta_def}
\end{align}
At fixed $y_0$ the fluctuation field has the standard Brownian bridge correlator
\begin{equation}
\langle\eta(\tau)\eta(\sigma)\rangle
=2T\bigl(\min(\tau,\sigma)-\tau\sigma\bigr).
\label{eq:app_bridge_kernel}
\end{equation}
The subsequent average over $y_0$ is important.  In particular, the circle average of the second derivative of the periodic warp weight vanishes, so that
\begin{equation}
\langle\delta I_2\rangle_{\rm tr}=0.
\end{equation}
The remaining quadratic contraction gives
\begin{equation}
\frac{\langle\delta I_1^2\rangle_{\rm tr}}{I_0^2}
=\frac{T}{2\ell^2}\left(\alpha\coth\alpha-1\right),
\qquad
\alpha\equiv k\ell,
\end{equation}
and therefore
\begin{equation}
\Delta_{\rm tr}^{(2)}
=\frac{3T}{2\ell^2}\left(\alpha\coth\alpha-1\right).
\label{eq:app_Delta2_final}
\end{equation}

At fourth order, the same trace average is applied to the Gaussian contractions of $\delta I_4$, $\delta I_1\delta I_3$, $\delta I_2^2$, $\delta I_1^2\delta I_2$, and $\delta I_1^4$.  Combining these terms gives
\begin{equation}
\Delta_{\rm tr}^{(4)}
=\frac{T^2}{4\ell^4}
\left[
3\alpha^2\coth^2\alpha
-3\alpha\coth\alpha
-4\alpha^2
\right].
\label{eq:app_Delta4_final}
\end{equation}
Thus, through fourth order,
\begin{align}
\left\langle\frac{I_0^2}{I[y]^2}\right\rangle_{\rm tr}
={}&1
+\frac{3T}{2\ell^2}\left(\alpha\coth\alpha-1\right)
\nonumber\\
&+\frac{T^2}{4\ell^4}
\left[
3\alpha^2\coth^2\alpha
-3\alpha\coth\alpha
-4\alpha^2
\right]
+{\cal R}_{\geq6}.
\label{eq:app_trace_result}
\end{align}

For the large-mass winding contribution, the leading proper-time saddle is $T_*=\ell/M_{\rm eff}$.  Defining $u\equiv M_{\rm eff}\ell$, the nearly flat expansion $\alpha\ll1$ gives
\begin{align}
\Delta_{\rm tr}^{(2)}
&=\frac{\alpha^2}{2u}+{\cal O}\!\left(\frac{\alpha^4}{u}\right),
\nonumber\\
\Delta_{\rm tr}^{(4)}
&=-\frac{3\alpha^2}{4u^2}+{\cal O}\!\left(\frac{\alpha^4}{u^2}\right).
\label{eq:app_trace_weak}
\end{align}
The leading correction is therefore further suppressed near the flat limit; equivalently, $\Delta_{\rm tr}^{(2)}\simeq (k/M_{\rm eff})(k\ell)/2$.

In the strongly warped regime $\alpha\gg1$,
\begin{align}
\Delta_{\rm tr}^{(2)}
&=\frac{3k}{2M_{\rm eff}}
-\frac{3}{2M_{\rm eff}\ell}
+{\cal O}\!\left(\frac{k}{M_{\rm eff}}e^{-2\alpha}\right),
\nonumber\\
\Delta_{\rm tr}^{(4)}
&=-\frac{k^2}{4M_{\rm eff}^2}
-\frac{3k}{4M_{\rm eff}^2\ell}
+{\cal O}\!\left(\frac{k^2}{M_{\rm eff}^2}e^{-2\alpha}\right).
\label{eq:app_trace_strong}
\end{align}
Hence $k/M_{\rm eff}$ is the leading strong-warp expansion parameter, while for general $k\ell$ the correction is more precisely organized as a large-$u$ expansion with coefficients depending on $\alpha$.  The fixed-point terms generated by the warp-dependent field redefinition are a separate effect and are treated in Appendix~\ref{app:common_determinant}.
\endgroup

\section{Kaluza-Klein spectral functions}
\label{app:Kaluza-Klein_spectral_functions}

In this appendix, we summarize the spectral functions used in Secs.~3 and~4 and assemble the basic building blocks for the KK spectrum analysis, based on the results of \cite{Choi:2002ps,Choi:2010xn}.
{The derivation here keeps the orbifold boundary conditions and, where relevant, the explicit fixed-point operators used in Sec.~\ref{sec:fixedpoint}.  Appendix~\ref{app:common_determinant} sets the explicit $U(1)_C$-violating operators to zero and compares the KK result with the proper-time calculation including the fixed-point terms generated by the field redefinitions.  Appendix~\ref{app:completed_monodromy} gives the corresponding transfer-matrix calculation.}
The spectral functions of the C-twisted or P-type bulk matter fields,
$\Phi=(\phi,\psi)$ or $\tilde\Phi=(\tilde\phi,\tilde\psi)$,
are defined such that their zeros determine the KK spectrum of the
associated bulk fields:
\begin{equation}
    {\cal N}_{\Phi,\tilde\Phi}(z;\theta)=0
    \qquad \Longleftrightarrow \qquad
    z^2=m_n^2(\theta).
\end{equation}

For a given bulk field in the RS background geometry on
$S^1/\mathbb{Z}_2$, the spectral function can be constructed from the
following building blocks encoding the bulk solutions and boundary
conditions:
\begin{align}
N^{\alpha}_{++}(z;r_0,r_\pi)
&\equiv -\Bigl(\mathcal{J}_{\alpha}^{0+} - r_0 \mathcal{J}_{\alpha}^{0-}\Bigr)
   \Bigl(\mathcal{Y}_{\alpha}^{\pi+} - r_\pi \mathcal{Y}_{\alpha}^{\pi-}\Bigr)
+\Bigl(\mathcal{J}_{\alpha}^{\pi+} - r_\pi \mathcal{J}_{\alpha}^{\pi-}\Bigr)
  \Bigl(\mathcal{Y}_{\alpha}^{0+} - r_0 \mathcal{Y}_{\alpha}^{0-}\Bigr),
\nonumber\\[4pt]
N^{\alpha}_{+-}(z;r_0,r_\pi)
&\equiv -\Bigl(\mathcal{J}_{\alpha}^{0+} - r_0 \mathcal{J}_{\alpha}^{0-}\Bigr)\,
    \mathcal{Y}_{\alpha}^{\pi-}
 + \mathcal{J}_{\alpha}^{\pi-}\,
   \Bigl(\mathcal{Y}_{\alpha}^{0+} - r_0 \mathcal{Y}_{\alpha}^{0-}\Bigr),
\nonumber\\[4pt]
N^{\alpha}_{-+}(z;r_0,r_\pi)
&\equiv \mathcal{J}_{\alpha}^{0-}\,
   \Bigl(\mathcal{Y}_{\alpha}^{\pi+} - r_\pi \mathcal{Y}_{\alpha}^{\pi-}\Bigr)
 - \Bigl(\mathcal{J}_{\alpha}^{\pi+} - r_\pi \mathcal{J}_{\alpha}^{\pi-}\Bigr)\,
   \mathcal{Y}_{\alpha}^{0-},
\nonumber\\[4pt]
N^{\alpha}_{--}(z)
&\equiv \mathcal{J}_{\alpha}^{0-}\,\mathcal{Y}_{\alpha}^{\pi-}
 - \mathcal{J}_{\alpha}^{\pi-}\,\mathcal{Y}_{\alpha}^{0-}.
 \label{eq:appen_building_block}
\end{align}
Here the functions appearing in these expressions are defined as
\begin{align}
\mathcal{J}_{\alpha}^{0-}(z) &\equiv J_{\alpha}\!\left(\frac{z}{k}\right),
&
\mathcal{J}_{\alpha}^{0+}(z) &\equiv \frac{s}{2}\,J_{\alpha}\!\left(\frac{z}{k}\right)
 + \frac{z}{k}\,J'_{\alpha}\!\left(\frac{z}{k}\right),
\nonumber\\
\mathcal{Y}_{\alpha}^{0-}(z) &\equiv Y_{\alpha}\!\left(\frac{z}{k}\right),
&
\mathcal{Y}_{\alpha}^{0+}(z) &\equiv \frac{s}{2}\,Y_{\alpha}\!\left(\frac{z}{k}\right)
 + \frac{z}{k}\,Y'_{\alpha}\!\left(\frac{z}{k}\right),
\nonumber\\
\mathcal{J}_{\alpha}^{\pi-}(z) &\equiv J_{\alpha}\!\left(\frac{z}{k e^{-k\pi R}}\right),
&
\mathcal{J}_{\alpha}^{\pi+}(z) &\equiv \frac{s}{2}\,
J_{\alpha}\!\left(\frac{z}{k e^{-k\pi R}}\right)
 + \frac{z}{k e^{-k\pi R}}\,J'_{\alpha}\!\left(\frac{z}{k e^{-k\pi R}}\right),
\nonumber\\
\mathcal{Y}_{\alpha}^{\pi-}(z) &\equiv Y_{\alpha}\!\left(\frac{z}{k e^{-k\pi R}}\right),
&
\mathcal{Y}_{\alpha}^{\pi+}(z) &\equiv \frac{s}{2}\,
Y_{\alpha}\!\left(\frac{z}{k e^{-k\pi R}}\right)
 + \frac{z}{k e^{-k\pi R}}\,Y'_{\alpha}\!\left(\frac{z}{k e^{-k\pi R}}\right),
\end{align}
where $J_\alpha(x)$ and $Y_\alpha(x)$ denote the Bessel functions of the
first and second kind, respectively. The parameter $\alpha$ denotes the
order of the Bessel functions, while the coefficient $s$ depends on the
type of bulk field:
\begin{equation}
s=4 \quad \text{for a scalar},
\qquad
s=1 \quad \text{for a fermion}.
\end{equation}
The coefficients $r_0$ and $r_\pi$ parameterize Robin boundary
conditions at the orbifold fixed points $y=0$ and $y=\pi R$, and a prime
denotes differentiation with respect to the argument.
 
\subsection{Spectral functions of  scalar fields}

We begin with the charged bulk scalar fields, the C-twisted field $\phi$
and the P-type field $\tilde\phi$, with the bulk action given in
Eq.~\eqref{eq:S_bulk} and the $U(1)_C$-violating fixed-point operators in
Eq.~\eqref{eq:appC_brane}.
In order for $\tilde\phi$ to have a nonzero field value at the fixed
points, we take $\tilde\phi$ to be $\mathbb{Z}_2$-even,
$\tilde\phi(x,-y)=\tilde\phi(x,y)$.
For definiteness, we also consider a $\mathbb{Z}_2$-even C-twisted scalar
satisfying $\phi(x,-y)=\phi^*(x,y)$ (the case of a $\mathbb{Z}_2$-odd
$\phi$ leads to essentially the same results).

For the discussion of the KK modes, it is convenient to remove
$C_5=\theta/2\pi R$ from the covariant derivative by the $y$-dependent
field redefinitions
\begin{equation}
\phi \rightarrow e^{iq\theta y/2\pi R}\phi, \qquad
\tilde\phi \rightarrow e^{iq\theta |y|/2\pi R}\tilde\phi .
\label{eq:appen_field_redef}
\end{equation}
In the redefined field basis, the $\theta$ dependence is transferred to
the fixed-point interactions,
\begin{align}
\Delta \mathcal{L}_0
&= J_0 \, \phi + \tilde J_0\, \tilde\phi
+ \frac{1}{2} b_0\, \phi^2 + \frac{1}{2} \tilde b_0\, \tilde\phi^2
+ \mathrm{h.c.},
\nonumber\\
\Delta \mathcal{L}_\pi 
&= e^{i q \theta/2} J_\pi\, \phi + e^{i q\theta/2} \tilde J_\pi\,\tilde\phi
+ \frac{1}{2} e^{i q \theta} b_{\pi} \phi^2
+ \frac{1}{2} e^{i q \theta}\tilde b_{\pi} \tilde \phi^2
+ \mathrm{h.c.}
\label{eq:scalar_brane_potential_shifted}
\end{align}
and to the twisted boundary condition for the C-twisted field $\phi$,
\begin{equation}
\phi(x,y+2\pi R)=e^{iq\theta}\phi(x,y).
\end{equation}

Since the main consequence of the linear terms is the tree-level axion
potential discussed in Sec.~4.1, in the following we set
\begin{equation}
J_i=\tilde J_i = 0,
\end{equation}
and focus only on the effects of the $U(1)_C$-violating fixed-point mass
terms on the KK spectrum.

Let us first consider the case of the C-twisted scalar field $\phi$.
Varying the action with respect to $\phi^\ast$, one obtains the equation
of motion
\begin{align}
\frac{1}{\sqrt{-g}}\,\partial_M\!\left(\sqrt{-g}\,g^{MN}\partial_N\phi\right)-M^2\,\phi
= \delta(y)\,\frac{\partial (\Delta \mathcal{L}_0)}{\partial \phi^\ast}
+ \delta(y-\pi R)\,\frac{\partial (\Delta \mathcal{L}_\pi)}{\partial \phi^\ast},
\label{eq:eom_covariant}
\end{align}
where we set $\sqrt{g_{55}}=1$. In the background RS metric this equation becomes
\begin{align}
&\Bigl[\partial_y^2 - 4k\,\epsilon(y)\,\partial_y
+ e^{2 k|y|}\eta^{\mu\nu}\partial_\mu\partial_\nu - M^2\Bigr]\phi(x,y)
\nonumber\\
&\quad =
\delta(y)\,\Bigl[J_0^{\ast}+b_0^{\ast}\,\phi^{\ast}(x,y)\Bigr]
+ \delta(y-\pi R)\,\Bigl[e^{-i q\theta/2}\,J_\pi^{\ast}
+e^{-i q\theta}\,b_\pi^{\ast}\,\phi^{\ast}(x,y)\Bigr].
\label{eq:eom_explicit}
\end{align}

We expand $\phi$ in Kaluza--Klein modes,
\begin{equation}
\phi(x,y)
= \frac{1}{\sqrt{2\pi R}}\sum_{n}\phi_n(x)\,f_n(y),
\end{equation}
in a basis in which each KK mode satisfies the four-dimensional
Klein--Gordon equation with mass $m_n$,
\begin{equation}
\Bigl[-\eta^{\mu\nu}\partial_\mu\partial_\nu + m_n^2\Bigr]\phi_n(x)=0.
\label{eq:KG_4d}
\end{equation}
The KK profiles satisfy the orthonormality condition
\begin{equation}
\frac{1}{2\pi R}\int_{-\pi R}^{\pi R}dy\,
e^{-2k|y|}\,f_m^{\ast}(y)\,f_n(y)=\delta_{mn}.
\label{eq:orthonormality}
\end{equation}
Substituting the mode expansion into Eq.~\eqref{eq:eom_explicit} yields the
mode equation for $f_n(y)$,
\begin{align}
&\Bigl[\partial_y^2 - 4k\,\epsilon(y)\,\partial_y
+ e^{2 k|y|} m_n^2 - M^2\Bigr]f_n(y)
\nonumber\\
&\quad =
\delta(y)\,\Bigl[b_0^{\ast}\,f_n^{\ast}(y)\Bigr]
+ \delta(y-\pi R)\,\Bigl[e^{-i q\theta}\,b_\pi^{\ast}\,f_n^{\ast}(y)\Bigr].
\label{eq:mode_equation}
\end{align}

To implement the boundary conditions at the fixed points, we solve the
equation of motion piecewise on the domains
\begin{equation}
f_n(y)=
\begin{cases}
f^{(1)}_{n}(y) & \quad (-\pi R<y<0),\\
f^{(2)}_{n}(y) & \quad (0<y<\pi R),\\
f^{(3)}_{n}(y) & \quad (\pi R<y<2\pi R),
\end{cases}
\label{eq:domains}
\end{equation}
where each $f^{(I)}_{n}(y)$ ($I=1,2,3$) satisfies the homogeneous bulk
equation away from the fixed points,
\begin{equation}
\Bigl[e^{2k|y|}m_n^2
+ e^{4 k|y|}\partial_y\!\bigl(e^{-s k|y|}\partial_y\bigr)
- M^2\Bigr]f^{(I)}_{n}(y)=0,
\qquad (I=1,2,3).
\label{eq:bulk_eq_piecewise}
\end{equation}
The general solution to Eq.~\eqref{eq:bulk_eq_piecewise} can be written as
\begin{equation}
f^{(I)}_{n}(y)
= e^{2 k|y|}
\left[
 A^{(I)}_{n}\,
J_{\alpha_\phi}\!\left(\frac{m_n}{k}e^{k|y|}\right)
+ B^{(I)}_{n}\,
Y_{\alpha_\phi}\!\left(\frac{m_n}{k}e^{k|y|}\right)
\right],
\label{eq:general_solution}
\end{equation}
where $J_\alpha$ and $Y_\alpha$ denote the Bessel functions of the first
and second kinds, respectively, and
\begin{equation}
\alpha_\phi = \sqrt{\frac{M^2}{k^2} + 4} = \frac{M_{\rm eff}}{k}.
\label{eq:bessel_order}
\end{equation}
The coefficients $A^{(I)}_{n}$ and $B^{(I)}_{n}$ are constants within
each domain and are determined by the boundary and matching conditions.

We now derive the spectral function ${\cal N}_{\phi}(z;\theta)$ of $\phi$ in the presence of the fixed-point masses
\begin{equation}
    b_{0,\pi} = b_{0,\pi}^* .
\end{equation}
The orbifold boundary conditions relate the solutions across the domains.
In particular, they require
\begin{equation}
A^{(1)}_{n} = A^{(2)*}_{n}= e^{iq\theta}\,A^{(3)}_{n},
\qquad
B^{(1)}_{n} = B^{(2)*}_{n} = e^{iq\theta}\,B^{(3)}_{n}.
\label{eq:phi_coeff_relations}
\end{equation}
Imposing the boundary and matching conditions at the fixed points
$y=0$ and $y=\pi R$ yields a homogeneous linear system for the
independent coefficients.
It is convenient to write these conditions in matrix form:
\begin{align}
\begin{pmatrix}
\mathcal{J}_{\alpha_\phi}^{0-}  & -\mathcal{J}_{\alpha_\phi}^{0-}   & \mathcal{Y}_{\alpha_\phi}^{0-}  & - \mathcal{Y}_{\alpha_\phi}^{0-}  \\
\bigl(\mathcal{J}_{\alpha_\phi}^{0+}- \mu_0^\ast \mathcal{J}_{\alpha_\phi}^{0-}\bigr)  & \mathcal{J}_{\alpha_\phi}^{0+}  & \bigl(\mathcal{Y}_{\alpha_\phi}^{0+}- \mu_0^\ast \mathcal{Y}_{\alpha_\phi}^{0-}\bigr)  & \mathcal{Y}_{\alpha_\phi}^{0+} \\
e^{-iq\theta/2}\,\mathcal{J}_{\alpha_\phi}^{\pi-}  & -e^{iq\theta/2}\,\mathcal{J}_{\alpha_\phi}^{\pi-} & e^{-iq\theta/2}\,\mathcal{Y}_{\alpha_\phi}^{\pi-}  & -e^{iq\theta/2}\,\mathcal{Y}_{\alpha_\phi}^{\pi-}  \\
e^{-iq\theta/2}\,\bigl(\mathcal{J}_{\alpha_\phi}^{\pi+}+ \mu_\pi^\ast \mathcal{J}_{\alpha_\phi}^{\pi-}\bigr)  & e^{iq\theta/2}\,\mathcal{J}_{\alpha_\phi}^{\pi+}  & e^{-iq\theta/2}\,\bigl(\mathcal{Y}_{\alpha_\phi}^{\pi+}+ \mu_\pi^\ast \mathcal{Y}_{\alpha_\phi}^{\pi-}\bigr)  & e^{iq\theta/2}\,\mathcal{Y}_{\alpha_\phi}^{\pi+}
\end{pmatrix}
\begin{pmatrix}
A^{(1)}_{n} \\
A^{(1)*}_{n} \\
B^{(1)}_{n} \\
B^{(1)*}_{n}
\end{pmatrix}
=0,
\label{eq:phi_matrix}
\end{align}
where we have introduced the dimensionless fixed-point parameters
\begin{equation}
\mu_{0,\pi}\equiv \frac{b_{0,\pi}}{k}.
\end{equation}
A nontrivial solution exists only if the determinant of the coefficient
matrix vanishes.
This motivates defining the spectral function ${\cal N}_{\phi}(z;\theta)$ as the
determinant of the matrix in Eq.~\eqref{eq:phi_matrix}, so that its zeros
reproduce the KK spectrum.
Evaluating the determinant and choosing a convenient overall normalization, we find
\begin{align}
{\cal N}_\phi (z; \theta) &= N_\phi(z) +{\cal A}(\theta) = \frac{\pi^2}{2}\,N^{\alpha_\phi}_{++}\!\left(z;\frac{\mu_0 }{2},-\frac{\mu_\pi }{2}\right)\,
        N^{\alpha_\phi}_{--}(z)
   + 2 \sin^2\!\left(\frac{q\theta}{2}\right),
\label{eq:appen_Nphi_result}
\end{align}
where  $N^{\alpha}_{++}\!\left(z;\frac{\mu_0 }{2},-\frac{\mu_\pi }{2}\right)$ and
        $N^{\alpha}_{--}(z)$ are given in Eq.~(\ref{eq:appen_building_block}).

The spectral function of the P-type scalar field $\tilde{\phi}$ can be
derived analogously. As in the C-twisted case, we solve the bulk
equation in each domain and impose the matching conditions at the
orbifold fixed points.
Even after the field redefinition~Eq.~(\ref{eq:appen_field_redef}), the
redefined $\tilde\phi$ satisfies the same orbifold boundary conditions
as the original field:
\bea
\tilde\phi(x,y+2\pi R)=\tilde\phi(x,y)=\tilde\phi(x,-y).
\eea
Such a field $\tilde\phi$ can be expanded as
\begin{equation}
\tilde\phi(x,y)
= \frac{1}{\sqrt{2\pi R}}\sum_{n}\tilde \phi_n(x)\,\tilde f_n(y),
\end{equation}
with KK profiles satisfying
\begin{equation}
\tilde f_n(y)=
\begin{cases}
\tilde f^{(1)}_{n}(y) & \quad (-\pi R<y<0),\\
\tilde f^{(2)}_{n}(y) & \quad (0<y<\pi R),\\
\tilde f^{(3)}_{n}(y) & \quad (\pi R<y<2\pi R).
\end{cases}
\label{eq:tilde_domains}
\end{equation}

Solving the equations of motion and imposing the orbifold boundary
conditions, the KK profiles can again be written as
\begin{equation}
\tilde f^{(I)}_{n}(y)
= e^{2 k|y|}
\left[
 \tilde A^{(I)}_{n}\,
J_{\alpha_\phi}\!\left(\frac{m_n}{k}e^{k|y|}\right)
+ \tilde B^{(I)}_{n}\,
Y_{\alpha_\phi}\!\left(\frac{m_n}{k}e^{k|y|}\right)
\right],
\label{eq:general_solution_tilde}
\end{equation}
where the coefficients in the KK expansion satisfy
\begin{equation}
\tilde A^{(1)}_{n}=\tilde A^{(2)}_{n}=\tilde A^{(3)}_{n},
\qquad
\tilde B^{(1)}_{n}=\tilde B^{(2)}_{n}=\tilde B^{(3)}_{n}.
\label{eq:tphi_coeff_equal}
\end{equation}
Note that, unlike in the $\phi$ case where $\theta$ enters through both
the fixed-point mass terms and the orbifold boundary conditions, here the
$\theta$ dependence arises only through the fixed-point mass terms.

As in the case of $\phi$, imposing the boundary conditions at
$y=0$ and $y=\pi R$ yields a homogeneous linear system for the
independent coefficients, which can be written in matrix form as
\begin{align}
\begin{pmatrix}
2 \mathcal{J}_{\alpha_{\tilde\phi}}^{0+}  & -\tilde{\mu}_0^\ast\,\mathcal{J}_{\alpha_{\tilde\phi}}^{0-}
& 2\mathcal{Y}_{\alpha_{\tilde\phi}}^{0+}  & -\tilde{\mu}_0^\ast\,\mathcal{Y}_{\alpha_{\tilde\phi}}^{0-}  \\
\tilde{\mu}_0 \,\mathcal{J}_{\alpha_{\tilde\phi}}^{0-}  & -2 \mathcal{J}_{\alpha_{\tilde\phi}}^{0+}
& \tilde{\mu}_0\,\mathcal{Y}_{\alpha_{\tilde\phi}}^{0-}  & -2 \mathcal{Y}_{\alpha_{\tilde\phi}}^{0+} \\
2 \mathcal{J}_{\alpha_{\tilde\phi}}^{\pi +}  & e^{-i q \theta}\,\tilde{\mu}_{\pi}^\ast\,\mathcal{J}_{\alpha_{\tilde\phi}}^{\pi -}
& 2 \mathcal{Y}_{\alpha_{\tilde\phi}}^{\pi +}  & e^{-i q \theta}\,\tilde{\mu}_{\pi}^\ast \,\mathcal{Y}_{\alpha_{\tilde\phi}}^{\pi -}  \\
-e^{i q \theta}\,\tilde{\mu}_{\pi}\,\mathcal{J}_{\alpha_{\tilde\phi}}^{\pi -}  & -2 \mathcal{J}_{\alpha_{\tilde\phi}}^{\pi +}
& -e^{i q \theta}\,\tilde{\mu}_{\pi}\,\mathcal{Y}_{\alpha_{\tilde\phi}}^{\pi -} & -2 \mathcal{Y}_{\alpha_{\tilde\phi}}^{\pi +}
\end{pmatrix}
\begin{pmatrix}
\tilde A^{(1)}_{n} \\
\tilde A^{(1)*}_{n} \\
\tilde B^{(1)}_{n} \\
\tilde B^{(1)*}_{n}
\end{pmatrix}
=0,
\label{eq:tphi_matrix}
\end{align}
where
\begin{equation}
\tilde{\mu}_{0,\pi}\equiv\frac{\tilde b_{0,\pi}}{k}.
\end{equation}
A nontrivial solution exists only if the determinant of the coefficient
matrix in Eq.~\eqref{eq:tphi_matrix} vanishes. Evaluating this
determinant and choosing a convenient overall normalization, we obtain
\begin{align}
{ \cal N}_{\tilde \phi}(z; \theta)
&\equiv  N_{\tilde\phi}(z)
+ \left(\tilde{b}_0\tilde{b}_\pi^{\ast}(1- e^{-i q\theta})+\mathrm{h.c.}\right)
\nonumber \\
&=  \bigl(\pi k N_{++}^{\alpha_{\tilde\phi}}(z; 0,0)\bigr)^2
- \frac{|\tilde {b}_0|^2}{4}\bigl( \pi N_{-+}^{\alpha_{\tilde\phi}}(z; 0,0)\bigr)^2
-\frac{|\tilde{b}_\pi|^2}{4} \bigl(\pi N_{+-}^{\alpha_{\tilde\phi}}(z; 0,0)\bigr)^2
\nonumber\\
&\quad
+ \frac{|\tilde{b}_0|^2|\tilde{b}_\pi|^2 }{16k^2} \bigl(\pi N_{--}^{\alpha_{\tilde\phi}}(z)\bigr)^2
-  \left(\tilde{b}_0\tilde{b}_\pi^{\ast}e^{-i q\theta}+\mathrm{h.c.}\right),
\label{eq:Ntphi_split_appendix}
\end{align}
where the entire $\theta$ dependence enters through the nonlocal
products of the $U(1)_C$-violating fixed-point masses, $\tilde b_0\tilde b_\pi^*$, as anticipated.
{In particular,
\begin{equation}
\left.{\cal N}_{\tilde\phi}(z;\theta)\right|_{\tilde b_0=\tilde b_\pi=0}
=
\left.{\cal N}_{\tilde\phi}(z;0)\right|_{\tilde b_0=\tilde b_\pi=0}
=
\bigl(\pi k N_{++}^{\alpha_{\tilde\phi}}(z;0,0)\bigr)^2,
\label{eq:appB_Ptype_zero}
\end{equation}
which gives the independent KK check of $V_{\tilde\phi}(\theta)=0$ used in Sec.~\ref{subsec:KKmethod}.}

\subsection{Spectral functions of fermion fields}\label{app:fermion}

Here we present the spectral function of a $C$-twisted charged bulk fermion field in the absence of fixed-point operators. Although one may introduce $U(1)_C$-violating fixed-point fermion masses as in Eq.~\eqref{eq:appC_brane}, the resulting KK spectra become considerably more involved, while their qualitative effects closely parallel those in the scalar case. We therefore restrict our attention to the case without fixed-point mass terms.

Consider a $C$-twisted bulk fermion $\psi$ with mass $M$ and $U(1)_C$ charge $q$ in the RS background, cf.\ Eq.~\eqref{eq:S_bulk}. The Dirac equation is given by
\begin{equation}
\Big(\Gamma^M \mathcal{D}_M + M\Big)\psi \;=\; 0,
\label{eq:dirac_eom}
\end{equation}
where the covariant derivative
\[
\mathcal{D}_M = \partial_M - i q\, C_M + S_M
\]
includes the spin connection~\cite{Gherghetta:2000qt},
\begin{equation}
S_\mu = \frac{1}{4}k\,\epsilon(y)\,[\Gamma_5,\Gamma_\mu],
\qquad
S_5=0.
\end{equation}
The gamma matrices in the RS geometry are given by
\[
\Gamma_\mu = e^{-k|y|}\gamma_\mu, \qquad \Gamma_5=\gamma_5,
\]
where the flat-space gamma matrices satisfy
\[
\{\gamma^M,\gamma^N\}=2\eta^{MN}, \qquad
\gamma_5=-i\gamma^0\gamma^1\gamma^2\gamma^3.
\]

After performing the field redefinition (as in Eq.~\eqref{eq:field_red_Phi}) that removes $C_5=\theta/2\pi R$ from the covariant derivative, and introducing the chiral projections $\psi_{L,R}\equiv P_{L,R}\psi$ with $P_{L,R}=\frac{1}{2}(1\pm\gamma^5)$, each chiral component satisfies a second-order differential equation,
\begin{equation}
\Big[ e^{2k|y|}\eta^{\mu\nu}\partial_\mu\partial_\nu
+ e^{k|y|}\partial_y\big(e^{-k|y|}\partial_y\big)
- \mathcal{M}^2_{L,R}(y)
\Big]\psi_{L,R}=0,
\label{eq:fermion_second_order}
\end{equation}
where
\begin{equation}
\mathcal{M}^2_{L,R}(y)\equiv M^2 \pm k \epsilon(y) M.
\label{eq:MLR_def}
\end{equation}
The left- and right-handed components are not independent, but are related through the first-order Dirac equation Eq.~\eqref{eq:dirac_eom}.

One can expand the 5D fermion field $\psi$ in Kaluza--Klein (KK) modes as
\begin{equation}
\psi(x,y)
=\frac{1}{\sqrt{2\pi R}}\sum_{\lambda=L,R}\sum_{n} \psi_n^\lambda(x)\,g_n^\lambda(y),
\label{eq:fermion_KK}
\end{equation}
where the 4D fields $\psi_n^\lambda(x)$ and the corresponding KK profiles $g_n^\lambda(y)$ satisfy
\begin{equation}
\Bigl[-\eta^{\mu\nu}\partial_\mu\partial_\nu + m_n^2\Bigr]\psi_n^\lambda(x)=0, \qquad
\frac{1}{2\pi R}\int_{-\pi R}^{\pi R}dy\,e^{k|y|}\,g_m^{\lambda\ast}(y)\,g_n^\lambda(y)=\delta_{mn}.
\label{eq:KG_4d_fermion}
\end{equation}
Substituting Eq.~\eqref{eq:fermion_KK} into Eq.~\eqref{eq:fermion_second_order}, we obtain the mode equation for the profiles $g_n^\lambda(y)$ ($\lambda=L,R$):
\begin{equation}
\Big[
e^{2k|y|}m_n^2
+ e^{k|y|}\partial_y\!\big(e^{-k|y|}\partial_y\big)
- \mathcal{M}^2_\lambda(y)
\Big]g_n^\lambda(y)=0.
\label{eq:mode_eom}
\end{equation}

As in the scalar case, the implementation of boundary and matching conditions at the orbifold fixed points is facilitated by solving the bulk equation piecewise over three domains,
\begin{equation}
g_n^\lambda(y)=
\begin{cases}
g_{n(1)}^\lambda(y) & \quad (-\pi R<y<0),\\
g_{n(2)}^\lambda(y) & \quad (0<y<\pi R),\\
g_{n(3)}^\lambda(y) & \quad (\pi R<y<2\pi R),
\end{cases}
\label{eq:domains_fermion}
\end{equation}
where each component $g_{n(I)}^\lambda(y)$ ($I=1,2,3$) satisfies the homogeneous bulk equation away from the fixed points,
\begin{equation}
\Big[
e^{2k|y|}m_n^2
+ e^{k|y|}\partial_y\big(e^{-k|y|}\partial_y\big)
- M_{\lambda}^2(y)
\Big]g_{n(I)}^\lambda(y)=0,
\qquad (I=1,2,3).
\label{eq:bulk_piecewise_fermion}
\end{equation}
In the present setup, unlike in the scalar case, the effective bulk mass is domain-dependent:
\begin{equation}
M_{\lambda}^2(y)=
\begin{cases}
M^2_{\lambda(1)}= M^2 \mp kM & \quad (-\pi R<y<0),\\
M^2_{\lambda(2)}= M^2 \pm kM & \quad (0<y<\pi R),\\
M^2_{\lambda(3)}= M^2 \mp kM & \quad (\pi R<y<2\pi R),
\end{cases}
\end{equation}
where the upper (lower) sign corresponds to $\lambda=L$ ($\lambda=R$).
The general solution to Eq.~\eqref{eq:bulk_piecewise_fermion} can be written as
\begin{equation}
g^\lambda_{n(I)}(y)
= e^{\frac{1}{2} k|y|}
\left[
A^\lambda_{n(I)}\,J_{\alpha_{\lambda(I)}}\!\left(\frac{m_n}{k}e^{k|y|}\right)
+ B^\lambda_{n(I)}\,Y_{\alpha_{\lambda(I)}}\!\left(\frac{m_n}{k}e^{k|y|}\right)
\right],
\label{eq:general_solution_fermion}
\end{equation}
with
\begin{equation}
\alpha_{\lambda(I)}=\sqrt{\frac{M_{\lambda(I)}^2}{k^2}+\frac14}.
\label{eq:bessel_order_fermion}
\end{equation}
The coefficients $A^\lambda_{n(I)}$ and $B^\lambda_{n(I)}$ are constant within each domain and are determined by the orbifold boundary conditions and the matching relations implied by the first-order Dirac equation Eq.~\eqref{eq:dirac_eom}.

The orbifold boundary conditions, together with the Dirac equation Eq.~\eqref{eq:dirac_eom}, relate the solutions across different domains. In particular, the coefficients satisfy
\begin{align}
&A^{L/R}_{n(1)}=\big[A^{R/L}_{n(2)}\big]^\ast,
\qquad
B^{L/R}_{n(1)}=\big[B^{R/L}_{n(2)}\big]^\ast,
\nonumber\\
&A^{L/R}_{n(3)}=e^{-iq\theta}\,A^{L/R}_{n(1)},
\qquad
B^{L/R}_{n(3)}=e^{-iq\theta}\,B^{L/R}_{n(1)}.
\end{align}
The phase factor $e^{-iq\theta}$ originates from the twisted boundary condition
$\psi(x,y+2\pi R)=e^{-iq\theta}\psi(x,y)$
after performing the field redefinition that removes $C_5=\theta/2\pi R$ from the covariant derivative.
Moreover, one may choose a basis for the bulk solutions on the central domain $(2)$ such that the remaining relations between the left- and right-handed coefficients take the form
\begin{align}
A^R_{n(2)} &=
\begin{cases}
A^L_{n(2)} & \left( M>\dfrac{k}{2}\right),\\[4pt]
\sin\!\left(\dfrac{M\pi}{k}\right)A^L_{n(2)}+\cos\!\left(\dfrac{M\pi}{k}\right)B^L_{n(2)}
&  \left( 0\le M\le \dfrac{k}{2}\right),
\end{cases}
\nonumber\\
B^R_{n(2)} &=
\begin{cases}
B^L_{n(2)}  & \left( M>\dfrac{k}{2}\right),\\[4pt]
-\cos\!\left(\dfrac{M\pi}{k}\right)A^L_{n(2)}+\sin\!\left(\dfrac{M\pi}{k}\right)B^L_{n(2)} 
& \left(0\le M\le \dfrac{k}{2}\right).
\end{cases}
\label{eq:LR_basis_relations}
\end{align}
Although the relations in Eq.~\eqref{eq:LR_basis_relations} take different forms in the regimes $0\le M\le k/2$ and $M>k/2$, they differ only by a choice of basis for the bulk solution coefficients. We therefore adopt the parametrization for $M>k/2$ to streamline the intermediate steps, while presenting final results that are valid in both regimes.
With this choice, the coefficient relations can be expressed entirely within a single chiral sector:
\begin{equation}
A^{L/R}_{n(1)}=\big[A^{L/R}_{n(2)}\big]^\ast=e^{iq\theta}\,A^{L/R}_{n(3)},
\qquad
B^{L/R}_{n(1)}=\big[B^{L/R}_{n(2)}\big]^\ast=e^{iq\theta}\,B^{L/R}_{n(3)}.
\label{eq:coeff_rel_simplified}
\end{equation}

Imposing the boundary/matching conditions at $y=0$ and $y=\pi R$ yields a homogeneous linear system for the
independent coefficients, which can be written as  
\begin{equation}
\begin{pmatrix}
\mathcal{J}_{\bar{\alpha}_{\lambda}}^{0-}  & -\mathcal{J}_{\alpha_{\lambda}}^{0-}   & \mathcal{Y}_{\bar{\alpha}_{\lambda}}^{0-}  & - \mathcal{Y}_{\alpha_{\lambda}}^{0-}  \\
\mathcal{J}_{\bar{\alpha}_\lambda}^{0+}  &  \mathcal{J}_{\alpha_{\lambda}}^{0+}   & \mathcal{Y}_{\bar{\alpha}_{\lambda}}^{0+}  &  \mathcal{Y}_{\alpha_{\lambda}}^{0+} \\
e^{-iq\theta} \mathcal{J}_{\bar{\alpha}_{\lambda}}^{\pi -}  & -\mathcal{J}_{\alpha_{\lambda}}^{\pi -} & e^{-iq\theta} \mathcal{Y}_{\bar{\alpha}_{\lambda}}^{\pi -}  & -\mathcal{Y}_{\alpha_{\lambda}}^{\pi -}  \\
e^{-iq\theta} \mathcal{J}_{\bar{\alpha}_{\lambda}}^{\pi +}  &  \mathcal{J}_{\alpha_{\lambda}}^{\pi +}  & e^{-iq\theta} \mathcal{Y}_{\bar{\alpha}_{\lambda}}^{\pi +}  &  \mathcal{Y}_{\alpha_{\lambda}}^{\pi +}
\end{pmatrix}
\begin{pmatrix}
A^{\lambda}_{n(1)} \\
\big[A^{\lambda}_{n(1)}\big]^\ast \\
B^{\lambda}_{n(1)} \\
\big[B^{\lambda}_{n(1)}\big]^\ast
\end{pmatrix}
=0,
\label{eq:matrix}
\end{equation}
where  
\begin{equation}
\alpha_{L,R} 
= \left|\frac{M}{k}\pm\frac12\right|
\equiv \frac{M_{L,R}}{k}, \qquad \bar\alpha_{L,R}=\alpha_{R,L}.
\label{eq:alpha_pm_def}
\end{equation}

Taking into account the fermion spin degeneracy, the spectral function of $\psi$ is given by the square of the determinant of the matrix in Eq.~\eqref{eq:matrix}. We then find
\begin{equation}
{\cal N}_\psi(z;\theta)= \left[\hat {\cal N}_\psi (z;\theta)\right]^2,
\end{equation}
where
\begin{equation}
\hat {\cal N}_\psi (z;\theta)=N_\psi(z) +{\cal A}_\psi(\theta),
\end{equation}
with
\begin{align}
N_\psi(z)+1
&= \frac{\pi^2}{8} N^{\alpha_\lambda}_{++}(z;0,0)\,N^{\bar{\alpha}_\lambda}_{--}(z)
+\frac{\pi^2}{8} N^{\alpha_\lambda}_{+-}(z;0,0)\,N^{\bar{\alpha}_\lambda}_{-+}(z;0,0)
+\Big(\alpha_\lambda\leftrightarrow\bar{\alpha}_\lambda\Big), \nonumber\\
{\cal A}_\psi(\theta)
&= 1 - \cos(q\theta).
\label{eq:Npsi_det}
\end{align}
Since $N_\psi(z)$ is symmetric under $\alpha_\lambda\leftrightarrow\bar{\alpha}_\lambda$, the left- and right-handed components yield identical spectral functions.
{In Sec.~\ref{subsec:KKmethod} and Appendix~\ref{app:common_determinant} we use the single-block function $\hat{\cal N}_\psi$ and account for the two identical spin blocks through the multiplicity $g_\psi=2$ rather than by squaring the determinant a second time.}

\begingroup
\section{Fixed-point terms in the worldline and KK calculations}
\label{app:common_determinant}

The warp-dependent field redefinitions used in Sec.~\ref{worldline} generate delta-function terms at the orbifold fixed points in the scalar and fermion quadratic equations.  These terms preserve $U(1)_C$.  They are therefore different from the explicit $U(1)_C$-violating fixed-point operators studied in Sec.~\ref{sec:fixedpoint}, and also from the curvature-dependent scalar boundary masses required by the supersymmetric hypermultiplet action.  In Sec.~\ref{worldline} we first isolated the propagation between the fixed points in order to display the winding action.  Here we include the fixed-point terms and determine how they change the one-loop prefactors.  As shown below, they do not change the winding exponents.

Appendix~\ref{app:Kaluza-Klein_spectral_functions} derives the KK spectral functions with general fixed-point operators.  In this appendix the explicit $U(1)_C$-violating operators are set to zero.  We first give the quadratic equations and their fixed-point matching conditions, then evaluate the first winding in the proper-time calculation, and finally obtain the same quantities from the KK spectral functions.  Appendix~\ref{app:completed_monodromy} gives the corresponding transfer-matrix calculation.

\subsection{Quadratic operators and fixed-point matching}

After Fourier transformation along the four-dimensional directions and the warp-dependent field redefinitions used in Sec.~\ref{worldline}, the one-dimensional quadratic operators are
\begin{align}
{\cal O}_\phi(p)
&=-\partial_y^2+M_{\rm eff}^2+e^{2k|y|}p^2
-2k\,\partial_y\epsilon(y),
\qquad M_{\rm eff}^2=M^2+4k^2,
\label{eq:appC_Ophi}\\
{\cal O}_\psi(p)
&=-\partial_y^2+
\left(M+\frac{k}{2}\epsilon(y)\right)^2
+e^{2k|y|}p^2
-\frac{k}{2}\,\partial_y\epsilon(y).
\label{eq:appC_Opsi}
\end{align}
With the periodic convention for $\epsilon(y)$ used in Sec.~2.1,
\begin{equation}
\partial_y\epsilon(y)
=2\sum_{m\in\mathbb Z}
\left[\delta(y-2m\pi R)-\delta\bigl(y-(2m+1)\pi R\bigr)\right].
\label{eq:appC_depsilon}
\end{equation}
Integrating the equations across a fixed point gives the matching conditions for the rescaled profiles $u_\Phi$,
\begin{equation}
u_\Phi'(y_i^+)-u_\Phi'(y_i^-)
=\lambda_i^\Phi u_\Phi(y_i),
\qquad
(\lambda_0^\phi,\lambda_\pi^\phi)=(-4k,+4k),
\qquad
(\lambda_0^\psi,\lambda_\pi^\psi)=(-k,+k).
\label{eq:appC_jumps}
\end{equation}
The profiles used in the KK calculation of Appendix~\ref{app:Kaluza-Klein_spectral_functions} are related to these rescaled variables by
\begin{equation}
u_{\phi,n}=e^{-2k|y|}f_n,
\qquad
u_{\psi,n}=e^{-k|y|/2}g_n^L.
\label{eq:appC_profile_relation}
\end{equation}

We remove the constant background $C_5=\theta/(2\pi R)$ from the differential operator by the same field redefinition used in the main text.  The axion dependence then appears in the boundary condition around the covering circle,
\begin{equation}
u_\Phi(y+2\pi R)=e^{-iq\theta}u_\Phi(y).
\label{eq:appC_twist}
\end{equation}
Thus Eqs.~\eqref{eq:appC_Ophi}--\eqref{eq:appC_twist} specify the scalar and fermion problems used below, including the fixed-point terms generated by the warp-dependent field redefinitions.

\subsection{Worldline calculation including the fixed-point terms}
\label{appC:worldline_matching}

For either $\Phi=\phi$ or $\psi$, the $\theta$-dependent part of the one-loop potential can be written as
\begin{equation}
V_\Phi(\theta)-V_\Phi(0)
=(-1)^{n_F}\frac{g_\Phi}{2}
\int\frac{d^4p}{(2\pi)^4}
\ln\frac{\det_\theta {\cal O}_\Phi(p)}{\det_0 {\cal O}_\Phi(p)},
\label{eq:appC_relative_det}
\end{equation}
where $(n_F,g_\Phi)=(0,1)$ for the complex scalar and $(1,2)$ for the Dirac fermion.  Here $\det_\theta$ denotes the determinant evaluated with the $\theta$-dependent boundary condition in Eq.~\eqref{eq:appC_twist}, while $\det_0$ denotes the periodic case.  For the fermion, ${\cal O}_\psi$ is one of the two identical four-dimensional spin blocks of the second-order equation; the factor $g_\psi=2$ accounts for their multiplicity.

Using the Schwinger proper-time formula,
\begin{equation}
\ln\frac{\det_\theta {\cal O}_\Phi(p)}{\det_0 {\cal O}_\Phi(p)}
=-\int_0^\infty\frac{dT}{T}
\left[
{\rm Tr}_\theta e^{-T{\cal O}_\Phi(p)}
-{\rm Tr}_0 e^{-T{\cal O}_\Phi(p)}
\right].
\label{eq:appC_propertime}
\end{equation}
On the covering circle, the trace separates into paths satisfying $y(1)=y(0)+2n\pi R$.  For $n\geq1$ define
\begin{equation}
W_{\Phi,n}(p)
=\int_0^\infty\frac{dT}{T}
\int_0^{2\pi R}dy\,
\langle y+2n\pi R|e^{-T{\cal O}_\Phi(p)}|y\rangle .
\label{eq:appC_W_worldline}
\end{equation}
This is the same $n$-winding contribution that appears in the worldline path integral of Eq.~\eqref{eq:V_worldline_master}.  The two orientations carry phases $e^{\pm inq\theta}$, and therefore
\begin{equation}
\ln\frac{\det_\theta {\cal O}_\Phi(p)}{\det_0 {\cal O}_\Phi(p)}
=2\sum_{n=1}^{\infty}W_{\Phi,n}(p)\,[1-\cos(nq\theta)].
\label{eq:appC_Wdef}
\end{equation}
For a large bulk mass the $n=1$ term gives the leading harmonic,
\begin{equation}
V_\Phi^{(1)}(\theta)
=(-1)^{n_F+1}g_\Phi\cos(q\theta)
\int\frac{d^4p}{(2\pi)^4}W_{\Phi,1}(p),
\label{eq:appC_first_harmonic}
\end{equation}
up to a $\theta$-independent constant.

For small four-dimensional Euclidean momentum we write
\begin{equation}
W_{\Phi,1}(p)
=W_{\Phi,1}(0)
\exp\!\left[-B_\Phi p^2+{\cal O}(p^4)\right].
\label{eq:appC_Bdef}
\end{equation}
The coefficient $B_\Phi$ follows directly from the same proper-time kernel.  Since the $p^2$ dependence of the operator is $p^2e^{2k|y|}$, differentiating Eq.~\eqref{eq:appC_W_worldline} gives
\begin{equation}
B_\Phi
=\frac{1}{W_{\Phi,1}(0)}
\int_0^{2\pi R}dy\,e^{2k|y|}
\int_0^\infty dT\,
\langle y+2\pi R|e^{-T{\cal O}_\Phi(0)}|y\rangle .
\label{eq:appC_B_propertime}
\end{equation}
Thus $W_{\Phi,1}(0)$ fixes the zero-momentum one-winding contribution, while $B_\Phi$ gives its leading momentum dependence.  If the quadratic term in Eq.~\eqref{eq:appC_Bdef} is used for the momentum integral, one obtains
\begin{equation}
V_\Phi^{(1)}(\theta)
\simeq
(-1)^{n_F+1}\frac{g_\Phi}{16\pi^2}
\frac{W_{\Phi,1}(0)}{B_\Phi^2}\cos(q\theta).
\label{eq:appC_gaussian_potential}
\end{equation}
Keeping only the first winding and approximating its momentum dependence by Eq.~\eqref{eq:appC_Bdef} are independent approximations.  The first is controlled by the large winding action; the second is the Gaussian momentum approximation tested numerically in Sec.~\ref{subsec:KKmethod}.

For the scalar, solving the zero-momentum equation between the two fixed points and imposing the matching conditions in Eq.~\eqref{eq:appC_jumps} gives, for $M_{\rm eff}\pi R\gg1$,
\begin{align}
W_{\phi,1}(0)
&\simeq
\frac{M_{\rm eff}^2}{M_{\rm eff}^2-4k^2}
\,e^{-2M_{\rm eff}\pi R},
\label{eq:appC_Wphi}\\[2mm]
B_\phi
&\simeq
\frac{e^{2k\pi R}-1}
{2k(M_{\rm eff}^2-4k^2)(M_{\rm eff}^2-k^2)}
\left[
M_{\rm eff}(M_{\rm eff}^2-7k^2)
+6k^3\coth(k\pi R)
\right].
\label{eq:appC_Bphi}
\end{align}
If the fixed-point term $-2k\,\partial_y\epsilon$ is omitted, the bulk equation instead gives
\begin{equation}
W_{\phi,1}^{\rm bulk}(0)=e^{-2M_{\rm eff}\pi R},
\qquad
B_\phi^{\rm bulk}
=\frac{e^{2k\pi R}-1}{2M_{\rm eff}k}.
\label{eq:appC_scalar_bulk}
\end{equation}
Thus the fixed-point term changes the prefactor but not the exponential $e^{-2M_{\rm eff}\pi R}$.  At fixed $k\pi R$ and for $k/M\ll1$,
\begin{align}
\frac{W_{\phi,1}(0)}{W_{\phi,1}^{\rm bulk}(0)}
&=1+4\left(\frac{k}{M}\right)^2,
\nonumber\\
\frac{B_\phi}{B_\phi^{\rm bulk}}
&=1-2\left(\frac{k}{M}\right)^2
+6\coth(k\pi R)\left(\frac{k}{M}\right)^3
+{\cal O}\!\left(\frac{k^4}{M^4}\right).
\label{eq:appC_scalar_smallk}
\end{align}
The coefficient obtained from Eq.~\eqref{eq:appC_gaussian_potential} therefore changes by
\begin{equation}
\frac{V_{\phi,\,\mathrm{Gauss}}^{(1)}}{V_{\phi,\,\mathrm{bulk}}^{(1)}}
=1+8\left(\frac{k}{M}\right)^2
-12\coth(k\pi R)\left(\frac{k}{M}\right)^3
+{\cal O}\!\left(\frac{k^4}{M^4}\right).
\label{eq:appC_scalar_prefactor_ratio}
\end{equation}
This is the result quoted in Sec.~\ref{worldline}: the fixed-point term gives no correction linear in $k/M$ to the coefficient obtained with the Gaussian momentum approximation.  This effect is distinct from the fluctuation correction to the bulk warp-weight functional studied in Appendix~\ref{app:path_integral_expansion}; the latter approaches order $k/M_{\rm eff}$ in the strongly warped limit but is further suppressed near the flat limit.

For the fermion, the zero-momentum operator including the fixed-point term factorizes as
\begin{equation}
{\cal O}_\psi(0)
=\left[-\partial_y+M+\frac{k}{2}\epsilon(y)\right]
\left[\partial_y+M+\frac{k}{2}\epsilon(y)\right].
\label{eq:appC_fermion_factor}
\end{equation}
The two first-order factors can be written as
\[
\partial_y+M+\frac{k}{2}\epsilon(y)
=e^{-k|y|/2}(\partial_y+M)e^{k|y|/2},
\qquad
-\partial_y+M+\frac{k}{2}\epsilon(y)
=e^{k|y|/2}(-\partial_y+M)e^{-k|y|/2}.
\]
Because the rescaling factors are periodic on the covering circle, they do not change either the $\theta$-dependent boundary condition or the periodic one.  The zero-momentum determinant therefore has the same winding dependence as the constant-mass problem, and
\begin{equation}
W_{\psi,1}(0)=e^{-2M\pi R}
\label{eq:appC_Wpsi}
\end{equation}
is exact.  For $M\pi R\gg1$, the quadratic momentum coefficient is
\begin{equation}
B_\psi\simeq
\frac{2(e^{2k\pi R}-1)
\left[M(4M^2-3k^2)+k^3\coth(k\pi R)\right]}
{k(4M^2-k^2)^2}.
\label{eq:appC_Bpsi_largeM}
\end{equation}
This fixed-$k/M$ large-mass expression is valid on both sides of $M=k/2$.  At the isolated point $k=2M$, the finite-$M\pi R$ expression must be evaluated before taking the large-mass limit, as in Eq.~\eqref{eq:fermion_k2M}.  The winding exponent remains $2M\pi R$.  To isolate the fixed-point term itself, the appropriate comparison is with the same piecewise mass profile after removing the $-(k/2)\partial_y\epsilon$ term.  For $k/M\ll1$, the ratio of the Gaussian coefficients is
\begin{equation}
\frac{V_{\psi,\,\mathrm{full}}^{(1)}}
{V_{\psi,\,\partial\epsilon=0}^{(1)}}
=1+\frac54\left(\frac{k}{M}\right)^2
-\frac32\coth(k\pi R)\left(\frac{k}{M}\right)^3
+{\cal O}\!\left(\frac{k^4}{M^4}\right).
\label{eq:appC_fermion_prefactor_ratio}
\end{equation}
There is again no correction linear in $k/M$ from the fixed-point term.  The denominator in Eq.~\eqref{eq:appC_fermion_prefactor_ratio} is obtained from the same second-order operator, with the $\partial_y\epsilon$ term removed, using the same Gaussian momentum expansion.  This isolates the fixed-point contribution.  It should be distinguished from a direct ratio to Eq.~\eqref{eq:worldline_fermion_warped_pot}, which also uses the classical approximation for $I[y]$.

\subsection{KK spectral-function calculation}
\label{appC:KK_matching}

The same fixed-point terms are included in the boundary and matching conditions used to derive the spectral functions in Appendix~\ref{app:Kaluza-Klein_spectral_functions}.  Setting the explicit $U(1)_C$-violating fixed-point operators to zero, a single spectral block can be normalized so that
\begin{equation}
\frac{\det_\theta {\cal O}_\Phi(p)}{\det_0 {\cal O}_\Phi(p)}
=
\frac{N_\Phi(ip)+1-\cos(q\theta)}{N_\Phi(ip)}.
\label{eq:appC_det_spectral}
\end{equation}
For small Euclidean momentum, write
\begin{equation}
N_\Phi(ip)=N_{\Phi,0}+N_{\Phi,2}p^2+{\cal O}(p^4).
\label{eq:appC_Nexpand}
\end{equation}
Expanding the logarithm of Eq.~\eqref{eq:appC_det_spectral} in harmonics and comparing with Eq.~\eqref{eq:appC_Wdef} gives
\begin{equation}
W_{\Phi,1}(0)
=\frac{1}{N_{\Phi,0}+1+
\sqrt{N_{\Phi,0}(N_{\Phi,0}+2)}},
\qquad
B_\Phi
=\frac{N_{\Phi,2}}
{\sqrt{N_{\Phi,0}(N_{\Phi,0}+2)}}.
\label{eq:appC_WB_from_N}
\end{equation}
For $N_{\Phi,0}\gg1$,
\begin{equation}
W_{\Phi,1}(0)\simeq\frac{1}{2N_{\Phi,0}},
\qquad
B_\Phi\simeq\frac{N_{\Phi,2}}{N_{\Phi,0}},
\label{eq:appC_largeN}
\end{equation}
which is the large-mass approximation used in Sec.~\ref{subsec:KKmethod}.

For the scalar, the expansion of the spectral function gives
\begin{align}
N_{\phi,0}
&=
\frac{2(M_{\rm eff}^2-4k^2)}{M_{\rm eff}^2}
\sinh^2(M_{\rm eff}\pi R),
\label{eq:appC_Nphi0}\\
N_{\phi,2}
&=
\frac{(e^{2k\pi R}-1)\sinh^2(M_{\rm eff}\pi R)}
{kM_{\rm eff}^2(M_{\rm eff}^2-k^2)}
\nonumber\\[-1mm]
&\hspace{1.0cm}\times
\left[
M_{\rm eff}(M_{\rm eff}^2-7k^2)\coth(M_{\rm eff}\pi R)
+6k^3\coth(k\pi R)
\right].
\label{eq:appC_Nphi2}
\end{align}
Substitution into Eq.~\eqref{eq:appC_WB_from_N} gives Eqs.~\eqref{eq:appC_Wphi} and \eqref{eq:appC_Bphi} for $M_{\rm eff}\pi R\gg1$.

For the fermion,
\begin{align}
N_{\psi,0}
&=2\sinh^2(M\pi R),
\label{eq:appC_Npsi0}\\
N_{\psi,2}
&=\frac{4e^{k\pi R}}{k(4M^2-k^2)^2}
\Big[
M(4M^2-3k^2)\sinh(k\pi R)\sinh(2M\pi R)
\nonumber\\[-1mm]
&\hspace{3.0cm}
+k^3\{\cosh(k\pi R)\cosh(2M\pi R)-1\}
\Big].
\label{eq:appC_Npsi2}
\end{align}
At $k=2M$, Eq.~\eqref{eq:appC_Npsi2} is understood by taking the continuous limit before the large-$M\pi R$ expansion.  Equation~\eqref{eq:appC_WB_from_N} gives Eq.~\eqref{eq:appC_Wpsi} exactly and Eq.~\eqref{eq:appC_Bpsi_largeM} for $M\pi R\gg1$.

Substituting these coefficients into Eq.~\eqref{eq:appC_gaussian_potential} gives Eqs.~\eqref{eq:pot_general_warping_scalar} and \eqref{eq:pot_general_warping_fermion} in the main text.  No strong-warping expansion is needed: the flat-space and $k\pi R\gg1$ limits follow from the same expressions.  Equations~\eqref{eq:appC_WB_from_N} show explicitly that the proper-time and KK calculations give the same zero-momentum first-winding contribution and the same quadratic momentum dependence.

\section{Monodromy calculation}
\label{app:completed_monodromy}

Section~\ref{subsec:fermion_det} used a monodromy matrix for the first-order Dirac equation.  Here we apply the same transfer-matrix method to the second-order scalar and fermion equations of Appendix~\ref{app:common_determinant}, including their fixed-point matching conditions.  This gives an independent derivation of the determinant ratio and also shows explicitly how the second-order fermion equation is related to the first-order calculation in Sec.~\ref{subsec:fermion_det}.

\subsection{Monodromy matrix for the second-order equation}

Between the fixed points, write
\begin{equation}
{\bf u}_\Phi(y)=
\begin{pmatrix}u_\Phi(y)\\ u_\Phi'(y)\end{pmatrix},
\qquad
\partial_y{\bf u}_\Phi=A_\Phi(y;p){\bf u}_\Phi,
\label{eq:appD_first_order}
\end{equation}
with
\begin{equation}
A_\phi(y;p)=
\begin{pmatrix}
0&1\\ M_{\rm eff}^2+e^{2k|y|}p^2&0
\end{pmatrix},
\qquad
A_\psi(y;p)=
\begin{pmatrix}
0&1\\ \left(M+\frac{k}{2}\epsilon(y)\right)^2+e^{2k|y|}p^2&0
\end{pmatrix}.
\label{eq:appD_A_matrices}
\end{equation}
The transfer matrix from $a$ to $b$ is
\begin{equation}
T_\Phi(b,a;p)
={\cal P}\exp\!\left[\int_a^bdy\,A_\Phi(y;p)\right].
\label{eq:appD_transfer}
\end{equation}
The fixed-point matching condition in Eq.~\eqref{eq:appC_jumps} can be written as
\begin{equation}
{\bf u}_\Phi(y_i^+)=M_i^\Phi {\bf u}_\Phi(y_i^-),
\qquad
M_i^\Phi=
\begin{pmatrix}1&0\\ \lambda_i^\Phi&1\end{pmatrix}.
\label{eq:appD_jump_matrix}
\end{equation}
Starting at $0^+$, the transfer matrix for one circuit of the covering circle is
\begin{equation}
U_\Phi(p)
=M_0^\Phi T_\Phi(2\pi R,\pi R;p)
M_\pi^\Phi T_\Phi(\pi R,0;p),
\qquad \det U_\Phi(p)=1.
\label{eq:appD_monodromy}
\end{equation}
We parameterize its eigenvalues as $e^{\pm\ell_\Phi(p)}$, so that
\begin{equation}
\frac12\operatorname{tr}U_\Phi(p)=\cosh\ell_\Phi(p).
\label{eq:appD_ell_def}
\end{equation}
The boundary condition in Eq.~\eqref{eq:appC_twist} requires $e^{-iq\theta}$ to be an eigenvalue of the one-circuit transfer matrix at an eigenvalue of the differential operator.  Equivalently, the spectral condition is
\[
\det\!\left[U_\Phi(p)-e^{-iq\theta}{\bf 1}_2\right]=0.
\]
Using $\det U_\Phi=1$ and dividing by the result at $\theta=0$ gives
\begin{equation}
\frac{\det_\theta {\cal O}_\Phi(p)}{\det_0 {\cal O}_\Phi(p)}
=
\frac{\cosh\ell_\Phi(p)-\cos(q\theta)}
{\cosh\ell_\Phi(p)-1}.
\label{eq:appD_monodromy_det}
\end{equation}
Comparison with Eqs.~\eqref{eq:appC_Wdef} and \eqref{eq:appC_det_spectral} gives
\begin{equation}
W_{\Phi,n}(p)=\frac{e^{-n\ell_\Phi(p)}}{n},
\qquad
N_\Phi(ip)=\cosh\ell_\Phi(p)-1.
\label{eq:appD_equivalence}
\end{equation}
Hence, if $\ell_\Phi(p)=\ell_\Phi(0)+B_\Phi p^2+{\cal O}(p^4)$, Eq.~\eqref{eq:appD_equivalence} gives the same $B_\Phi$ as Eq.~\eqref{eq:appC_Bdef}.

At zero momentum the transfer matrices can be evaluated in closed form.  For the scalar and fermion,
\begin{align}
\frac12\operatorname{tr}U_\phi(0)
&=1+\frac{2(M_{\rm eff}^2-4k^2)}{M_{\rm eff}^2}
\sinh^2(M_{\rm eff}\pi R),
\label{eq:appD_scalar_trace}\\
\frac12\operatorname{tr}U_\psi(0)
&=\cosh(2M\pi R),
\nonumber\\[-1mm]
\ell_\psi(0)&=2M\pi R.
\label{eq:appD_fermion_trace}
\end{align}
Together with Eq.~\eqref{eq:appD_equivalence}, these expressions reproduce Eqs.~\eqref{eq:appC_Nphi0} and \eqref{eq:appC_Npsi0}; for the fermion they also give Eq.~\eqref{eq:appC_Wpsi}.  Expanding the one-circuit transfer matrix to first order in $p^2$ gives Eqs.~\eqref{eq:appC_Nphi2} and \eqref{eq:appC_Npsi2}.  Thus the transfer-matrix calculation gives the same $W_{\Phi,1}(0)$ and $B_\Phi$ as the proper-time and KK calculations.

\subsection{Relation to the first-order fermion calculation}

Section~\ref{subsec:fermion_det} starts from the first-order Dirac equation.  Equation~\eqref{eq:appD_monodromy} instead follows from the second-order equation in Appendix~\ref{app:common_determinant}.  The two are related by a periodic change of variables.  For $|p|>0$, let $\chi$ be the two-component state used in Sec.~\ref{subsec:fermion_det}.  It is related to the second-order state by
\begin{equation}
\chi(y)=S(y;p){\bf u}_\psi(y),
\qquad
S(y;p)=
\begin{pmatrix}
e^{k|y|/2} & 0\\[1mm]
\dfrac{e^{-k|y|/2}}{|p|}\left(M+\dfrac{k}{2}\epsilon(y)\right)
& \dfrac{e^{-k|y|/2}}{|p|}
\end{pmatrix}.
\label{eq:appD_S_matrix}
\end{equation}
The matching condition at a fixed point implies
\begin{equation}
S(y_i^+;p)M_i^\psi=S(y_i^-;p).
\label{eq:appD_S_jump}
\end{equation}
Thus the fixed-point matching is already contained in the first-order variables.  Since $S(y;p)$ is periodic on the covering circle, the two one-circuit matrices satisfy
\begin{equation}
U_h(p)=S(0^+;p)\,U_\psi(p)\,S(0^+;p)^{-1}.
\label{eq:appD_similarity}
\end{equation}
They therefore have the same eigenvalues.  With the notation of Sec.~\ref{subsec:fermion_det},
\begin{equation}
\ell_\psi(p)=\ell(p),
\qquad
B_\psi=C_2.
\label{eq:appD_fermion_equivalence}
\end{equation}
This shows that the first-order fermion calculation in Sec.~\ref{subsec:fermion_det} and the second-order calculation above give the same determinant ratio, including the fixed-point term generated by the warp-dependent field redefinition.

\endgroup

\bibliographystyle{JHEP}
\bibliography{Refs}

@article{Peccei:1977hh,
  author       = {Peccei, R. D. and Quinn, Helen R.},
  title        = {CP Conservation in the Presence of Instantons},
  journal      = {Phys. Rev. Lett.},
  volume       = {38},
  pages        = {1440--1443},
  year         = {1977},
  doi          = {10.1103/PhysRevLett.38.1440}
}

@article{Peccei:1977ur,
  author       = {Peccei, R. D. and Quinn, Helen R.},
  title        = {Constraints Imposed by CP Conservation in the Presence of Instantons},
  journal      = {Phys. Rev. D},
  volume       = {16},
  pages        = {1791--1797},
  year         = {1977},
  doi          = {10.1103/PhysRevD.16.1791}
}

@article{Weinberg:1977ma,
  author       = {Weinberg, Steven},
  title        = {A New Light Boson?},
  journal      = {Phys. Rev. Lett.},
  volume       = {40},
  pages        = {223--226},
  year         = {1978},
  doi          = {10.1103/PhysRevLett.40.223}
}

@article{Csaki:2026qjl,
    author = "Cs{\'a}ki, Csaba and Kuflik, Eric and Xue, Wei and Youn, Taewook",
    title = "{The Holographic QCD Axion in Five Dimensions}",
    eprint = "2604.02411",
    archivePrefix = "arXiv",
    primaryClass = "hep-ph",
    month = "4",
    year = "2026"
}

@article{Babu:2024qzb,
    author = "Babu, K. S. and Dutta, Bhaskar and Mohapatra, Rabindra N.",
    title = "{Accidental Peccei-Quinn symmetry from gauged U(1) and a high quality axion}",
    eprint = "2412.21157",
    archivePrefix = "arXiv",
    primaryClass = "hep-ph",
    doi = "10.1007/JHEP03(2026)084",
    journal = "JHEP",
    volume = "03",
    pages = "084",
    year = "2026"
}

@article{Babu:2026yqp,
    author = "Babu, K. S. and Chandrasekar, Sai Charan and Tavartkiladze, Zurab",
    title = "{Fermion Mass Hierarchy and a High Quality Axion From Gauged U(1) Flavor Symmetry}",
    eprint = "2602.24253",
    archivePrefix = "arXiv",
    primaryClass = "hep-ph",
    month = "2",
    year = "2026"
}

@article{Wilczek:1977pj,
  author       = {Wilczek, Frank},
  title        = {Problem of Strong $P$ and $T$ Invariance in the Presence of Instantons},
  journal      = {Phys. Rev. Lett.},
  volume       = {40},
  pages        = {279--282},
  year         = {1978},
  doi          = {10.1103/PhysRevLett.40.279}
}

@article{Benabou:2023npn,
    author = "Benabou, Joshua N. and Bonnefoy, Quentin and Buschmann, Malte and Kumar, Soubhik and Safdi, Benjamin R.",
    title = "{Cosmological dynamics of string theory axion strings}",
    eprint = "2312.08425",
    archivePrefix = "arXiv",
    primaryClass = "hep-ph",
    doi = "10.1103/PhysRevD.110.035021",
    journal = "Phys. Rev. D",
    volume = "110",
    number = "3",
    pages = "035021",
    year = "2024"
}

@article{Adachi:2021rjw,
    author = "Adachi, Yuki and Lim, C. S. and Maru, Nobuhito",
    title = "{The strong CP problem and higher-dimensional gauge theories}",
    eprint = "2108.07367",
    archivePrefix = "arXiv",
    primaryClass = "hep-th",
    reportNumber = "OCU-PHYS-543, NITEP-111",
    doi = "10.1093/ptep/ptac070",
    journal = "PTEP",
    volume = "2022",
    number = "5",
    pages = "053B06",
    year = "2022"
}

@article{Arkani-Hamed:2006emk,
    author = "Arkani-Hamed, Nima and Motl, Lubos and Nicolis, Alberto and Vafa, Cumrun",
    title = "{The String landscape, black holes and gravity as the weakest force}",
    eprint = "hep-th/0601001",
    archivePrefix = "arXiv",
    reportNumber = "HUTP-05-A0057",
    doi = "10.1088/1126-6708/2007/06/060",
    journal = "JHEP",
    volume = "06",
    pages = "060",
    year = "2007"
}

@article{Choi:2014uaa,
    author = "Choi, Kiwoon and Jeong, Kwang Sik and Seo, Min-Seok",
    title = "{String theoretic QCD axions in the light of PLANCK and BICEP2}",
    eprint = "1404.3880",
    archivePrefix = "arXiv",
    primaryClass = "hep-th",
    reportNumber = "DESY-14-051, CTPU-14-01",
    doi = "10.1007/JHEP07(2014)092",
    journal = "JHEP",
    volume = "07",
    pages = "092",
    year = "2014"
}

@article{Hosotani:1983xw,
    author = "Hosotani, Yutaka",
    title = "{Dynamical Mass Generation by Compact Extra Dimensions}",
    reportNumber = "UPR-0216T",
    doi = "10.1016/0370-2693(83)90170-3",
    journal = "Phys. Lett. B",
    volume = "126",
    pages = "309--313",
    year = "1983"
}

@article{Azatov:2025mep,
    author = "Azatov, Aleksandr and Mahdi Khalil, Mohamed and Suzuki, Motoo",
    title = "{Towards a post-inflationary composite axion model}",
    eprint = "2510.18538",
    archivePrefix = "arXiv",
    primaryClass = "hep-ph",
    reportNumber = "SISSA 13/2025/FISI",
    doi = "10.1007/JHEP03(2026)143",
    journal = "JHEP",
    volume = "03",
    pages = "143",
    year = "2026"
}

@article{Flacke:2006ad,
    author = "Flacke, Thomas and Gripaios, Ben and March-Russell, John and Maybury, David",
    title = "{Warped axions}",
    eprint = "hep-ph/0611278",
    archivePrefix = "arXiv",
    reportNumber = "OUTP-06-16-P",
    doi = "10.1088/1126-6708/2007/01/061",
    journal = "JHEP",
    volume = "01",
    pages = "061",
    year = "2007"
}

@article{GrootNibbelink:2001bx,
    author = "Groot Nibbelink, S.",
    title = "{Dimensional regularization of a compact dimension}",
    eprint = "hep-th/0108185",
    archivePrefix = "arXiv",
    doi = "10.1016/S0550-3213(01)00539-9",
    journal = "Nucl. Phys. B",
    volume = "619",
    pages = "373--384",
    year = "2001"
}

@article{Kim:1979if,
  author       = {Kim, Jihn E.},
  title        = {Weak-Interaction Singlet and Strong CP Invariance},
  journal      = {Phys. Rev. Lett.},
  volume       = {43},
  pages        = {103--107},
  year         = {1979},
  doi          = {10.1103/PhysRevLett.43.103}
}

@article{Shifman:1979if,
  author       = {Shifman, M. A. and Vainshtein, A. I. and Zakharov, V. I.},
  title        = {Can Confinement Ensure Natural CP Invariance of Strong Interactions?},
  journal      = {Nucl. Phys. B},
  volume       = {166},
  pages        = {493--506},
  year         = {1980},
  doi          = {10.1016/0550-3213(80)90209-6}
}

@article{Dine:1981rt,
  author       = {Dine, Michael and Fischler, Willy and Srednicki, Mark},
  title        = {A Simple Solution to the Strong CP Problem with a Harmless Axion},
  journal      = {Phys. Lett. B},
  volume       = {104},
  pages        = {199--202},
  year         = {1981},
  doi          = {10.1016/0370-2693(81)90590-6}
}

@article{Zhitnitsky:1980tq,
  author       = {Zhitnitsky, A. R.},
  title        = {On Possible Suppression of the Axion Hadron Interactions},
  journal      = {Sov. J. Nucl. Phys.},
  volume       = {31},
  pages        = {260},
  year         = {1980},
  note         = {[Yad. Fiz. 31 (1980) 497]}
}

@article{Barr:1992qq,
  author       = {Barr, S. M. and Seckel, D.},
  title        = {Planck-Scale Corrections to Axion Models},
  journal      = {Phys. Rev. D},
  volume       = {46},
  pages        = {539--549},
  year         = {1992},
  doi          = {10.1103/PhysRevD.46.539}
}

@article{Kamionkowski:1992mf,
  author       = {Kamionkowski, Marc and March-Russell, John},
  title        = {Planck-Scale Physics and the Peccei-Quinn Mechanism},
  journal      = {Phys. Lett. B},
  volume       = {282},
  pages        = {137--141},
  year         = {1992},
  doi          = {10.1016/0370-2693(92)90492-M}
}

@article{Holman:1992us,
  author       = {Holman, R. and Hsu, S. D. H. and Kephart, T. W. and Kolb, E. W. and Watkins, R. and Widrow, L. M.},
  title        = {Solutions to the Strong CP Problem in a World with Gravity},
  journal      = {Phys. Lett. B},
  volume       = {282},
  pages        = {132--136},
  year         = {1992},
  eprint       = {hep-ph/9203206},
  archivePrefix= {arXiv},
  primaryClass = {hep-ph},
  doi          = {10.1016/0370-2693(92)90491-L}
}

@article{Kim:2008hd,
  author       = {Kim, Jihn E. and Carosi, Gianpaolo},
  title        = {Axions and the Strong CP Problem},
  journal      = {Rev. Mod. Phys.},
  volume       = {82},
  pages        = {557--602},
  year         = {2010},
  eprint       = {0807.3125},
  archivePrefix= {arXiv},
  primaryClass = {hep-ph},
  doi          = {10.1103/RevModPhys.82.557}
}

@article{Choi:2020rgn,
  author       = {Choi, Kiwoon and Im, Sang Hui and Shin, Chang Sub},
  title        = {Recent Progress in the Physics of Axions and Axion-Like Particles},
  journal      = {Ann. Rev. Nucl. Part. Sci.},
  volume       = {71},
  pages        = {225--252},
  year         = {2021},
  eprint       = {2012.05029},
  archivePrefix= {arXiv},
  primaryClass = {hep-ph},
  doi          = {10.1146/annurev-nucl-120720-031147}
}

@article{Randall:1992ut,
  author       = {Randall, Lisa},
  title        = {Composite Axion Models and Planck Scale Physics},
  journal      = {Phys. Lett. B},
  volume       = {284},
  pages        = {77--80},
  year         = {1992},
  doi          = {10.1016/0370-2693(92)91928-3}
}

@article{Redi:2016esr,
  author       = {Redi, Michele and Sato, Ryosuke},
  title        = {Composite Accidental Axions},
  journal      = {JHEP},
  volume       = {05},
  pages        = {104},
  year         = {2016},
  eprint       = {1602.05427},
  archivePrefix= {arXiv},
  primaryClass = {hep-ph},
  doi          = {10.1007/JHEP05(2016)104}
}

@article{Burgess:2023dow,
  author       = {Burgess, C. P. and Choi, Gongjun and Quevedo, Fernando},
  title        = {UV and IR effects in axion quality control},
  journal      = {JHEP},
  volume       = {03},
  pages        = {051},
  year         = {2024},
  eprint       = {2301.00549},
  archivePrefix= {arXiv},
  primaryClass = {hep-th},
  doi          = {10.1007/JHEP03(2024)051}
}

@article{Gherghetta:2025uds,
  author       = {Gherghetta, Tony and Murayama, Hitoshi and Qu{\'i}lez, Pablo},
  title        = {A High-Quality Composite Pati-Salam Axion},
  eprint       = {2505.08866},
  archivePrefix= {arXiv},
  primaryClass = {hep-ph},
  year         = {2025},
  doi          = {10.48550/arXiv.2505.08866}
}

@article{Loladze:2025uvf,
    author = "Loladze, Vazha and Platschorre, Arthur and Reig, Mario",
    title = "{Higher axion strings}",
    eprint = "2503.18707",
    archivePrefix = "arXiv",
    primaryClass = "hep-ph",
    doi = "10.1007/JHEP08(2025)182",
    journal = "JHEP",
    volume = "08",
    pages = "182",
    year = "2025"
}

@article{Choi:2003wr,
  author       = {Choi, Kiwoon},
  title        = {A QCD Axion from Higher Dimensional Gauge Field},
  journal      = {Phys. Rev. Lett.},
  volume       = {92},
  pages        = {101602},
  year         = {2004},
  eprint       = {hep-ph/0308024},
  archivePrefix= {arXiv},
  primaryClass = {hep-ph},
  doi          = {10.1103/PhysRevLett.92.101602}
}

@article{Reece:2024qv,
  author       = {Reece, Matthew},
  title        = {Extra-Dimensional Axion Expectations},
  journal      = {JHEP},
  volume       = {07},
  pages        = {130},
  year         = {2025},
  eprint       = {2406.08543},
  archivePrefix= {arXiv},
  primaryClass = {hep-ph},
  doi          = {10.1007/JHEP07(2025)130}
}

@article{Craig:2024cqs,
  author       = {Craig, Nathaniel and Kongsore, Marius},
  title        = {High-Quality Axions from Higher-Form Symmetries in Extra Dimensions},
  journal      = {Phys. Rev. D},
  volume       = {111},
  number       = {1},
  pages        = {015047},
  year         = {2025},
  eprint       = {2408.10295},
  archivePrefix= {arXiv},
  primaryClass = {hep-ph},
  doi          = {10.1103/PhysRevD.111.015047}
}

@article{Cheng:2001ys,
  author       = {Cheng, Hsin-Chia and Kaplan, David E.},
  title        = {Axions and a Gauged Peccei-Quinn Symmetry},
  eprint       = {hep-ph/0103346},
  archivePrefix= {arXiv},
  primaryClass = {hep-ph},
  year         = {2001}
}

@article{Choi:2002ps,
    author = "Choi, Ki-woon and Kim, Ian-Woo",
    title = "{One loop gauge couplings in AdS(5)}",
    eprint = "hep-th/0208071",
    archivePrefix = "arXiv",
    reportNumber = "KAIST-TH-02-18",
    doi = "10.1103/PhysRevD.67.045005",
    journal = "Phys. Rev. D",
    volume = "67",
    pages = "045005",
    year = "2003"
}

@article{Hebecker:2018ofv,
    author = "Hebecker, Arthur and Mikhail, Thomas and Soler, Pablo",
    title = "{Euclidean wormholes, baby universes, and their impact on particle physics and cosmology}",
    eprint = "1807.00824",
    archivePrefix = "arXiv",
    primaryClass = "hep-th",
    doi = "10.3389/fspas.2018.00035",
    journal = "Front. Astron. Space Sci.",
    volume = "5",
    pages = "35",
    year = "2018"
}

@article{Rey:1989mg,
    author = "Rey, Soo-Jong",
    title = "{The Axion Dynamics in Wormhole Background}",
    reportNumber = "PRINT-89-0095 (UC,SANTA-BARBARA)",
    doi = "10.1103/PhysRevD.39.3185",
    journal = "Phys. Rev. D",
    volume = "39",
    pages = "3185",
    year = "1989"
}

@article{Honecker:2013mya,
    author = "Honecker, Gabriele and Staessens, Wieland",
    title = "{On axionic dark matter in Type IIA string theory}",
    eprint = "1312.4517",
    archivePrefix = "arXiv",
    primaryClass = "hep-th",
    reportNumber = "MITP-13-082",
    doi = "10.1002/prop.201300036",
    journal = "Fortsch. Phys.",
    volume = "62",
    pages = "115--151",
    year = "2014"
}

@article{Buchbinder:2014qca,
    author = "Buchbinder, Evgeny I. and Constantin, Andrei and Lukas, Andre",
    title = "{Heterotic QCD axion}",
    eprint = "1412.8696",
    archivePrefix = "arXiv",
    primaryClass = "hep-th",
    doi = "10.1103/PhysRevD.91.046010",
    journal = "Phys. Rev. D",
    volume = "91",
    number = "4",
    pages = "046010",
    year = "2015"
}

@article{Haba:2008sd,
  author       = {Haba, Naoyuki and Matsumoto, Shigeki and Okada, Nobuchika and Yamashita, Toshifumi},
  title        = {Effective Potential of Higgs Field in Warped Gauge-Higgs Unification},
  journal      = {Prog. Theor. Phys.},
  volume       = {120},
  pages        = {77--98},
  year         = {2008},
  eprint       = {0802.3431},
  archivePrefix= {arXiv},
  primaryClass = {hep-ph},
  doi          = {10.1143/PTP.120.77}
}

@article{Choi:2010xn,
  author       = {Choi, Kiwoon and Kim, Ian-Woo and Shin, Chang Sub},
  title        = {Gauge Threshold Corrections in Warped Geometry},
  journal      = {New J. Phys.},
  volume       = {12},
  pages        = {075014},
  year         = {2010},
  eprint       = {1001.1473},
  archivePrefix= {arXiv},
  primaryClass = {hep-ph},
  doi          = {10.1088/1367-2630/12/7/075014}
}

@article{ArkaniHamed:2003wu,
  author       = {Arkani-Hamed, Nima and Cheng, Hsin-Chia and Creminelli, Paolo and Randall, Lisa},
  title        = {Extranatural Inflation},
  journal      = {Phys. Rev. Lett.},
  volume       = {90},
  pages        = {221302},
  year         = {2003},
  eprint       = {hep-th/0301218},
  archivePrefix= {arXiv},
  primaryClass = {hep-th},
  doi          = {10.1103/PhysRevLett.90.221302}
}

@article{Cox:2019rro,
  author       = {Cox, Peter and Gherghetta, Tony and Nguyen, Minh D.},
  title        = {A Holographic Perspective on the Axion Quality Problem},
  journal      = {JHEP},
  volume       = {01},
  pages        = {188},
  year         = {2020},
  eprint       = {1911.09385},
  archivePrefix= {arXiv},
  primaryClass = {hep-ph},
  doi          = {10.1007/JHEP01(2020)188}
}

@article{Petrossian-Byrne:2025mto,
    author = "Petrossian-Byrne, Rudin and Villadoro, Giovanni",
    title = "{Open string axiverse}",
    eprint = "2503.16387",
    archivePrefix = "arXiv",
    primaryClass = "hep-ph",
    doi = "10.1007/JHEP07(2025)049",
    journal = "JHEP",
    volume = "07",
    pages = "049",
    year = "2025"
}

@article{Randall:1999ee,
  author       = {Randall, Lisa and Sundrum, Raman},
  title        = {A Large Mass Hierarchy from a Small Extra Dimension},
  journal      = {Phys. Rev. Lett.},
  volume       = {83},
  pages        = {3370--3373},
  year         = {1999},
  eprint       = {hep-ph/9905221},
  archivePrefix= {arXiv},
  primaryClass = {hep-ph},
  doi          = {10.1103/PhysRevLett.83.3370}
}

@article{Gherghetta:2000qt,
  author       = {Gherghetta, Tony and Pomarol, Alex},
  title        = {Bulk Fields and Supersymmetry in a Slice of AdS},
  journal      = {Nucl. Phys. B},
  volume       = {586},
  pages        = {141--162},
  year         = {2000},
  eprint       = {hep-ph/0003129},
  archivePrefix= {arXiv},
  primaryClass = {hep-ph},
  doi          = {10.1016/S0550-3213(00)00392-8}
}

@article{Strassler:1992zr,
  author       = {Strassler, M. J.},
  title        = {Field Theory Without Feynman Diagrams: One-Loop Effective Actions},
  eprint       = {hep-ph/9205205},
  archivePrefix= {arXiv},
  primaryClass = {hep-ph},
  year         = {1992}
}

@article{Schubert:2001he,
  author       = {Schubert, Christian},
  title        = {Perturbative Quantum Field Theory in the String-Inspired Formalism},
  journal      = {Phys. Rept.},
  volume       = {355},
  pages        = {73--234},
  year         = {2001},
  eprint       = {hep-th/0101036},
  archivePrefix= {arXiv},
  primaryClass = {hep-th},
  doi          = {10.1016/S0370-1573(01)00013-8}
}

@article{Bastianelli:2002fv,
  author       = {Bastianelli, Fiorenzo and Zirotti, Alessandro},
  title        = {Worldline Formalism in a Gravitational Background},
  journal      = {Nucl. Phys. B},
  volume       = {642},
  pages        = {372--388},
  year         = {2002},
  eprint       = {hep-th/0205182},
  archivePrefix= {arXiv},
  primaryClass = {hep-th},
  doi          = {10.1016/S0550-3213(02)00605-1}
}

@article{Forman:1987xk,
  author       = {Forman, Robin},
  title        = {Functional Determinants and Geometry},
  journal      = {Invent. Math.},
  volume       = {88},
  pages        = {447--493},
  year         = {1987}
}

@article{Kirsten:2003py,
  author       = {Kirsten, Klaus and McKane, Alan J.},
  title        = {Functional Determinants by Contour Integration Methods},
  journal      = {Annals Phys.},
  volume       = {308},
  pages        = {502--527},
  year         = {2003},
  eprint       = {math-ph/0305010},
  archivePrefix= {arXiv},
  primaryClass = {math-ph},
  doi          = {10.1016/S0003-4916(03)00149-0}
}

@article{Witten:1996qb,
  author       = {Witten, Edward},
  title        = {Phase Transitions in M-Theory and F-Theory},
  journal      = {Nucl. Phys. B},
  volume       = {471},
  pages        = {195--216},
  year         = {1996},
  eprint       = {hep-th/9603150},
  archivePrefix= {arXiv},
  primaryClass = {hep-th},
  doi          = {10.1016/0550-3213(96)00161-3}
}

@article{Abe:2016tfq,
    author = "Abe, Yugo and Goto, Yuhei and Kawamura, Yoshiharu and Nishikawa, Yasunari",
    title = "{Conjugate boundary condition, hidden particles, and gauge-Higgs inflation}",
    eprint = "1608.06393",
    archivePrefix = "arXiv",
    primaryClass = "hep-ph",
    doi = "10.1142/S0217732316502084",
    journal = "Mod. Phys. Lett. A",
    volume = "31",
    number = "35",
    pages = "1650208",
    year = "2016"
}

@article{Agrawal:2025mke,
    author = "Agrawal, Prateek and Hook, Anson and Loladze, Vazha and Reig, Mario",
    title = "{Axion quality problem: keep calm and baryon}",
    eprint = "2510.07366",
    archivePrefix = "arXiv",
    primaryClass = "hep-ph",
    doi = "10.1007/JHEP03(2026)041",
    journal = "JHEP",
    volume = "03",
    pages = "041",
    year = "2026"
}

@article{Kim:1984pt,
    author = "Kim, Jihn E.",
    title = "{A COMPOSITE INVISIBLE AXION}",
    reportNumber = "SNUHE-84-02",
    doi = "10.1103/PhysRevD.31.1733",
    journal = "Phys. Rev. D",
    volume = "31",
    pages = "1733",
    year = "1985"
}

@article{Choi:1985cb,
    author = "Choi, Kiwoon and Kim, Jihn E.",
    title = "{DYNAMICAL AXION}",
    reportNumber = "SNUHE-84-05-REV, SNUHE-84-05",
    doi = "10.1103/PhysRevD.32.1828",
    journal = "Phys. Rev. D",
    volume = "32",
    pages = "1828",
    year = "1985"
}

@article{Kallosh:1995hi,
    author = "Kallosh, Renata and Linde, Andrei D. and Linde, Dmitri A. and Susskind, Leonard",
    title = "{Gravity and global symmetries}",
    eprint = "hep-th/9502069",
    archivePrefix = "arXiv",
    reportNumber = "SU-ITP-95-2",
    doi = "10.1103/PhysRevD.52.912",
    journal = "Phys. Rev. D",
    volume = "52",
    pages = "912--935",
    year = "1995"
}

@article{DiLuzio:2020wdo,
    author = "Di Luzio, Luca and Giannotti, Maurizio and Nardi, Enrico and Visinelli, Luca",
    title = "{The landscape of QCD axion models}",
    eprint = "2003.01100",
    archivePrefix = "arXiv",
    primaryClass = "hep-ph",
    reportNumber = "DESY 20-036, DESY-20-036",
    doi = "10.1016/j.physrep.2020.06.002",
    journal = "Phys. Rept.",
    volume = "870",
    pages = "1--117",
    year = "2020"
}

@article{Kim:1988ix,
    author = "Kim, Jihn E. and Lee, Ki-Myeong",
    title = "{The Scale Problem in Wormhole Physics}",
    reportNumber = "FERMILAB-PUB-88-095-T",
    doi = "10.1103/PhysRevLett.63.20",
    journal = "Phys. Rev. Lett.",
    volume = "63",
    pages = "20",
    year = "1989"
}

@article{Lillard:2017cwx,
  author       = {Lillard, Benjamin and Tait, Tim M. P.},
  title        = {A Composite Axion from a Supersymmetric Product Group},
  journal      = {JHEP},
  volume       = {11},
  pages        = {005},
  year         = {2017},
  eprint       = {1707.04261},
  archivePrefix= {arXiv},
  primaryClass = {hep-ph},
  doi          = {10.1007/JHEP11(2017)005}
}

@article{Gavela:2018paw,
  author       = {Gavela, M. B. and Ibe, M. and Quilez, P. and Yanagida, T. T.},
  title        = {Automatic Peccei-Quinn Symmetry},
  journal      = {Eur. Phys. J. C},
  volume       = {79},
  number       = {6},
  pages        = {542},
  year         = {2019},
  eprint       = {1812.08174},
  archivePrefix= {arXiv},
  primaryClass = {hep-ph},
  doi          = {10.1140/epjc/s10052-019-7079-2}
}

@article{Ardu:2020fck,
  author       = {Ardu, Marco and Di Luzio, Luca and Landini, Gabriele and Strumia, Alessandro and Teresi, Daniele and Wang, Jiong-Wei},
  title        = {Axion Quality from the (Anti)Symmetric of SU($N$)},
  journal      = {JHEP},
  volume       = {11},
  pages        = {090},
  year         = {2020},
  eprint       = {2007.12663},
  archivePrefix= {arXiv},
  primaryClass = {hep-ph},
  doi          = {10.1007/JHEP11(2020)090}
}

@article{Contino:2021udf,
  author       = {Contino, Roberto and Podo, Alessandro and Revello, Filippo},
  title        = {Chiral Models of Composite Axions and Accidental Peccei-Quinn Symmetry},
  journal      = {JHEP},
  volume       = {04},
  pages        = {180},
  year         = {2022},
  eprint       = {2112.09635},
  archivePrefix= {arXiv},
  primaryClass = {hep-ph},
  doi          = {10.1007/JHEP04(2022)180}
}

@article{Cox:2023lcv,
  author       = {Cox, Peter and Gherghetta, Tony and Paul, Arpon},
  title        = {A Common Origin for the QCD Axion and Sterile Neutrinos from SU(5) Strong Dynamics},
  journal      = {JHEP},
  volume       = {12},
  pages        = {180},
  year         = {2023},
  eprint       = {2310.08557},
  archivePrefix= {arXiv},
  primaryClass = {hep-ph},
  doi          = {10.1007/JHEP12(2023)180}
}

@article{Gherghetta:2025kff,
    author = "Gherghetta, T. and Murayama, H. and Noether, B. and Qu{\'\i}lez, P.",
    title = "{A High-Quality Axion from Exact SUSY Chiral Dynamics}",
    eprint = "2508.21813",
    archivePrefix = "arXiv",
    primaryClass = "hep-ph",
    month = "8",
    year = "2025"
}

@article{Babu:2002ic,
  author       = {Babu, K. S. and Gogoladze, Ilia and Wang, Kai},
  title        = {Stabilizing the Axion by Discrete Gauge Symmetries},
  journal      = {Phys. Lett. B},
  volume       = {560},
  pages        = {214--222},
  year         = {2003},
  eprint       = {hep-ph/0212339},
  archivePrefix= {arXiv},
  primaryClass = {hep-ph},
  doi          = {10.1016/S0370-2693(03)00490-7}
}

@article{Lee:2011dya,
  author       = {Lee, Hyun Min and Raby, Stuart and Ratz, Michael and Ross, Graham G. and Schieren, Rolf and Schmidt-Hoberg, Kai and Vaudrevange, Patrick K. S.},
  title        = {Discrete R Symmetries for the MSSM and Its Singlet Extensions},
  journal      = {Nucl. Phys. B},
  volume       = {850},
  pages        = {1--30},
  year         = {2011},
  eprint       = {1102.3595},
  archivePrefix= {arXiv},
  primaryClass = {hep-ph},
  doi          = {10.1016/j.nuclphysb.2011.04.009}
}

@article{Harigaya:2013vwa,
  author       = {Harigaya, Keisuke and Ibe, Masahiro and Schmitz, Kai and Yanagida, Tsutomu T.},
  title        = {Peccei-Quinn Symmetry from a Gauged Discrete R Symmetry},
  journal      = {Phys. Rev. D},
  volume       = {88},
  number       = {7},
  pages        = {075022},
  year         = {2013},
  eprint       = {1308.1227},
  archivePrefix= {arXiv},
  primaryClass = {hep-ph},
  doi          = {10.1103/PhysRevD.88.075022}
}

@article{Bhattiprolu:2021vdu,
  author       = {Bhattiprolu, Prudhvi N. and Martin, Stephen P.},
  title        = {High-Quality Axions in Solutions to the $\mu$ Problem},
  journal      = {Phys. Rev. D},
  volume       = {104},
  number       = {5},
  pages        = {055014},
  year         = {2021},
  eprint       = {2106.14964},
  archivePrefix= {arXiv},
  primaryClass = {hep-ph},
  doi          = {10.1103/PhysRevD.104.055014}
}

@article{Witten:1984dg,
  author       = {Witten, Edward},
  title        = {Some Properties of O(32) Superstrings},
  journal      = {Phys. Lett. B},
  volume       = {149},
  pages        = {351--356},
  year         = {1984},
  doi          = {10.1016/0370-2693(84)90422-2}
}

@article{Choi:1985je,
  author       = {Choi, Kiwoon and Kim, Jihn E.},
  title        = {Harmful Axions in Superstring Models},
  journal      = {Phys. Lett. B},
  volume       = {154},
  pages        = {393--396},
  year         = {1985},
  doi          = {10.1016/0370-2693(85)90416-5},
  note         = {[Erratum: Phys. Lett. B 156 (1985) 452]}
}

@article{Barr:1985hk,
  author       = {Barr, S. M.},
  title        = {Harmless Axions in Superstring Theories},
  journal      = {Phys. Lett. B},
  volume       = {158},
  pages        = {397--400},
  year         = {1985},
  doi          = {10.1016/0370-2693(85)90440-X}
}

@article{Svrcek:2006yi,
  author       = {Svrcek, Peter and Witten, Edward},
  title        = {Axions In String Theory},
  journal      = {JHEP},
  volume       = {06},
  pages        = {051},
  year         = {2006},
  eprint       = {hep-th/0605206},
  archivePrefix= {arXiv},
  primaryClass = {hep-th},
  doi          = {10.1088/1126-6708/2006/06/051}
}

@inproceedings{Cheng:2010pt,
    author = "Cheng, Hsin-Chia",
    title = "{Introduction to Extra Dimensions}",
    booktitle = "{Theoretical Advanced Study Institute in Elementary Particle Physics}: {Physics of the Large and the Small}",
    eprint = "1003.1162",
    archivePrefix = "arXiv",
    primaryClass = "hep-ph",
    doi = "10.1142/9789814327183_0003",
    pages = "125--162",
    year = "2011"
}

@article{Grossman:1999ra,
  author        = {Grossman, Yuval and Neubert, Matthias},
  title         = {Neutrino masses and mixings in non-factorizable geometry},
  journal       = {Phys. Lett. B},
  volume        = {474},
  number        = {3-4},
  pages         = {361--371},
  year          = {2000},
  doi           = {10.1016/S0370-2693(00)00054-X},
  eprint        = {hep-ph/9912408},
  archivePrefix = {arXiv},
  primaryClass  = {hep-ph}
}

@article{Bergshoeff:2000zn,
  author        = {Bergshoeff, Eric and Kallosh, Renata and Van Proeyen, Antoine},
  title         = {Supersymmetry in Singular Spaces},
  journal       = {JHEP},
  volume        = {10},
  pages         = {033},
  year          = {2000},
  doi           = {10.1088/1126-6708/2000/10/033},
  eprint        = {hep-th/0007044},
  archivePrefix = {arXiv},
  primaryClass  = {hep-th}
}

@article{Fujita:2001bd,
  author        = {Fujita, Tomoyuki and Kugo, Taichiro and Ohashi, Keisuke},
  title         = {Off-Shell Formulation of Supergravity on Orbifold},
  journal       = {Prog. Theor. Phys.},
  volume        = {106},
  pages         = {671--690},
  year          = {2001},
  doi           = {10.1143/PTP.106.671},
  eprint        = {hep-th/0106051},
  archivePrefix = {arXiv},
  primaryClass  = {hep-th}
}

@article{ArkaniHamed:2001tb,
    author = "Arkani-Hamed, Nima and Gregoire, Thomas and Wacker, Jay G.",
    title = "{Higher dimensional supersymmetry in 4D superspace}",
    eprint = "hep-th/0101233",
    archivePrefix = "arXiv",
    reportNumber = "HUTP-01/A004, LBNL-47410, UCB-PTH-01/02",
    doi = "10.1088/1126-6708/2002/03/055",
    journal = "JHEP",
    volume = "03",
    pages = "055",
    year = "2002"
}

@article{Marti:2001iw,
    author = "Marti, Daniel and Pomarol, Alex",
    title = "{Supersymmetric theories with compact extra dimensions in N=1 superfields}",
    eprint = "hep-th/0106256",
    archivePrefix = "arXiv",
    reportNumber = "UAB-FT-518",
    doi = "10.1103/PhysRevD.64.105025",
    journal = "Phys. Rev. D",
    volume = "64",
    pages = "105025",
    year = "2001"
}

@article{Linch:2002wg,
    author = "Linch, William D. and Luty, Markus A. and Phillips, Joseph",
    title = "{Five dimensional supergravity in N=1 superspace}",
    eprint = "hep-th/0209060",
    archivePrefix = "arXiv",
    reportNumber = "UMD-PP-03-008",
    doi = "10.1103/PhysRevD.68.025008",
    journal = "Phys. Rev. D",
    volume = "68",
    pages = "025008",
    year = "2003"
}

@article{Bigazzi:2019hav,
    author = "Bigazzi, Francesco and Caddeo, Alessio and Cotrone, Aldo L. and Di Vecchia, Paolo and Marzolla, Andrea",
    title = "{The Holographic QCD Axion}",
    eprint = "1906.12117",
    archivePrefix = "arXiv",
    primaryClass = "hep-th",
    doi = "10.1007/JHEP12(2019)056",
    journal = "JHEP",
    volume = "12",
    pages = "056",
    year = "2019"
}

@article{Choi:2025wog,
    author = "Choi, Gongjun and Gherghetta, Tony",
    title = "{An Extra-Dimensional Axion in a 5D Warped Orbifold GUT}",
    eprint = "2512.15666",
    archivePrefix = "arXiv",
    primaryClass = "hep-ph",
    doi = "10.1007/JHEP03(2026)169",
    journal = "JHEP",
    volume = "03",
    pages = "169",
    year = "2026"
}

@article{Choi:2022jqy,
    author = "Choi, Gongjun and Yanagida, Tsutomu T.",
    title = "{High Quality Axion in Supersymmetric Models}",
    eprint = "2209.09290",
    archivePrefix = "arXiv",
    primaryClass = "hep-ph",
    doi = "10.1007/JHEP12(2022)067",
    journal = "JHEP",
    volume = "12",
    pages = "067",
    year = "2022"
}

\end{document}